\documentclass[11pt,a4paper]{article}              

\usepackage[margin=25mm]{geometry}
\usepackage{colortbl}
\usepackage{fancyhdr}
\usepackage{amssymb,amsfonts,amsmath,amsthm}
\usepackage{natbib}
\usepackage{subfig}
\usepackage{graphicx}
\usepackage{multirow}
\usepackage{threeparttable}
\usepackage{booktabs, caption, makecell}

\usepackage[section]{placeins}
\usepackage{hyperref}
\usepackage{authblk}
	
\makeatletter
\AtBeginDocument{%
	\expandafter\renewcommand\expandafter\subsection\expandafter{%
		\expandafter\@fb@secFB\subsection
	}%
}
\makeatother

\graphicspath{{./},{./Figures/}}

\newcommand{\ve}[1]{\boldsymbol{#1}}			

\newcommand{\mat}[1]{\boldsymbol{#1}}			

\newcommand{\enum}{ , \, \dots \,,}

\newcommand{\prt}[1]{\left(#1\right)}			
\newcommand{\acc}[1]{\left\{#1\right\}}			
\newcommand{\bra}[1]{\left[ #1 \right]}			
\newcommand{\abs}[1]{\left| #1 \right|}			

\newcommand{\Prob}[1]{{\mathbb P}\left( #1 \right)}	

\newcommand{\ie}{{\em i.e.} }

\newcommand{\rev}[1]{{\color{black}#1}}

\begin{document}
\title{Active learning for structural reliability: survey, general framework and benchmark} 
	
\author[1]{M. Moustapha} \author[1]{S. Marelli}  \author[1]{B. Sudret}
	
\affil[1]{Chair of Risk, Safety and Uncertainty Quantification,
		
		ETH Zurich, Stefano-Franscini-Platz 5, 8093 Zurich, Switzerland}
	
\date{}
\maketitle
	
\abstract{Active learning methods have recently surged in the literature due to their ability to solve complex structural reliability problems within an affordable computational cost. These methods are designed by adaptively building an inexpensive surrogate of the original limit-state function. Examples of such surrogates include Gaussian process models which have been adopted in many contributions, the most popular ones being the efficient global reliability analysis (EGRA) and the active Kriging Monte Carlo simulation (AK-MCS), two milestone contributions in the field. In this paper, we first conduct a survey of the recent literature, showing that most of the proposed methods actually span from modifying one or more aspects of the two aforementioned methods. We then propose a generalized modular framework to build on-the-fly efficient active learning strategies by combining the following four ingredients or modules: surrogate model, reliability estimation algorithm, learning function and stopping criterion. Using this framework, we devise 39 strategies for the solution of $20$ reliability benchmark problems. The results of this extensive benchmark (more than $12,000$ reliability problems solved) are analyzed under various criteria leading to a synthesized set of recommendations for practitioners. These may be refined with a priori knowledge about the feature of the problem to solve, \ie dimensionality and magnitude of the failure probability.  This benchmark has eventually highlighted the importance of using surrogates in conjunction with sophisticated reliability estimation algorithms as a way to enhance the efficiency of the latter.\\[1em]

}
		
\maketitle
	
\section{Introduction}
Structural reliability analysis is a central tool for the design and assessment of complex engineering systems. Such systems are affected by uncertainties, which may arise from natural variability in their physical properties (\emph{e.g.}, material strength, manufacturing tolerances), operating conditions (\emph{e.g.}, variable loads, environmental conditions) or simply because of an incomplete or lack of knowledge (\emph{e.g.}, in the non-destructive assessment of existing structures). Structural reliability analysis aims at assessing the effects of such uncertainties, by estimating the associated failure probability with respect to some relevant limit states. In this paper, we consider a probabilistic setting, in which the uncertainties are represented through a set of random parameters $\ve{X} \in \mathcal{D}_{\ve{X}} \subset \mathbb{R}^{M}$ completely defined by their joint probability distribution function (PDF) $f_{\ve{X}}$. These parameters represent the state of the system, which can be evaluated through a so-called performance function (a.k.a. limit-state function), herein denoted by $g\prt{\ve{X}}$. By convention, the system is assumed to be in a failure (resp. safe) state when $g\prt{\ve{x}} \leq 0$ (resp. $g\prt{\ve{x}} > 0$). The probability of failure of the system can then be defined as
\begin{equation}\label{eq:Pf}
	P_f = \mathbb{P}\prt{g(\ve{X}) \leq 0} = \int_{\mathcal{D}_f} f_{\ve{X}}\prt{\ve{x}} \textrm{d}\ve{x}.
\end{equation}

This integration over an implicitly defined domain ${\mathcal{D}_f = \acc{\ve{x}:g\prt{\ve{x}}\leq 0}}$ is not straightforward to solve and has motivated the development of a rich variety of techniques \rev{\citep{Ditlevsen1996,LeMaire2009,Melchers2018}}. 
These techniques can be broadly grouped in several classes. These include \emph{approximation methods}, where the limit-state function is linearized (or otherwise approximated) around a so-called design point, \emph{e.g.}, the most probable failure point (MPFP) in a suitably transformed probabilistic input space. This step allows one to then derive (semi-)analytically an approximation of the failure probability. This class includes the well-known first-order and second-order reliability methods (FORM and SORM) \citep{Hasofer1974,Rackwitz78}. This family, however, is known to suffer severe limitations when the limit-state function is strongly non linear, or in the presence of multiple failure modes. 
A second class of methods, namely that of \emph{simulation techniques}, is widely used for the solution of Eq.~\ref{eq:Pf}. 
Monte Carlo simulation is certainly among the most widely-used methods in this category. 
It is known to be robust and unbiased, yet its convergence rate is extremely slow, especially when the target failure probability is small. 
This is problematic when the computational model used in the evaluation of the limit-state function is costly, a common occurrence when \emph{e.g.}, finite element analysis is involved. 
More advanced methods, based on variance-reduction techniques, are constantly being developed. A non-exhaustive list of the latter include importance sampling \citep{Melchers1989}, subset simulation \citep{Au2001}, directional simulation \citep{Ditlevsen1990}, line sampling \citep{Koutsourelakis2004} and asymptotic sampling \citep{Bucher2009}. 
Numerous variants of these methods have been introduced in the recent literature in an attempt to further accelerate their convergence rates, \emph{e.g.}, \citet{Papaioannou2016, Wang2019, GeyerSS2019}. However, the computational cost remains unaffordably high (\ie $\mathcal{O}(10^{3-4})$ model runs) when considering time-consuming computational models.

In the past decade, a different avenue that offers substantial savings in the computational budget, while retaining the favourable properties of simulation methods, has been explored in the reliability analysis literature: \textit{surrogate-model aided methods}. 
Surrogate models are inexpensive approximations of the original computational model, which have consistently shown superior performance when combined with the traditional simulation methods introduced earlier. Originally, simple polynomial response surface models (RSM) were built using a set of carefully designed computer experiments \citep{Faravelli1989, LeMaire1998}
These RSM were then used \emph{in lieu} of the original computational model to approximately solve Eq.~\ref{eq:Pf}. 
Borrowing from the machine learning community, this process has evolved into a more sophisticated methodology known as \emph{active learning} \citep{Bichon2008,Echard2011}. In active learning, the surrogate model is not used as a mere proxy of the original computational model, but as a tool to help explore the random input variable space efficiently. 
The idea is to start with an initial small set of model evaluations, known as the \emph{experimental design}, which is then sequentially enriched following a so-called \emph{learning function}. 
The latter aims at finding which model evaluation would bring the most useful information for the purpose of accurately assessing the failure probability of the system under consideration. 
This starts from the premise that in Eq.~\ref{eq:Pf}, only the sign of the limit-state is required to characterize the failure domain $\mathcal{D}_f$ in simulation-based reliability algorithms. 
The goal is then to approximate the limit-state surface as parsimoniously as possible (\emph{i.e.}, using the least number of model evaluations) to achieve the best possible accuracy for the estimated failure probability. 
In the past few years, an increasingly large number of contributions have been proposed in the field of active learning for reliability analysis. 
The most popular approaches are based on Kriging, a.k.a. Gaussian process modelling, owing to its built-in error measure. The most prominent examples are the efficient global reliability analysis (EGRA) proposed by \citet{Bichon2008} and the active Kriging Monte Carlo simulation (AK-MCS) developed by \citet{Echard2011}. 
The latter is a cornerstone of various methods derived incrementally by modifying one or another aspect of the AK-MCS algorithm \citep{Lelievre2018} and commonly referred to as \emph{AK methods}. For instance, replacing the Monte Carlo simulation part of the algorithm with importance sampling or subset simulation leads respectively to AK-IS \citep{Echard2013} or AK-SS \citep{Huang2016}. Similarly, other contributions have targeted the surrogate model type, introducing for instance support vector machines \citep{Hurtado2004a, Deheeger2007, Bourinet2011, BourinetHDR} or polynomial chaos expansions \citep{MarelliSS2018}.
A comprehensive overview of recent developments in active-learning based reliability analysis can be found in \citet{Teixeira2021}.

This paper aims at achieving two goals. The first is to provide an in-depth characterization of the current trends in active-learning-based reliability analysis through a comprehensive survey, following the footsteps of \citet{Teixeira2021}. In doing so, however we focus on highlighting common aspects in the numerous literature contributions. More specifically, we classify the methods with respect to the specific \textit{novelty} put forward in each contribution.

We then propose a generalized framework that summarizes the survey and puts the entire reviewed literature under a consistent formal umbrella.
This framework is built by combining non-intrusively four identified common ingredients of active learning-based methods: i. a surrogate model, ii. a reliability estimation algorithm, iii. a learning function and iv. a stopping criterion. 
In the second part of the paper, we then conduct the first-ever extensive benchmark of active learning methods considering, on the one hand, a collection of $20$ problems of diverse characteristics and on the other hand, a total of $39$ active learning schemes built by combining selected methods in each of the four aforementioned components. 

\rev{This benchmark is mainly aimed at illustrating how easily the proposed framework can be configured to reproduce a wide class of recently published studies, and hence only a limited number of methods per component was considered. The methods were selected for their maturity and ease of deployment, \emph{i.e.}, they do not require extensive tuning by the user and are relatively fast to run on standard workstations. For instance, only Kriging, polynomial chaos expansions and PC-Kriging surrogate models are considered in this study due to their prevalence in the reliability analysis literature and their off-the-shelf availability. Similarly the scope of the problems solved within the benchmark is limited to the ones typically considered in the reviewed active learning papers. More specifically, we do not consider time-variant (such as in \citet{Kroetz2020}), dynamic or extremely high-dimensional problems (\emph{i.e.}, in the order of hundreds) as they would require special treatment. For the former, dimensionality reduction techniques are often combined with surrogate modelling as in manifold learning \citep{Zheng2009}, active subspace method \citep{Constantine2014,Zhou2020} or in a more general setting as in \citet{Lataniotis2020}. Even though some of these methods may be used for mildly high dimensional problems, they are not considered in the survey or benchmark carried out in this paper.}

The large batch of analyses resulting from the selected methods is repeated $15$ times, to obtain statistically significant estimates on the stability of each method. This results in a set of over $12,000$ reliability analyses which allows us to validate, repeat and assess most of the methods introduced in the recent literature, and at the same time to explore a large portion of new methods and combinations that have not been published yet.
The results of this benchmark are used to give recommendations as to which type of methods performs the best generally or at least to be preferred given features of the reliability problem at hand. 

The remainder of the paper is organized as follows. Section~\ref{sec:LitRev} presents a literature review of the current state-of-the art in surrogate-modeling based reliability analysis.
Section~\ref{sec:Framework} introduces a generalized active learning reliability framework inferred from the literature review. 
In Section~\ref{sec:Benchmark}, an extensive comparative benchmark study is carried out on a wide class of methods and benchmark problems. Finally, recommendations and conclusions are given in Section~\ref{sec:Recommendations}~and~\ref{sec:Conclusion}.

\section{A short overview of recent literature}\label{sec:LitRev}
\subsection{Common rationale}
\begin{figure}[!ht]
	\centering
	\includegraphics[width=0.90\textwidth]{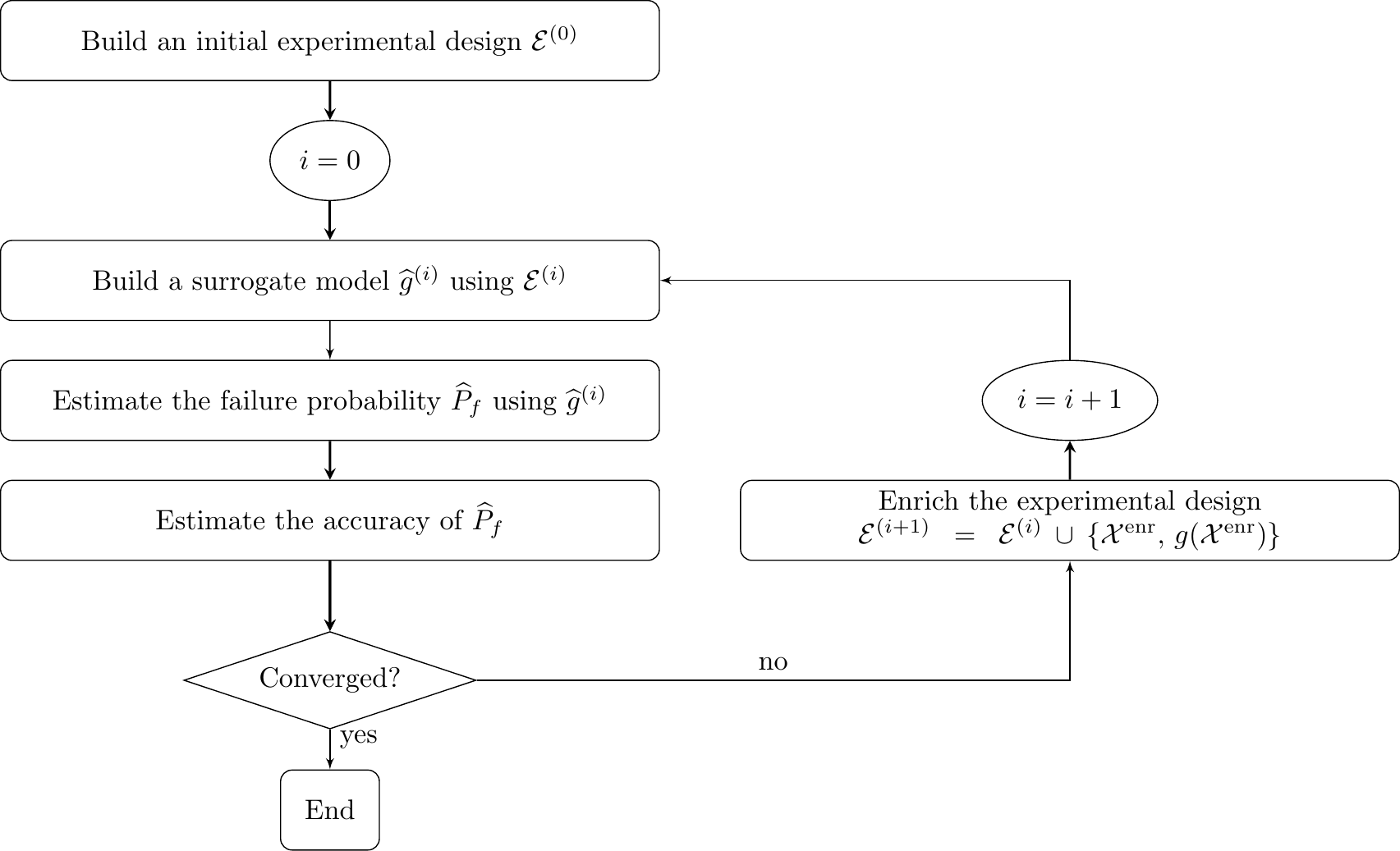}%
	\caption{General flowchart of active learning reliability.}
	\label{fig:ALRFlowchart}
\end{figure}

At the core of active-learning reliability lies the idea of reducing the cost of simulation algorithms by introducing a surrogate model as an inexpensive approximation of the  expensive-to-evaluate limit-state function. Surrogate models were first introduced in a static scheme to globally replace computer codes mainly for the purpose of visualization or optimization. Active learning pushes this concept further by aiming to an efficient allocation of resources, \emph{i.e.}, computer simulations are performed sparingly and only when most needed. Active learning reliability \rev{(ALR)} algorithms are practically devised using the general framework illustrated in Figure~\ref{fig:ALRFlowchart}. In the initialization step, a so-called \emph{experimental design} $\mathcal{E}^{(0)} = \acc{\prt{\ve{\mathcal{X}}^{\prt{i}}, \mathcal{Y}^{\prt{i}}}: \mathcal{Y}^{\prt{i}} = g\prt{\ve{\mathcal{X}}^{\prt{i}}} \in \mathbb{R}, \ve{\mathcal{X}}^{\prt{i}} \in \mathbb{X} \subset \mathbb{R}^M, i = 1 \enum m_0}$ is initially generated. The input sample set $\acc{\ve{\mathcal{X}}^{\prt{i}}, i = 1 \enum m_0}$ is often drawn using space-filling methods such as Latin hypercube sampling (LHS, \citet{McKay1979}) or randomized low-discrepancy sequences \citep{Sobol1967}. Typically $m_0$ is chosen small, \emph{i.e.}, in the order of tens of samples. Following initialization, the algorithm enters in a four-step loop where:
\begin{enumerate}
	\item A \textbf{surrogate model} is built using the current experimental design;
	\item The failure probability is estimated using the current surrogate model and an appropriate \textbf{reliability estimation} algorithm;
	\item The \textbf{convergence} of the algorithm is assessed;
	\item An \textbf{enrichment of the experimental design} is carried out by appropriately selecting at least one pair of sample points $\acc{\ve{\mathcal{X}}^{\prt{\textrm{enr}}},\, g\prt{\ve{\mathcal{X}}^{\prt{\textrm{enr}}}} }$, when convergence is not achieved. This is often achieved by evaluating a so-called \emph{learning function} which gives information as to which points are most likely to increase the accuracy of the surrogate (and subsequently of the estimated failure probability) when added to the experimental design.
\end{enumerate}

One of the first implementations of this flowchart was proposed by \citet{Bichon2008} in their efficient global reliability analysis method (EGRA). In this work, they used Gaussian process regression as a surrogate model, Monte Carlo simulation as reliability estimation algorithm and the so-called expected feasibility function (EFF) as a means to find points to enrich the experimental design. The latter is actually an adaptation to contour estimation of the well-known expected improvement (EI) function \citep{Ranjan2008} widely used in Bayesian optimization as first introduced in \citet{Jones1998}. A noticeable improvement of EGRA was introduced by \citet{Echard2011} in the widely known active Kriging - Monte Carlo simulation (AK-MCS) method. Contrary to EGRA, where reliability estimation is carried out only after the enrichment stage is completed, AK-MCS couples enrichment and reliability estimation. Furthermore, it introduces a new learning function, the so-called deviation number $U$, which is optimized with respect to a pre-defined sample set. This highly reduces the computational cost and the complexity of the active learning procedure. 

While this algorithm is at the time of writing already ten years old, it has aged surprisingly well, with a number of recent methods proposing only minor variations to one or more of the steps just reported. A comprehensive survey on recent developments on this topic was recently proposed by \citet{Teixeira2021}. In the following sections we identify and describe in more detail four key ingredients that are common to all of the aforementioned active-learning based methods.

\subsection{Surrogate models in structural reliability}\label{sec:LitRev:SM}
Various surrogate models were already used in adaptive schemes for the solution of reliability problems even before the emergence of the AK methods. Polynomial response surface models were arguably the first type of surrogates used in the context of structural reliability analysis. \citet{Faravelli1989} uses a second-order polynomial while \citet{Bucher1990} introduced a two-stage approach where a first quadratic response surface is used to locate the MPFP. A second response surface is then built close to that MPFP to refine the knowledge of the limit-state surface. A direct improvement of this approach which introduces an iterative procedure was proposed by \citet{Rajashekhar1993}. \rev{\citet{Leonel2011} considered various schemes using response surfaces for reliability analysis in crack propagation and concluded that a direct coupling (\emph{i.e.}, building the surrogate only once the MPFP has been located) is more efficient}. More recently, \citet{Roussouly2013} proposed an iterative scheme that combines trust regions, sparse response surface and bootstrap for the identification of regions where enrichment is necessary.  Radial basis functions (RBF), which have been popular in static surrogate-assisted reliability analysis, were introduced in a sequential approach as well. \citet{Li2018} presented an MCS-based approach where an RBF is sequentially updated through a constrained min-max optimization problem which aims at finding points close to the limit-state surface while keeping a minimum distance to the existing ED points. \citet{Shi2019} proposed two other learning schemes based on RBF considering either an ensemble of surrogates or cross-validation. In the former case, the interquartile range of the predictions using an ensemble of surrogates is considered as a measure of uncertainty to derive a learning function similar to the $U$-function of \citet{Echard2011}. Another learning function similar to $U$ was developped by \citet{MarelliSS2018} using bootstrap and polynomial chaos expansions (PCE). More recently, sparse Bayesian PCE was used by \citet{Cheng2020} where a new learning function relying on the Gaussian process variance was proposed. \citet{Pan2020} also used Bayesian regression PCE combined with the deviation number $U$. 

Popular methods from the machine learning community, such as support vector machines or neural networks, have also been steadily introduced in structural reliability. Support vector machines for classification was first introduced by \citet{Hurtado2004a}. \citet{Basudhar2010} proposed an adaptive scheme combining SVM classification and Monte Carlo simulation. The enrichment scheme is based on finding the point belonging to the limit-state surface approximation that is the furthest from the existing training points. This is a maximin problem solved using a general-purpose optimization algorithm. \citet{Lacaze2014} proposed an improvement of this maximin scheme by including a weight which accounts for the random variables joint PDF. Another improvement aiming at avoiding the optimization problem and relying on a candidate pool for enrichment has been proposed by \citet{Pan2017}. Combining SVM and subset simulation, \citet{Bourinet2011} proposed a learning scheme where a classifier is built in each iteration of the SuS algorithm. SVM has also been widely used in its regression form (SVR) for reliability analysis \citep{BourinetHDR}. \citet{Bourinet2017} proposed an SVR scheme with three novelties: i. the sample set size is kept constant, meaning some samples are withdrawn from the training set as the algorithm is proceeding, ii. intermediate thresholds are used to approximate the limit-state functions and iii. the surrogate models built in each stage are combined in a weighted ensemble to keep information of all training points without increasing the computation time. Another popular machine learning method widely used in structural reliability analysis is neural networks. Even though most of the contributions are in a static scheme, the most recent ones consider adaptivity \citep{Chojazyk2015,Kroetz2017}. \citet{Sundar2016} proposed a two-stage algorithm where an artificial neural network (ANN) is first used together with parallel Markov Chains to identify (possibly disjoint) failure regions. The ANN is then enriched to accurately represent the limit-state surface. Finally, \citet{Gomes2019} introduced an active scheme combining artificial neural networks and Monte Carlo simulation using the bootstrap-based learning function introduced in \citet{MarelliSS2018}.

Various other surrogate model types have been used to propose new active learning reliability algorithms, following similar schemes as introduced earlier, \emph{e.g.}, polynomial chaos-Kriging \citep{SchoebiASCE2016}, high-dimensional model reduction (HDMR) \citep{Sadoughi2017}, deep neural networks \citep{Li2020} or stochastic spectral embedding \citep{Wagner2021}, among others.

\subsection{Reliability estimation algorithm}\label{sec:LitRev:RA}
An immediate alternative strategy to AK-MCS can be devised by focusing on the reliability estimation algorithm. 
The benefits of replacing Monte Carlo simulation are two-fold. First, more sophisticated algorithms have been developed to reduce the variance of the failure probability estimate, and introducing them in active learning allows overcoming the pitfalls of MCS, \emph{i.e.}, its slow convergence rate.
Second, choosing another reliability estimation algorithm also allows one to modify the way sample candidates to enrichment are generated. 
In fact, for problems with low failure probability, the initial candidate set for enrichment may not contain any sample point at all in the actual failure domain. 
This can seriously reduce the chances of convergence of the active learning scheme. In contrast, more advanced reliability estimation algorithms may allow one to reach more easily areas associated with small probability densities, as well as disconnected failure regions.

Basically, almost all well-established simulation-based reliability estimation methods have been used together with active learning in the literature. 
A direct adaptation of AK-MCS, simply coined AK-IS, has been proposed by \citet{Echard2013} using importance sampling {\citep{Melchers2018}} in lieu of Monte Carlo simulation. In this contribution, they first find the design point using FORM and the original model. They then build an importance density sample set around this point which is used both for computing the failure probability and as candidate pool for enrichment. \citet{Gaspar2017} proposed to use a surrogate model even for the location of the design point, hence further reducing the computational cost. \citet{Zhao2015} did not rely on the design point but rather uses Monte Carlo Markov Chain (MCMC) to generate points in the failure domains. Importance sampling is then performed around those points together with enrichment. This allows overcoming a major shortcoming of importance sampling related to the presence of multiple design points. Another line of research involving Kriging combined with IS includes the meta-IS algorithm where \citet{Dubourg2012} proposed to use the Kriging model to approximate the optimal importance density in an iterative scheme. \citet{Cadini2014}  combined the work of \citet{Dubourg2012} and \citet{Echard2013} in a two-stage algorithm called metaAK-IS$^2$. Other sequential importance sampling methods have been adapted in an active Kriging strategy. For instance, \citet{Balesdent2013} sequentially built and enriched Kriging models in intermediate steps of a cross-entropy and non-parametric adaptive importance sampling algorithm. Other contributions using adaptive importance sampling include \citet{Gaspar2017, Razaaly2018, Yangetal2018, Zhang2018, Liu2019, Pan2020, Zhang2020}.

Another popular reliability estimation algorithm that has been used in an active learning scheme is subset simulation \citep{Au2003}. \citet{Huang2016} introduced AK-SS which, as its name suggests, is a declination of AK-MCS with the use of subset simulation for the computation of the failure probability. All other aspects are those of the original AK-MCS algorithm, including the candidate pool for enrichment which is obtained by an initial large Monte Carlo sample set. The obvious limitation here is that it may be difficult to find points in the failure region for problems where failure is an extremely rare event. \rev{\citet{Zhang2019b}} then proposed an improvement where the first and last levels of subset simulation are used as candidate pool for enrichment. The first level being a global Monte Carlo and the last one leading to points closest to the limit-state surface, this method allows both exploration and exploitation of the random input space. \citet{Ling2019} proposed an intermediate approach where a local Kriging model is built at each stage of subset simulation. Other similar methods include Bayesian subset simulation \citep{Li2012,Bect2017} which combines subset simulation, sequential Monte Carlo and Kriging and AK-SSIS \citep{Tong2015} which combines subset simulation and importance sampling in an active Kriging strategy.

Finally, we shall note that even though importance sampling and subset simulation have been widely exploited in active learning methods, the use of other variance-reduction simulation methods has been explored. Examples include algorithms such as directional importance sampling \citep{Guo2020}, radial basis importance sampling \citep{Bo2017} or line sampling \citep{Lv2015}.

\subsection{Enrichment of the experimental design}\label{sec:LitRev:LF}
A core feature of active-learning-based reliability methods is that the accuracy of the failure probability estimate is gradually increased by enriching the experimental design. A key component in this respect is the learning function (LF), which plays the central role of providing a measure of the information value of any experimental design enrichment candidates. 
Many authors have come up with new learning functions that can increase the efficiency of otherwise comparable methods. 
A direct improvement of the deviation number $U$ was given for instance by \citet{Peijuan2017}, where a line search step is added to get even closer to the limit-state surface once the best next point with respect to $U$ is found. Arguing that errors due to regions with small density would be negligible in the final estimate of the failure probability, \citet{Wen2016} also used the random variables joint PDF to constrain the EFF learning function. Similarly, \citet{Sun2017} proposed the least improvement function (LIF) which weights the probability of misclassification $\Phi\prt{-U \prt{\ve{x}}}$ with the joint probability density of the samples $f_{\ve{X}}\prt{\ve{x}}$, an idea already used in \citet{Dubourg2012}. 
\citet{Tong2019} followed this idea and, adding more terms related to global/local uncertainty, they created a new learning function.

From another perspective, \citet{Lv2015} introduced a new learning function based on the information theory with an analytical expression similar to EFF. 
\citet{Hu2016} introduced a method which relies on computing the sensitivities of the failure probability to add new points to the experimental design.
A new learning function using $K$-fold cross validation generalizable to any type of surrogate model was introduced by \citet{Xiao2018a}. 
More recently, \citet{Jiang2019} proposed an approach which is based on splitting the space using Voronoi cells and finding out those points with the largest sensitivities to the estimated failure probability. 
The use of Voronoi cells allows the authors to both reduce the computational cost and spread the sample points as much as possible.
The latter goal is also achieved through a pre-processing step by \citet{Zhang2019a} who then introduced a new LF inspired from the expected improvement.

\subsection{Stopping criterion}\label{sec:LitRev:SC}
The stopping criterion is an often overlooked yet crucial part of any active learning reliability algorithm. Three types of criteria have been proposed in the literature to halt the iterative enrichment scheme. The first one is directly based on the learning function. For instance, \citet{Bichon2008} proposed to stop the enrichment scheme when the value of the expected feasibility function is lower than $10^{-3}$. Similarly, \citet{Echard2011} stops AK-MCS iterations when $U>2$ for all candidate points. This actually means that the probability of misclassifying any point from the sample set used to evaluate the failure probability is below $2.28 \%$. This criterion has shown to be extremely conservative, leading to unnecessarily added points. Some authors have tried softening it either by considering the whole candidate set through an average, for instance \citet{Jian2017,Sun2017,Lelievre2018}, or by considering convergence when only a small proportion of the candidate set does not comply with $U>2$ \citep{MoustaphaSMO2016, Fauriat2014}. 

The second family of convergence criteria are those based on the accuracy of the failure probability. Using the Kriging variance, \citet{Dubourg2013} proposed a bound on the estimate $\widehat{P}_f$ which accounts for the Kriging epistemic uncertainty. Similarly, \citet{Sun2017,Jian2017} proposed an upper bound on $\abs{\widehat{P}_f - P_f}$ using the probability of misclassification $\Phi(-U)$. For surrogate models which do not possess a built-in error measure, similar bounds have been derived considering either cross-validation \citep{Shi2019} or bootstrap replicates \citep{MarelliSS2018}. 

Finally, the third family of stopping criteria has been built using the stabilization of either the limit-state surface or the failure probability estimates within enrichment iterations. \citet{Basudhar2010} tracked the fraction of some predefined convergence points that changed sign within two updates of an SVM model and assumed convergence when this fraction was relatively small. This criteria, often with slight adjustments, has been used in numbers of SVM-based active learning schemes \citep{BourinetHDR}. As for the failure probability, the obvious approach is to track its variation within iterations. Stabilization criteria may often lead to premature convergence when the initial surrogate model is extremely inaccurate. A workaround consists in tracking the convergence over several iterations, on average $2$ to $3$ and in some contributions and up to $10$ iterations \citep{Bourinet2017}. An alternative is to smooth out the convergence criterion as in \citet{Basudhar2010} by using an exponential curve fitted to the convergence criterion. 

\section{A generalized active learning reliability framework}\label{sec:Framework}

\subsection{Motivation}
As anticipated in the previous section, the state-of-the-art in active learning reliability can essentially be summarized into four basic components or modules. These modules are namely the surrogate model, the reliability estimation algorithm, the learning function and the stopping criterion. 
Most of the contributions in the recent literature can be reconstructed by combining various methods within each module. 
In a few cases, elaborate ad-hoc techniques are devised by taking advantage of some highly specific combination of such methods. 
The modules are in these cases not independent anymore but intrusively linked to each other. 
However, the advantages brought by such configurations are most often only marginal and not systematically justified by benchmarks.

In this section, we present a modular framework for active learning whose first aim is to unify and present the plethora of active learning reliability methods from a single and consistent viewpoint. 
The interest of such an approach can be seen through the prism of the no-free lunch principle. 
Despite their claimed advantages, methods proposed in the literature perform best under certain conditions and do not generalize so well as to provide consistently superior performance in the wide spectrum of structural reliability problems. 
By framing active learning reliability under a modular framework, we can then take advantage of each method to solve a wide class of reliability problems, possibly identifying guidelines based on limited prior information, such as the problem dimensionality.

A second advantage of the framework we propose is that it decouples the four modules. 
This independence means that there is no need to alter or adapt a given method/module to fit in the overall workflow. 
Methods can be used solely through their input/output structures, as "black boxes" even allowing for a seamless interconnection to third-party software.

\begin{figure}[!ht]
	\centering
	\includegraphics[width=0.90\textwidth]{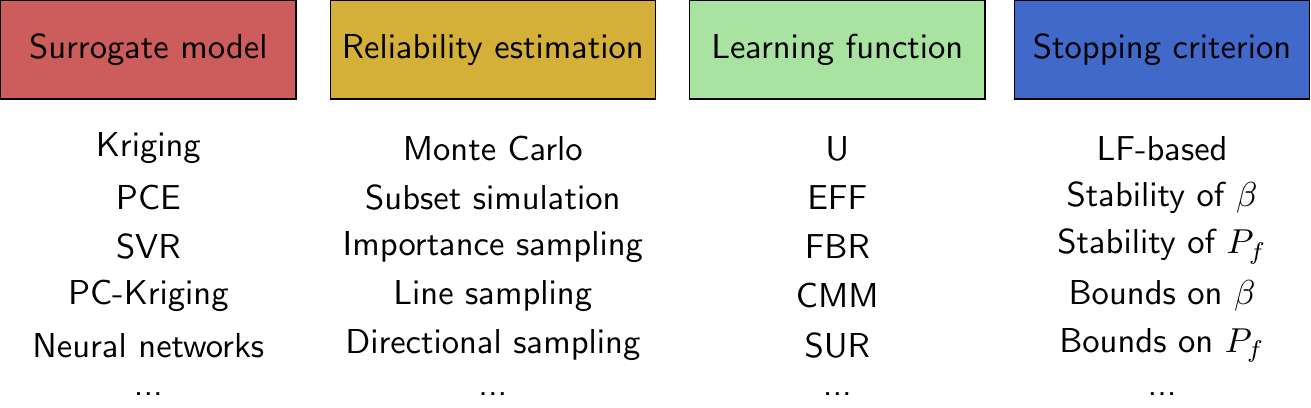}%
	\caption{Active learning reliability framework with example of methods.}
	\label{fig:ALRFramework}
\end{figure}

The core idea of the framework is depicted in Figure~\ref{fig:ALRFramework}. 
The four modules are shown in columns with examples of popular methods. Most of the contributions surveyed in the previous section can be retrieved by appropriately combining the methods. For instance, combining all methods on the first row, \emph{i.e.}, Kriging, Monte Carlo simulation, deviation number $U$ and the LF-based stopping criterion, leads to the well-known AK-MCS. 
In principle, all methods within each module block can be combined with any from the other blocks. 
The only exceptions are from the methods that specifically rely on the surrogate built-in error, \emph{e.g.}, the Kriging variance. 
It should be noted however that alternatives have been proposed in the literature to estimate comparable local error measures when not directly provided by the surrogate model itself.  

\subsubsection{Surrogate models}
Surrogate models lie at the core of any active learning reliability algorithm. 
A relevant aspect that needs to be stressed here is that they are merely used as a tool to explore the random input space in conjunction with the original computational model. 
They are not meant to replace the latter \emph{per se}. 
Many surrogates have been adopted in the literature for active learning and can be further classified on the basis of different properties. 
One way is to consider interpolation \emph{vs.} regression approaches. 
The former are often preferred in active learning schemes as they allow to precisely approximate the limit-state function in the vicinity of any point that belongs to the experimental design. 
One may also consider a classification with regards to the availability of a built-in error measure. 
Built-in errors have been instrumental in the proliferation of active learning schemes, because most enrichment schemes rely on them. 
Kriging for instance, with its local variance estimator, is arguably the most adopted surrogate in recent active learning contributions. 
However, as shown in the literature review above, alternatives also exist and have even shown to be quite efficient. Such alternatives may include a similar error measure (either through statistical methods such as bootstrap and cross-validation or plug-in methods) or the use of alternative learning functions, as will be explained shortly.

\subsubsection{Reliability estimation algorithm}\label{sec:Framework:RelAlg}
Despite approximation methods have been sometimes used in early surrogate-assisted reliability methods \citep{Bucher1990,Rajashekhar1993}, most of the recent contributions rely on simulation-based methods. 
Monte Carlo simulation, thanks to its generality, is naturally one of the most commonly-used methods. 
The use of variance-reduction techniques such as subset simulation or importance sampling has also been widely explored. 
The latter can reduce the error due to the random nature of the sampling algorithm while keeping the number of model evaluations as low as possible, especially when the probability of failure is low.

In this contribution, we advocate for going even one step further by \rev{using an ``overkill setting"} of the reliability estimation algorithm, further capitalizing on the negligible computational costs associated with the use of surrogate models. \rev{By ``overkill setting", we mean that the parameters of the algorithms are tuned to drastically reduce the coefficient of variation of the resulting failure probability estimate. This set of parameters depends on the reliability estimation algorithm. When using subset simulation for instance, the batch sample size and conditional failure probability are made larger than in the settings classically found in the literature: we choose here $10^5$ samples per simulation step and an intermediate probability of $0.25$ instead of the usual $0.10$ value.}

This approach has a two-fold benefit. First, the stochastic error due to the reliability estimation algorithm is reduced as much as possible, hence leaving only the surrogate-induced error to dominate the global estimation uncertainty on the failure probability. 
It should be noted that the overall computational time is of course increased, especially when Kriging is used. 
However, this overhead is expected to be marginal when compared to that of an actual computational model, \emph{e.g.}, a finite element analysis. 
Second, by over-calibrating the reliability algorithm, we allow for the random space to be even more thoroughly explored, hence increasing the likelihood of finding sample points in the failure regions when the latter is considerably small. 
Examples of such settings will be shown in the benchmark section and compared to more traditional settings. 

\subsubsection{Learning function}
The learning function is used as a driver to add new points in the experimental design. 
It is often intrinsically linked to both the surrogate model and the reliability estimation algorithm. 
Indeed, its very definition often draws from the characteristics of the surrogate model, \emph{e.g.}, variance or built-in error measure. 
This does not need to be the case systematically, as the same features can be replaced by statistical methods that provide comparable error metrics, such as bootstrap and cross-validation or even mere distance measures to the existing experimental design points.

Generally, new candidate enrichment points are obtained by minimizing (or maximizing) the learning function over the input domain. 
The optimization problem is most often simplified into a discrete approximation where the enrichment samples are chosen from a finite candidate pool. \rev{In most of the literature (starting from the original AK-MCS algorithm \citep{Echard2011} and in most subsequent variants), the candidate pool is defined prior to the analysis using Monte Carlo sampling. However this is not an optimal approach. For instance, it is difficult to find points in the failure domain when the real failure probability is very small (\emph{i.e.}, $P_f < 10^{-6}$) as this would require an extremely large candidate pool. 
	An alternative and more efficient approach proposed here is to define the candidate pool as the set of samples generated by the chosen reliability estimation algorithm, which can in principle be updated throughout the ALR iterations. This allows us to fully exploit the benefits of more advanced reliability estimation algorithms that are more likely to locate the (multiple) failure domains and sample more points where needed most.}
To further avoid being intrusive, we consider as candidate pool for enrichment all samples that were used to estimate the failure probability in the previous iteration of the algorithm. 
To accelerate the procedure, it is possible to statistically reduce the size of the candidate pool by simple down-sampling or clustering. 
The latter approach can also serve as a way of simultaneously identifying multiple enrichment points so as to take advantage of any available parallelization capability. 

\subsubsection{Stopping criterion}
The stopping criterion is an important part of the active learning scheme as the efficiency of the algorithm is ultimately and largely driven by its robustness. 
Too loose a stopping criterion can lead to premature convergence, while a too strict one would cause the unnecessarily addition of costly experimental design points. 
The criteria proposed in the literature can be classified into two groups.
First are those based on the learning function value, \emph{e.g.}, \citet{Bichon2008, Echard2011}. 
These have shown to often be extremely conservative. 
The second family is derived by directly monitoring the accuracy of the estimated failure probability.
Confidence bounds on the latter can be derived \citep{Dubourg2012, SchoebiASCE2016}, and convergence is assumed when such bounds are small enough. Alternatively, one may monitor their evolution and assume convergence when a certain degree of stability is observed. 
Finally, increased robustness may be achieved by either combining different stopping criteria and/or considering convergence only when the criteria are satisfied consistently within a given number of  consecutive iterations.

\section{Comparative study}\label{sec:Benchmark}
\subsection{Benchmark set-up}
The ingredients shown in the previous section can be assembled non-intrusively to build active learning schemes.
In this section we perform an extensive comparison of several framework configurations on a set of benchmark reliability problems representative of a wide range of real case applications.
We selected such configurations by considering some of the most widely-used methods in each module. Table~\ref{tab:Strategies} shows the different algorithms considered for the benchmark in this paper. The first part of the table deals with methods that use the built-in surrogate model variance, while the second is based on a regression method and bootstrap error estimation. All possible combinations resulting from the tensor product of each compatible ingredient are considered. This amounts in a total of $39$ strategies ($36$ for the first surrogate class and $3$ for the second). \textsc{UQLab} \citep{MarelliUQLab2014}, a \textsc{Matlab} framework for uncertainty quantification was used to run these analyses.
\begin{table}[!ht]
	\centering
	\caption{Methods selected in each module to create the $39$ solution strategies used in the benchmark. Further details about each method are given in Sections~\ref{sec:AlSet} and in \rev{the supplementary materials (Appendix~\ref{sec:Suppl})}.}
	\label{tab:Strategies}
	\begin{tabular}{cccc}
		\hline
		Reliability & Metamodel & Learning function & Stopping criterion  \\ \hline
		Monte Carlo simulation & \multirow{2}{*}{Kriging} & \multirow{2}{*}{U} & Beta bounds \\
		Subset simulation &  &  &  Beta stability \\
		Importance sampling & \multirow{-2}{*}{PC-Kriging} & \multirow{-2}{*}{EFF} & Combined \\ \hline
		Monte Carlo simulation & \multirow{3}{*}{PCE} & \multirow{3}{*}{FBR} & \multirow{3}{*}{Beta stability} \\
		Subset simulation  \\
		Importance sampling \\ \hline		
	\end{tabular}
\end{table}

\begin{figure}[!ht]
	\centering
	\includegraphics[width=0.65\textwidth]{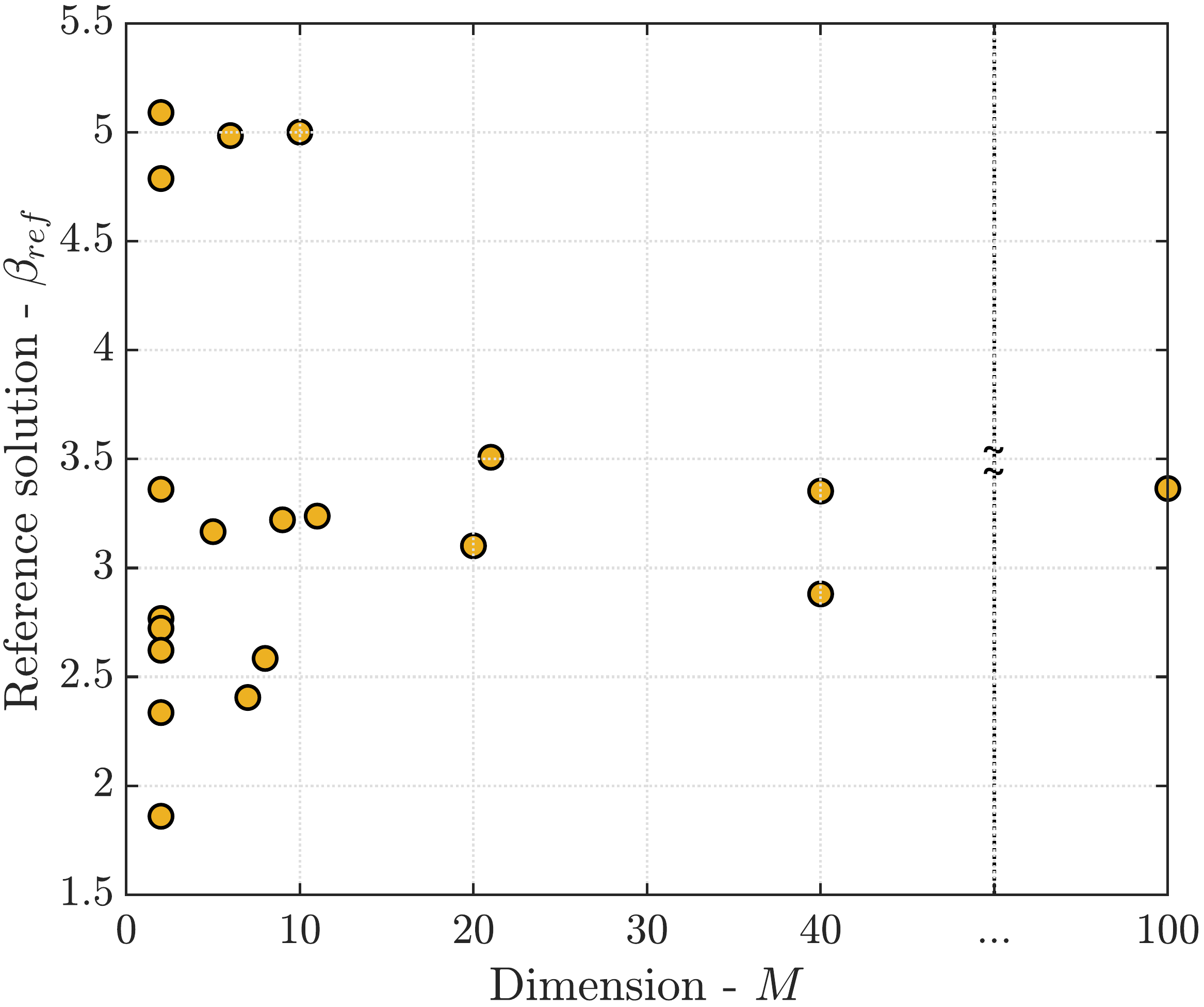}
	\caption{Collection of problems selected for the benchmark in terms of dimensions $M$ and reference reliability indices $\beta_{\textrm{ref}}$.}
	\label{fig:Dim_vs_Betaref}
\end{figure}

Using these 39 strategies, a collection of $20$ reliability problems are solved. $11$ of these problems were collected from the TNO reliability benchmark repository \citep{Rozsas2019}. The remaining were chosen from the literature with the aim of ensuring a large variety both in terms of limit-state function dimensionality and reference failure probability/reliability index. Most of the limit-state functions are analytical, except for two which are based on a truss finite element model. A list of the problems together with some references is given in \rev{the supplementary materials (Appendix~\ref{sec:Suppl})}. Figure~\ref{fig:Dim_vs_Betaref} summarizes these problems in terms of dimension against reference reliability index. They range from input dimension $M=2$ to $M=100$ and have a reliability index (resp. failure probability) that ranges from $\beta_{\textrm{ref}} \approx 1.86$ (resp. $P_{f,\textrm{ref}} \approx 3.14 \cdot 10^{-2}$) to $\beta_{\textrm{ref}} \approx 5.15$ (resp. $P_{f,\textrm{ref}} \approx 1.32 \cdot 10^{-7}$). The reference solutions are calculated using the original models and a large Monte Carlo set whose size is set adaptively until a coefficient of variation of $1\%$ is reached. For the four problems that have a failure probability smaller than $10^{-7}$, an overcalibrated subset simulation is used instead.

Throughout the benchmark, each analysis is repeated $15$ times. The only exception being benchmark $\#8$  of dimension $100$, which is repeated only $10$ times. It should be noted that within different strategies, the same initial random conditions/seeds are used for each of the repetitions. Hence, a total of $11,700$ reliability analyses ($39$ strategies x $20$ problems x $15$ repetitions) are carried out for this benchmark.

\subsection{Algorithmic settings}\label{sec:AlSet}
\begin{figure}[!ht]
	\centering
	\includegraphics[width=0.95\textwidth]{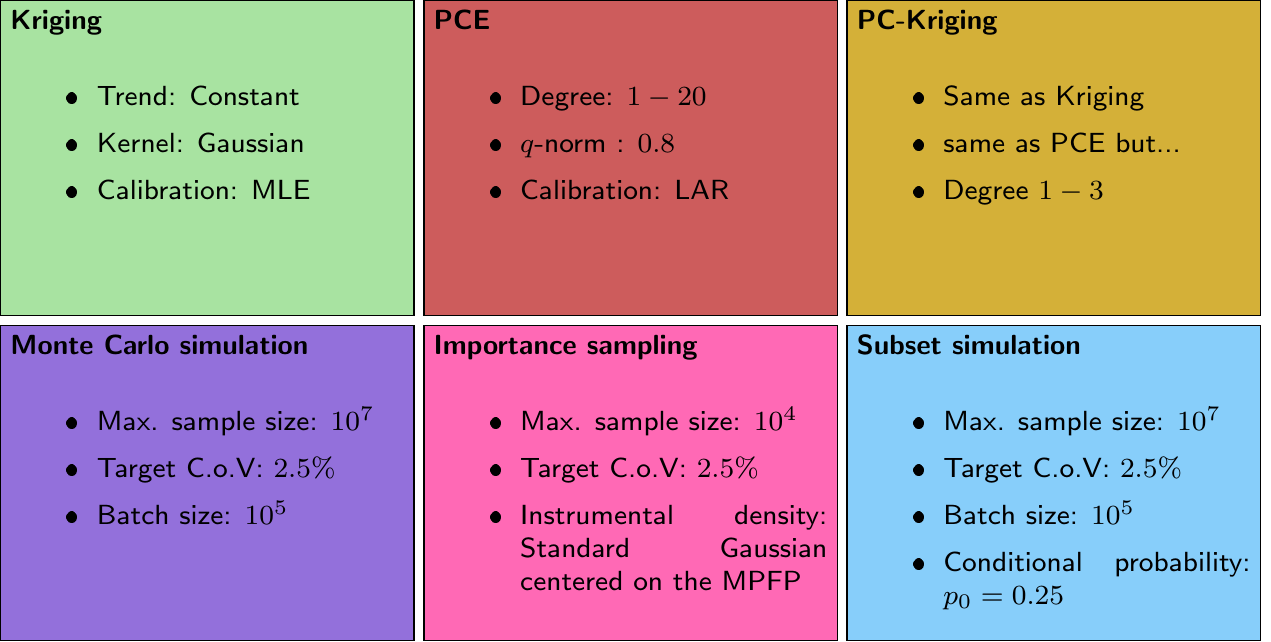}
	\caption{A summary of the most important settings for the surrogate models and reliability estimation algorithms considered in this paper. The meaning of each of these parameters can be found in \rev{the supplementary materials (Appendix~\ref{sec:Suppl})}.}
	\label{fig:AlSet}
\end{figure}

In this section, we will briefly review the algorithm settings used for each of the methods selected in Table~\ref{tab:Strategies}. 
The methods selected for the first three components, \emph{i.e.}, surrogate models, reliability estimation algorithm and learning functions are detailed in \rev{the supplementary materials (Appendix~\ref{sec:Suppl})}. 
A summary of the most important settings for the surrogate models and reliability estimation algorithms is given in Figure~\ref{fig:AlSet}. 
The learning functions however do not possess any special setting and are used exactly as described in \rev{the supplementary materials}. 

This section therefore focuses on the three stopping criteria mentioned in Table~\ref{tab:Strategies}:
\paragraph{Beta bounds:} This stopping criterion is based on the Kriging variance and reads:
\begin{equation}\label{eq:Eps_BB}
	\frac{\abs{\widehat{\beta}^{+} - \widehat{\beta}^{-}}}{\widehat{\beta}} \leq \bar{\varepsilon}_{BB}
\end{equation}
where $\widehat{\beta}^{+}$ and $\widehat{\beta}^{-}$ are the reliability indices respectively obtained using the limit-state functions $\mu_{\widehat{g}}\prt{\ve{x}} - 2 \sigma_{\widehat{g}}\prt{\ve{x}}$ and $\mu_{\widehat{g}}\prt{\ve{x}} + 2 \sigma_{\widehat{g}}\prt{\ve{x}}$, while $\widehat{\beta}$ is the reliability index obtained using the limit-state $\mu_{\widehat{g}}\prt{\ve{x}}$. 

The threshold $\bar{\varepsilon}_{BB}$ is set to $0.01$ which is arguably a relatively large value. However, convergence is assumed only when this criterion is respected three times in a row, hence ensuring some degree of robustness.

\paragraph{Beta stability}
This convergence criterion ensures the stability of the failure probability estimate assuming that convergence is achieved when adding new points do not noticeably modify the estimate. Using the reliability index, it reads:
\begin{equation}\label{eq:Eps_BS}
	\frac{\abs{\widehat{\beta}^{(i)} - \widehat{\beta}^{(i-1)}}}{\widehat{\beta}^{(i)}} \leq \bar{\varepsilon}_{BS},
\end{equation}
where $\widehat{\beta}^{(i)}$ represents the estimated reliability index at the $i$-th iteration. 

The threshold is set to $\bar{\varepsilon}_{BS} = 0.005$ and convergence is considered only when this criterion is respected within three consecutive iterations.

\paragraph{Combined stopping criterion:} This stopping criterion is simply a combination of the previous two. Convergence is assumed when the two criteria in Eq.~\ref{eq:Eps_BB}~and~\ref{eq:Eps_BS} are met within two consecutive iterations.

\subsubsection{Other common settings}
Beside the method-specific settings introduced in the previous paragraph, others, which are common to all methods, need to be defined. 
These are mainly related to the initial experimental design which is drawn using the Latin hypercube sampling  (LHS) method \citep{McKay1979}. 
The number of initial ED points is set to $\max(10, 2 M)$, where $M$ is the problem dimensionality. 
This allows one to ensure a minimum of $10$ points for low-dimensional problem while at the same time making sure that there are enough sample points w.r.t. the dimension when the latter increases. 
Similarly at the other end of the spectrum, the  number of sample points is limited and only a maximum of $100 + 10 M$ points can be added during the enrichment process. This number appears realistic for classical costly simulators used in engineering.

\subsection{Criteria for the evaluation of the strategies}
To properly compare different reliability analysis strategies, a performance measure needs to be defined. 
Due to the inherent complexity of the problem, we compare our benchmark results in terms of several different measures that take into account different performance metrics.
Perhaps the most straightforward measures are i. how close the reliability estimate is to the reference and ii. how many points are needed to reach it. 
Focusing on the reliability index rather than the failure probability, a first criterion can be simply computed using the following relative reliability error estimator:
\begin{equation}
	\label{eq:Relerr}
	\varepsilon_{\beta_{i,j}}^{\prt{k}} = \abs{\frac{\widehat{\beta}_{i,j}^{\prt{k}}  - \beta_{\textrm{ref},j}}{\beta_{\textrm{ref},j}}},
\end{equation}
where $\widehat{\beta}_{i,j}^{\prt{k}}$ denotes the reliability index resulting from the $k$-th replication of the $i$-th strategy applied to the $j$-th problem and $\beta_{\textrm{ref},j}$ is the reference solution for the $j$-th problem. As a reminder, this benchmark comprises a set of $20$ problems solved using $39$ strategies, each repeated $15$ times.

The criterion presented in Eq.~\ref{eq:Relerr} measures the accuracy of the resulting reliability index estimate. 
Additionally, we need to also assess the efficiency of the method, which simply relates to the number of model evaluations $N_{\textrm{eval}}$ necessary to converge. 
The lower $N_{\textrm{eval}}$, the better the strategy. 
However $N_{\textrm{eval}}$ alone is not a sufficient measure of the strategy efficiency, as premature convergence may occur. 
To avoid this, we will consider only solutions whose relative error, as computed in Eq.~\ref{eq:Relerr}, is below a threshold arbitrarily set at $0.05$ when ranking w.r.t. $N_{\textrm{eval}}$. 
Those with larger error will be automatically ranked in the last position, regardless of the number of model evaluations needed to converge. 

Ideally, both criteria should be as low as possible, but they are by construction conflicting. 
Finding the best approach would therefore mean finding a good trade-off between relative error and computational cost. 
We therefore propose here a third criterion that combines these two criteria in one, making the ranking easier:
\begin{equation}
	\label{eq:err_delta}
	\Delta_{i,j}^{\prt{k}} = \varepsilon_{\beta_{i,j}}^{\prt{k}} \frac{N_{{\textrm{eval}},{i,j}}^{\prt{k}}}{N_{\textrm{med},j}},
\end{equation}
where $N_{\textrm{med},j}$ is the median number of model evaluations considering all strategies, repetitions included, to solve the $j$-th problem (in  total, there are $15 \times 39 = 585$ runs for each problem).

All the $39$ strategies are compared with each other and a ranking is established based on the three criteria defined above. 
The main goal of such a ranking is to find out if one strategy is consistently better than the others. 
If no such strategy were to be found, next is to find whether there are methods within each module that are consistently better than the others.
Finally, the ranking will be used to assess whether given methods are better when applied to a specific feature of the problem at hand, \emph{i.e.}, dimensionality and failure probability magnitude.

\subsection{Methods ranking over different problems}

\subsubsection{Ranking of the strategies}
At first, we compare different strategies considering all the criteria previously defined, to which we add as reference two plain simulation methods, \emph{i.e.}, importance sampling and subset simulation (the total number of reliability analysis runs becomes then $12,300$). They  provide us with reference results without surrogates and serve as a benchmark baseline. For importance sampling, the MPFP is found using FORM and the MCS sample set is of size $10^3$. For subset simulation, the subset sample set is of size $10^3$ while $p_0$ is set to $0.1$. 

\begin{figure}[!ht]
	\centering
	\includegraphics[width=0.74\textwidth]{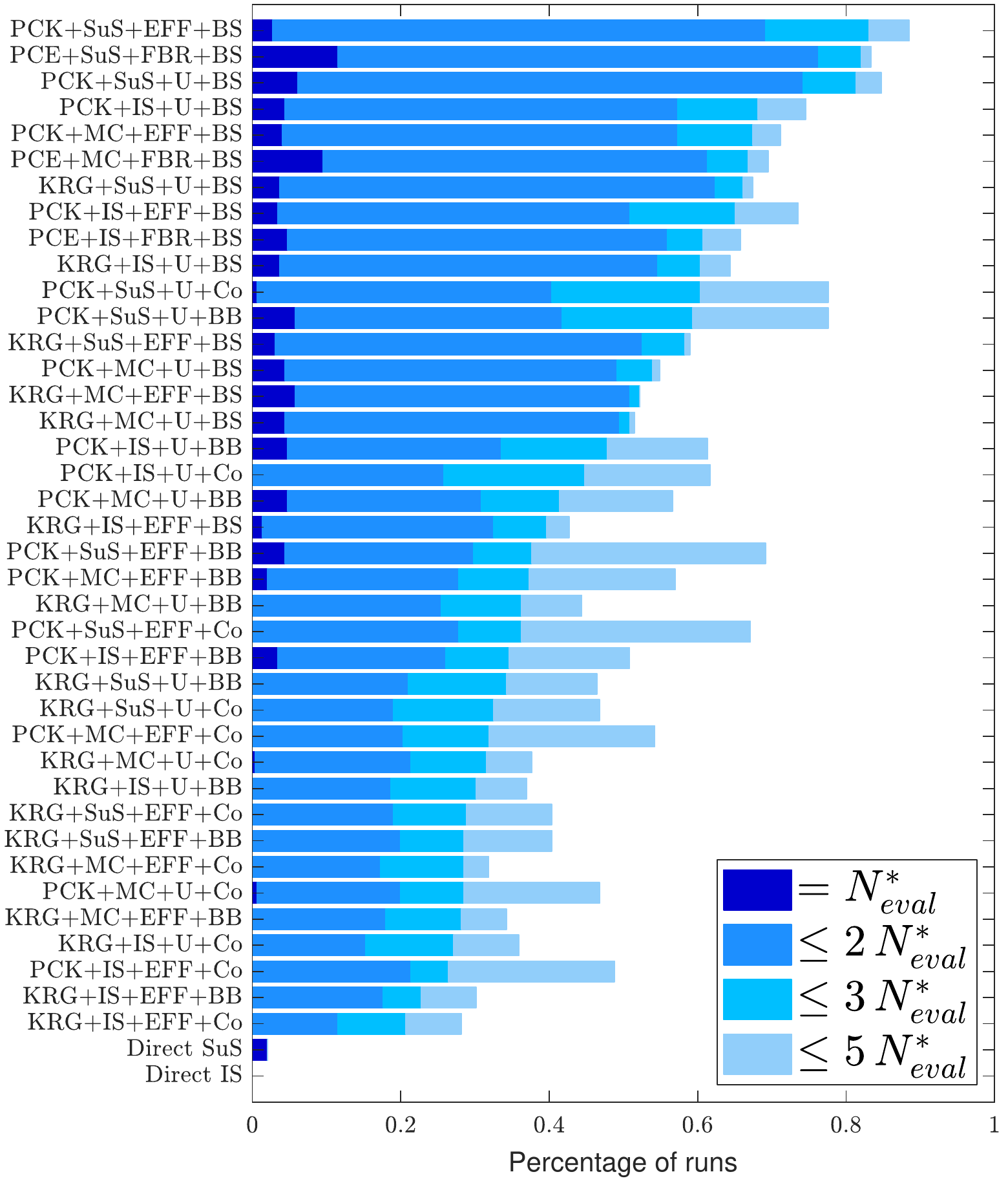}
	\caption{\rev{Ranking of the strategies w.r.t. $N_{\textrm{eval}}$. The overall ranking is based on the number of times the method performs within $3 \, N_{\textrm{eval}}^{\ast}$.}}
	\label{fig:RankingNeval}
\end{figure}

\begin{figure}[!ht]
	\centering
	\includegraphics[width=0.74\textwidth]{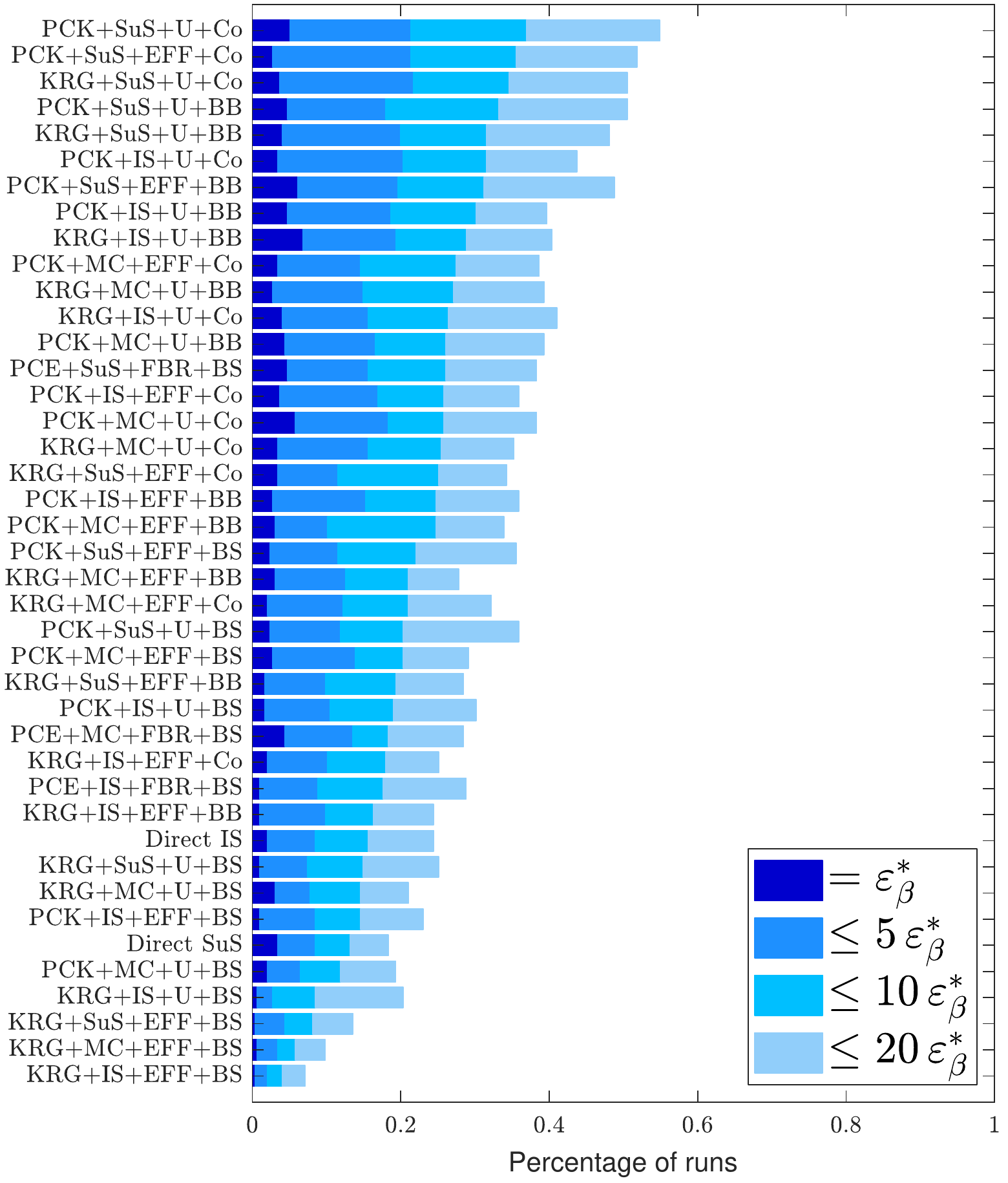}
	\caption{\rev{Ranking of the strategies w.r.t. $\varepsilon_{\beta}$. The overall ranking is based on the number of times the method performs within $10 \, \varepsilon_{\beta}^{\ast}$ (one order of magnitude).}}
	\label{fig:RankingRelerr}
\end{figure}

\begin{figure}[!ht]
	\centering
	\includegraphics[width=0.74\textwidth]{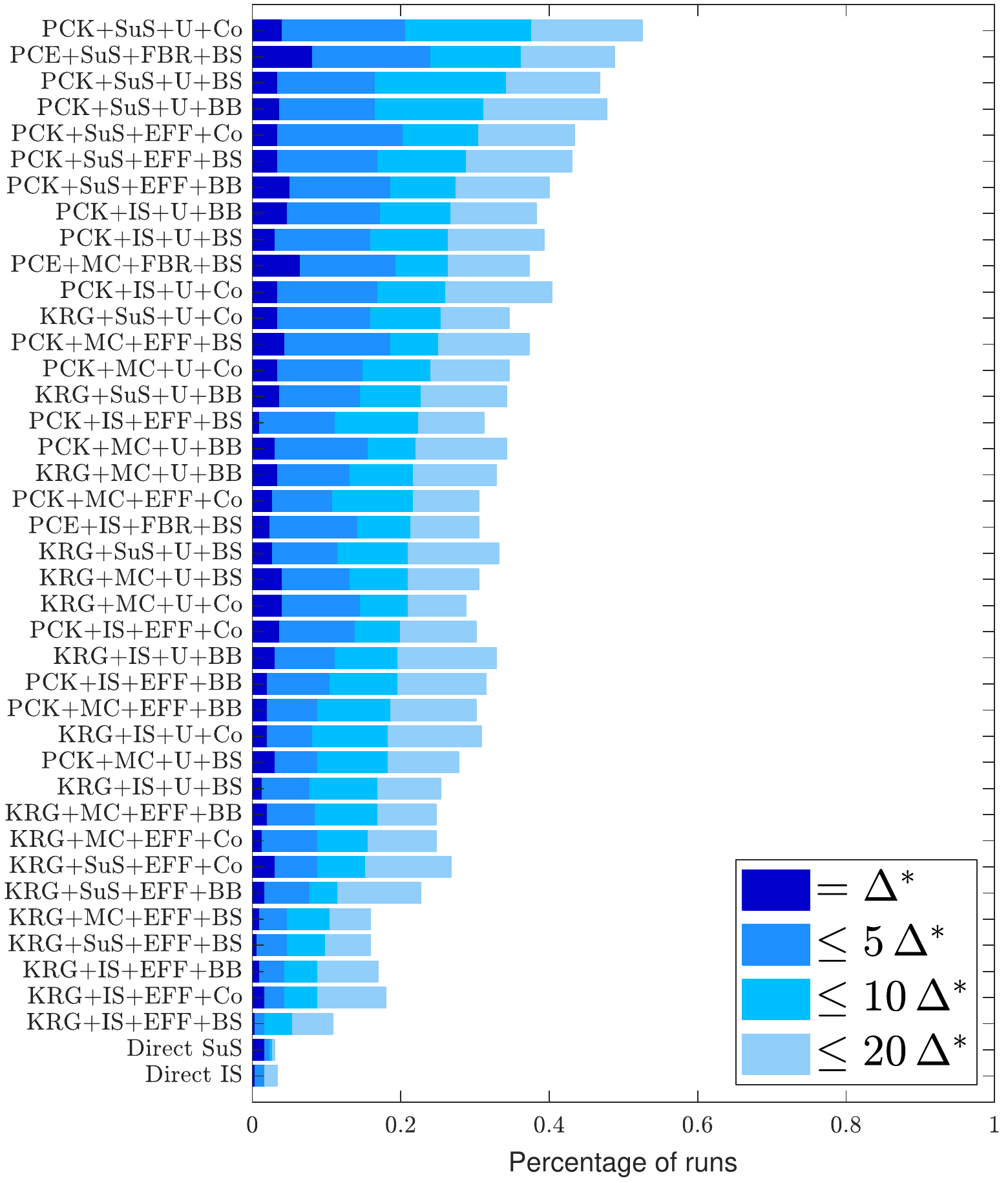}%
	\caption{\rev{Ranking of the strategies w.r.t. $\Delta$. The overall ranking is based on the number of times the method performs within $10 \, \Delta^\ast$ (one order of magnitude).}}
	\label{fig:RankingDelta}
\end{figure}

Additionally to the ranking, the robustness of each strategy is assessed. For each problem and replication, we observe whether a given strategy is within a certain distance from the best solution w.r.t. a chosen criterion. For the criterion ``number of model evaluations", the distance is set to \rev{$\acc{2, \, 3, \, 5} \times N_{\textrm{eval}}^{\ast}$} where $N_{\textrm{eval}}^{\ast}$ is the smallest number of model evaluations among strategies whose relative error is below the threshold of $0.05$. For the ``relative error" (resp. $\Delta$-criterion), the distance is measured as \rev{$\acc{5, \, 10, \, 20} \times \varepsilon_{\beta}^{\ast}$ (resp. $\acc{5, \, 10, \, 20} \times \Delta^{\ast}$)} where $\varepsilon_{\beta}^{\ast}$ (resp. $\Delta^{\ast}$) is the smallest relative error (resp. $\Delta$ value) among all strategies. This count is aggregated over all problems and replications (in total, $20 \times 15 = 300$ analyses) and given in terms of percentage as illustrated by the \rev{bars} in Figures~\ref{fig:RankingNeval},~\ref{fig:RankingRelerr}~and~\ref{fig:RankingDelta}. \rev{The mid-distance (\emph{i.e.}, $3 \, N_{\textrm{eval}}^{\ast}$, $10 \, \varepsilon_{\beta}^{\ast}$ and $10 \, \Delta^{\ast}$)} is used to rank the methods in these figures (the best solutions are in the upper positions).

In general, \rev{the larger the bars}, the more robust and accurate the associated method is. \rev{For instance looking at the first line of Figure~\ref{fig:RankingNeval}, the second bar shows that for the combination PCK + SuS + EFF + $\beta$-stability criterion and considering all the problems, the number of model evaluations required to converge in $69\%$ of the repetitions is below $2$ times the smallest number of model evaluations achieved for each given problem}. The largest bar shows that this ratio increases to $88\%$ when considering a threshold within $5$ times the best achieved number of model evaluations. \rev{Finally, in each of these figures, the smallest and darkest bar represents the percentage of times a strategy was ranked first for a given experimental design.} 

\paragraph{Ranking with respect to the number of model evaluations\\} The first criterion we consider is the number of model evaluations, whose results are shown in Figure~\ref{fig:RankingNeval}. As expected, when considering only the number of model evaluations, the direct solutions (\emph{i.e.}, subset simulation and importance sampling without the use of surrogates) rank last. The most robust and efficient solution w.r.t. the number of model evaluations is the combination of PCK with subset simulation, EFF and $\beta$-stability stopping criterion. Overall, considering the $10$ best solutions, PCE or PC-Kriging as surrogates, subset simulation as reliability estimation algorithm and $\beta$-stability as stopping criterion seem to dominate. Regarding the learning function, there is no clear top performer as they all appear at least twice in the first ten positions. 

\paragraph{Ranking with respect to the relative error\\}
The next criterion we consider is the relative error as illustrated in Figure~\ref{fig:RankingRelerr}. 
Here the direct solutions (without surrogates) are better ranked than with the previous criterion but they still rank worse than more than half of the methods considered. 
The better performance of surrogate-based methods is due to the ``overkill" setup of the reliability solvers used in conjunction with the surrogates. 
As explained in Section \ref{sec:Framework:RelAlg}, the computational efficiency of the surrogates allows one to use reliability solver configurations that maximize accuracy and minimize the stochastic uncertainty in the reliability estimator, without the traditional trade-offs associated.
This shows that the use of surrogate models as an instrument for the exploration of the random input space can lead to results at least as equally accurate as a direct solution, \emph{i.e.}, without the use of surrogates.

Regarding the best strategy, we can observe a few differences with the previous ranking. 
The overall most robust and efficient strategy is the combination of PC-Kriging with subset simulation, deviation number $U$ and the combined stopping criterion.
PCE is not so well classified when focus is put solely on accuracy. PC-Kriging is dominating the top ranking with a few occurences of Kriging.
As far as reliability estimation algorithm and learning function are concerned, subset simulation and the deviation number $U$ are preferred. 
Finally, regarding the stopping criterion, $\beta$-stability (Eq.~\ref{eq:Eps_BS}) seems not to favor accuracy, in sharp contrast to the combined criterion, which now appears in the top performing combinations.
\paragraph{Ranking with respect to the $\Delta$-criterion\\}
The last criterion considered is $\Delta$ (Figure~\ref{fig:RankingDelta}, Eq.~\ref{eq:err_delta}), which as expected results in a combination of the two previous rankings.
First the direct solutions are penalized by their relatively large number of model evaluations and rank again in the last two positions. 
Overall, this criterion favours solution accuracy, because the relative error in Eq.~\ref{eq:Relerr} can vary orders of magnitudes, while the range of variation of the number of model evaluations is not equally large (remember that the allowed number of model evaluations is limited to $100 + 10 M$). 
Therefore the latter can only help make a difference within strategies that already lead to roughly the same accuracy. 

As evidenced by  Figures~\ref{fig:RankingNeval},~\ref{fig:RankingRelerr}~and~\ref{fig:RankingDelta}, there is not a single strategy that outperforms all others in every benchmark.
However, some trends are emerging from these results. 
The next section dives deeper into the specific methods selected for each module.

\subsubsection{Ranking of the methods within each module}
\begin{figure}[!ht]
	\centering
	\subfloat[Surrogate model]{\label{fig:RankingStacked_Delta_a}\includegraphics[width=0.49\textwidth]{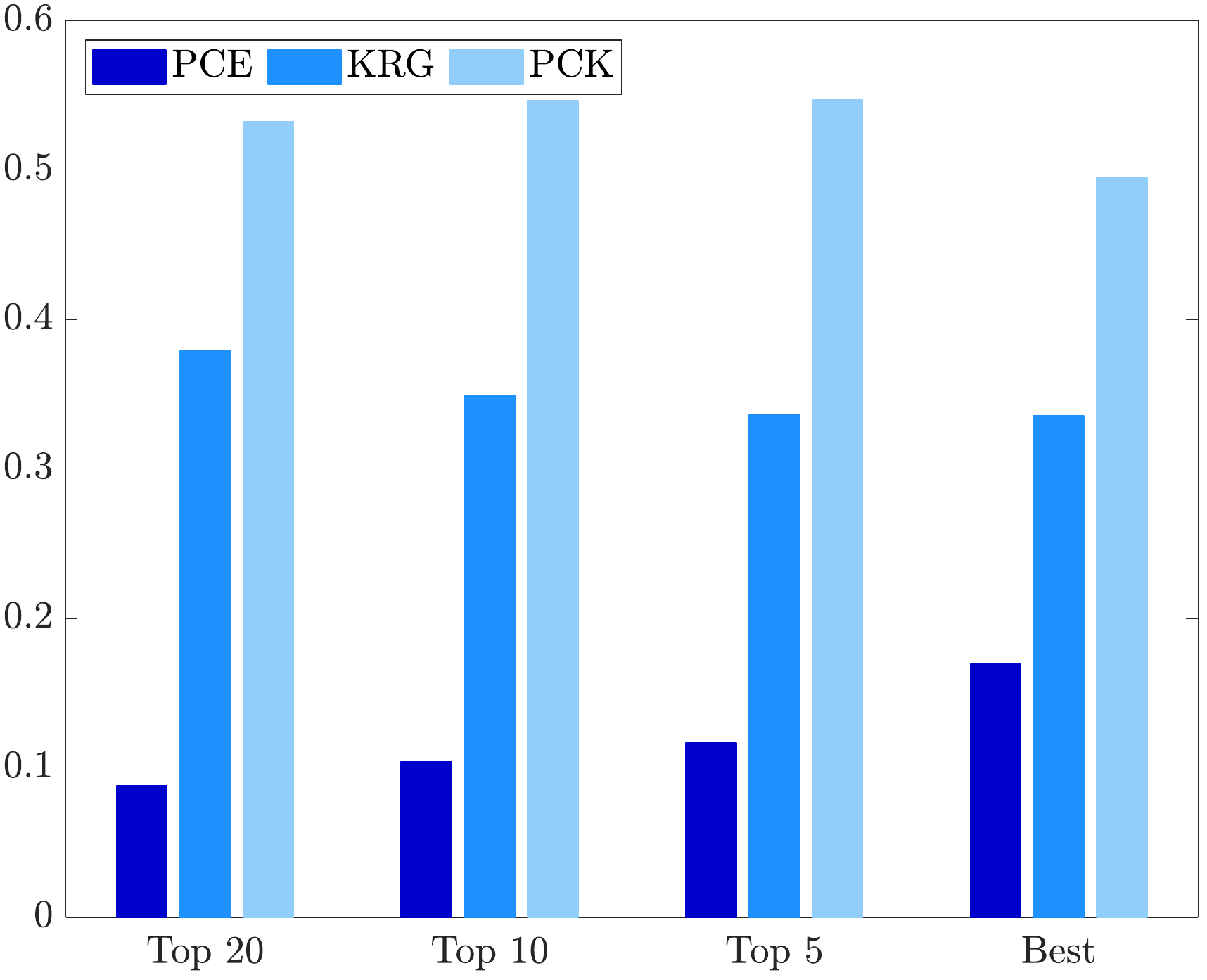}}%
	\hfill
	\subfloat[Reliability estimation algorithm]{\label{fig:RankingStacked_Delta_b}\includegraphics[width=0.49\textwidth]{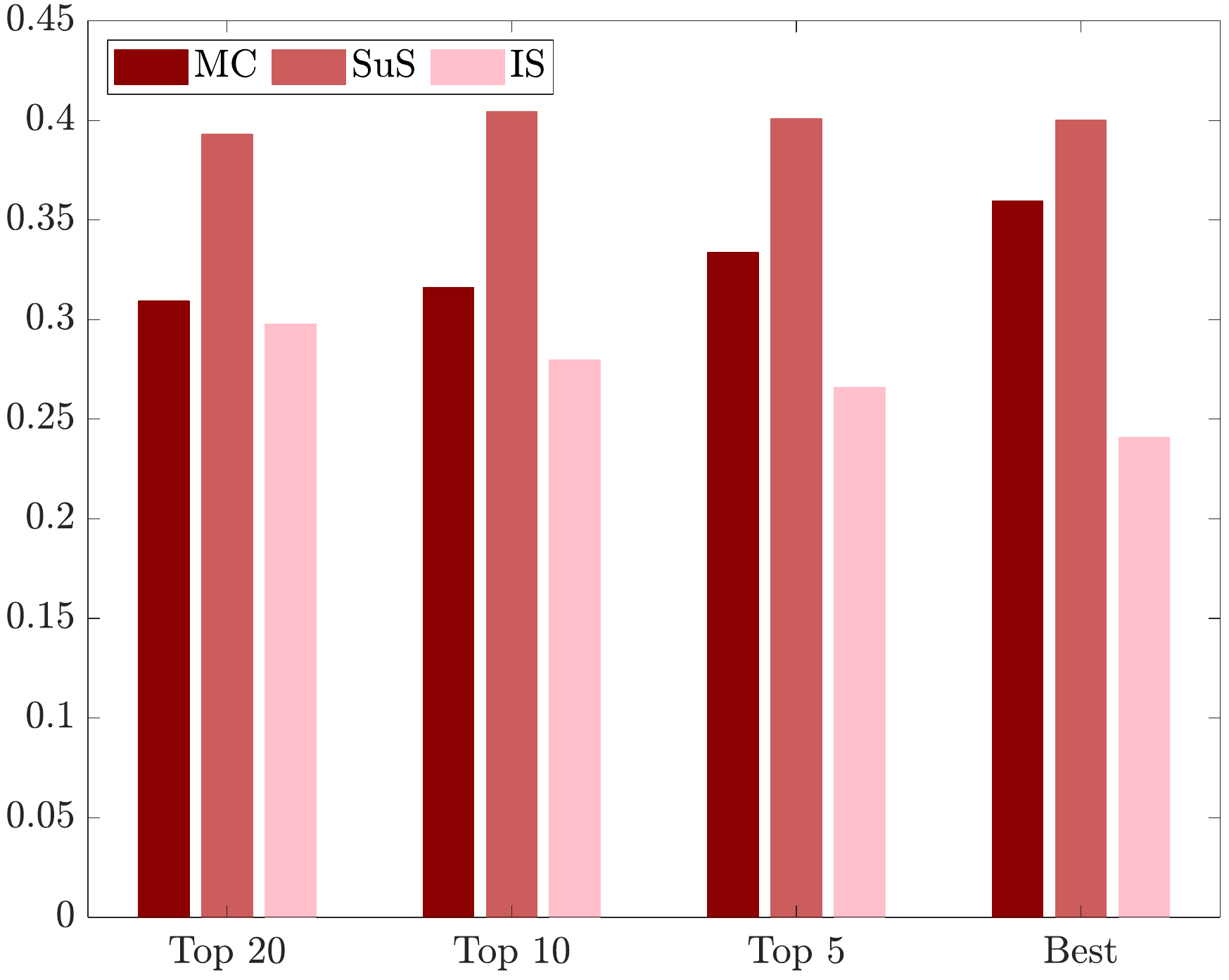}}%
	\\
	\subfloat[Learning function]{\label{fig:RankingStacked_Delta_c}\includegraphics[width=0.49\textwidth]{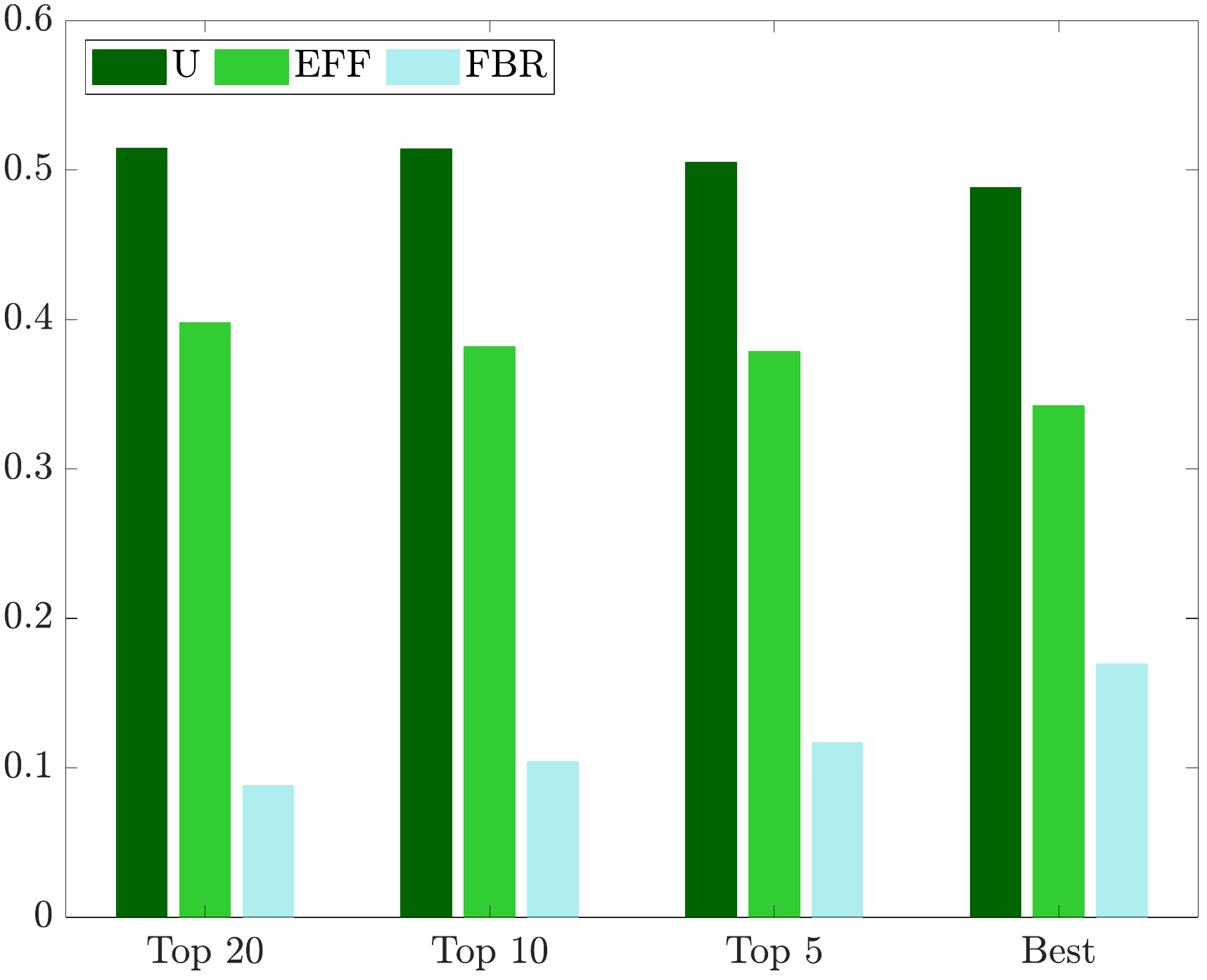}}%
	\hfill
	\subfloat[Stopping criterion]{\label{fig:RankingStacked_Delta_d}\includegraphics[width=0.49\textwidth]{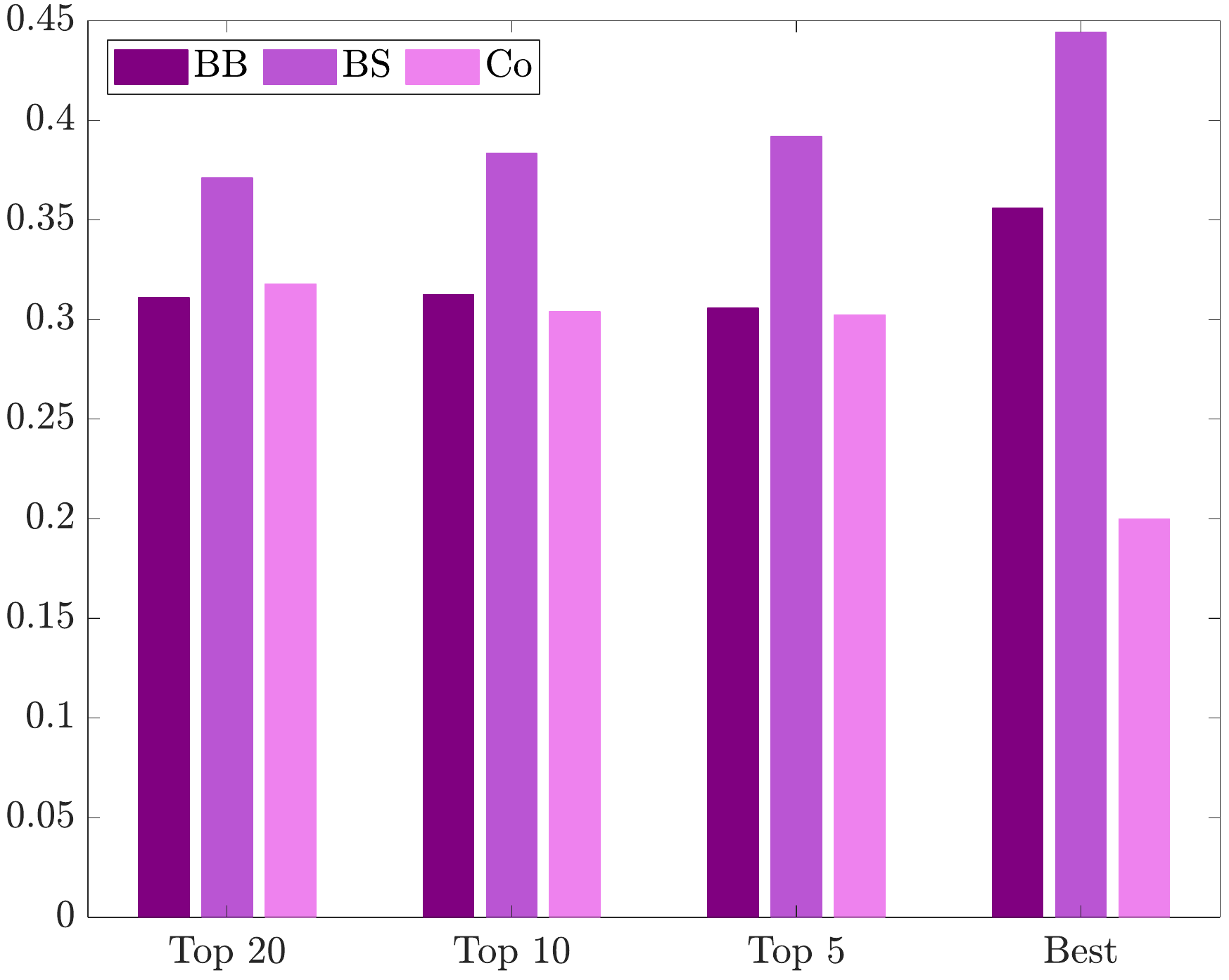}}%
	\caption{\rev{Relative number of times a given method is among the top 20, top 10, top 5 or is the best. Ranking is made w.r.t. to the $\Delta$-criterion.}}
	\label{fig:RankingStacked_Delta}
\end{figure}

\begin{figure}[!ht]
	\centering
	\subfloat[Surrogate model]{\label{fig:RankingStacked_Relerr_a}\includegraphics[width=0.49\textwidth]{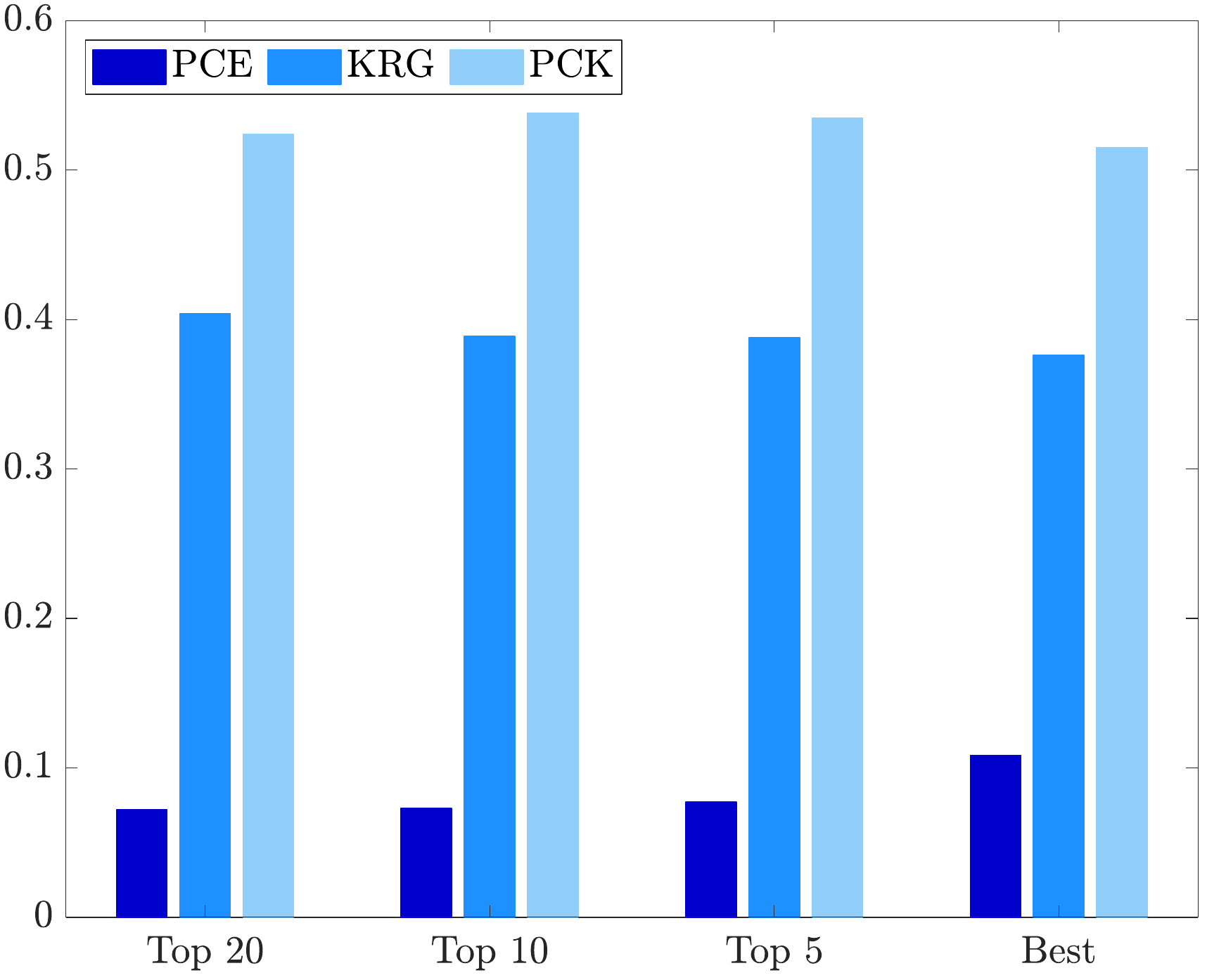}}%
	\hfill
	\subfloat[Reliability estimation algorithm]{\label{fig:RankingStacked_Relerr_b}\includegraphics[width=0.49\textwidth]{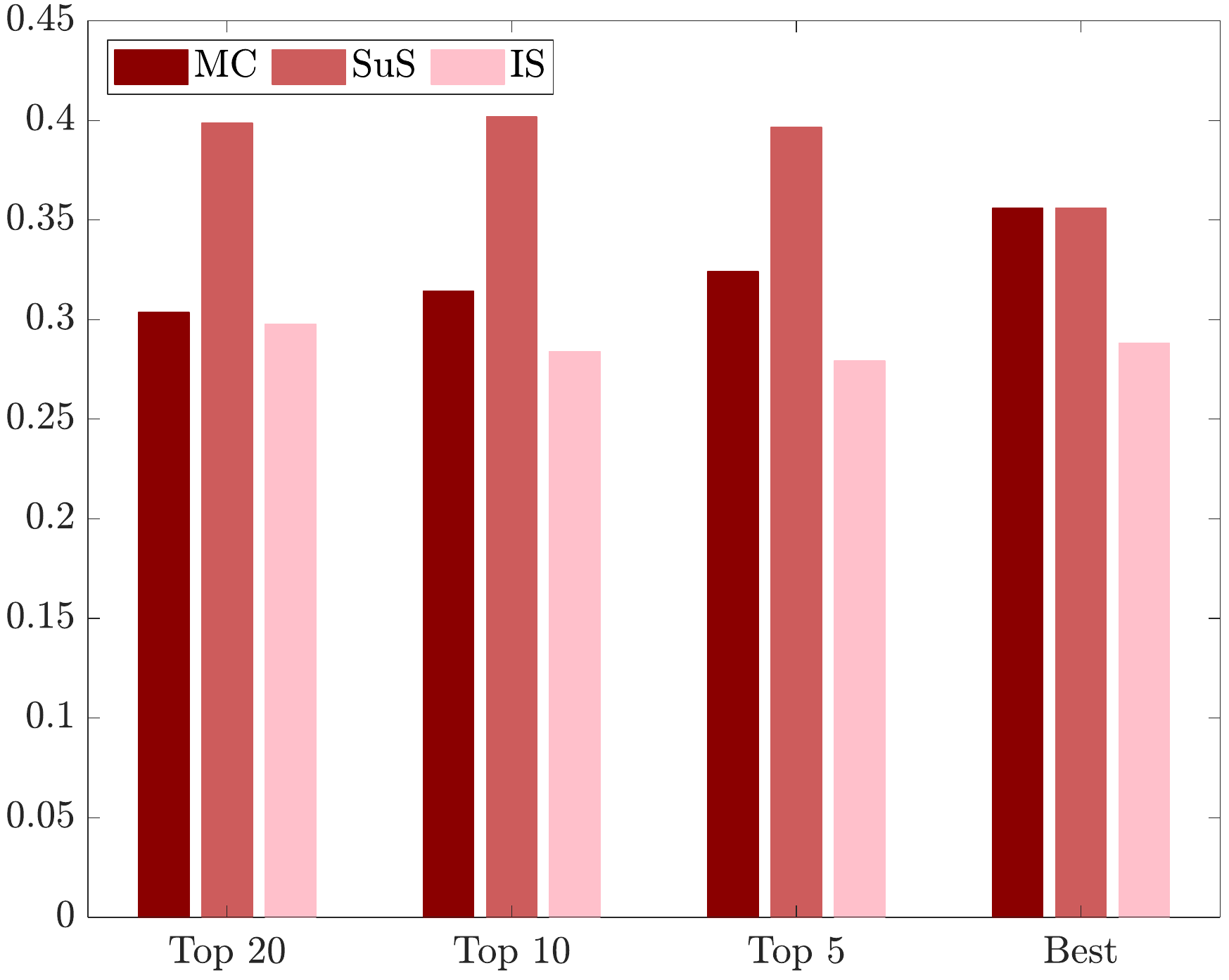}}%
	\\
	\subfloat[Learning function]{\label{fig:RankingStacked_Relerr_c}\includegraphics[width=0.49\textwidth]{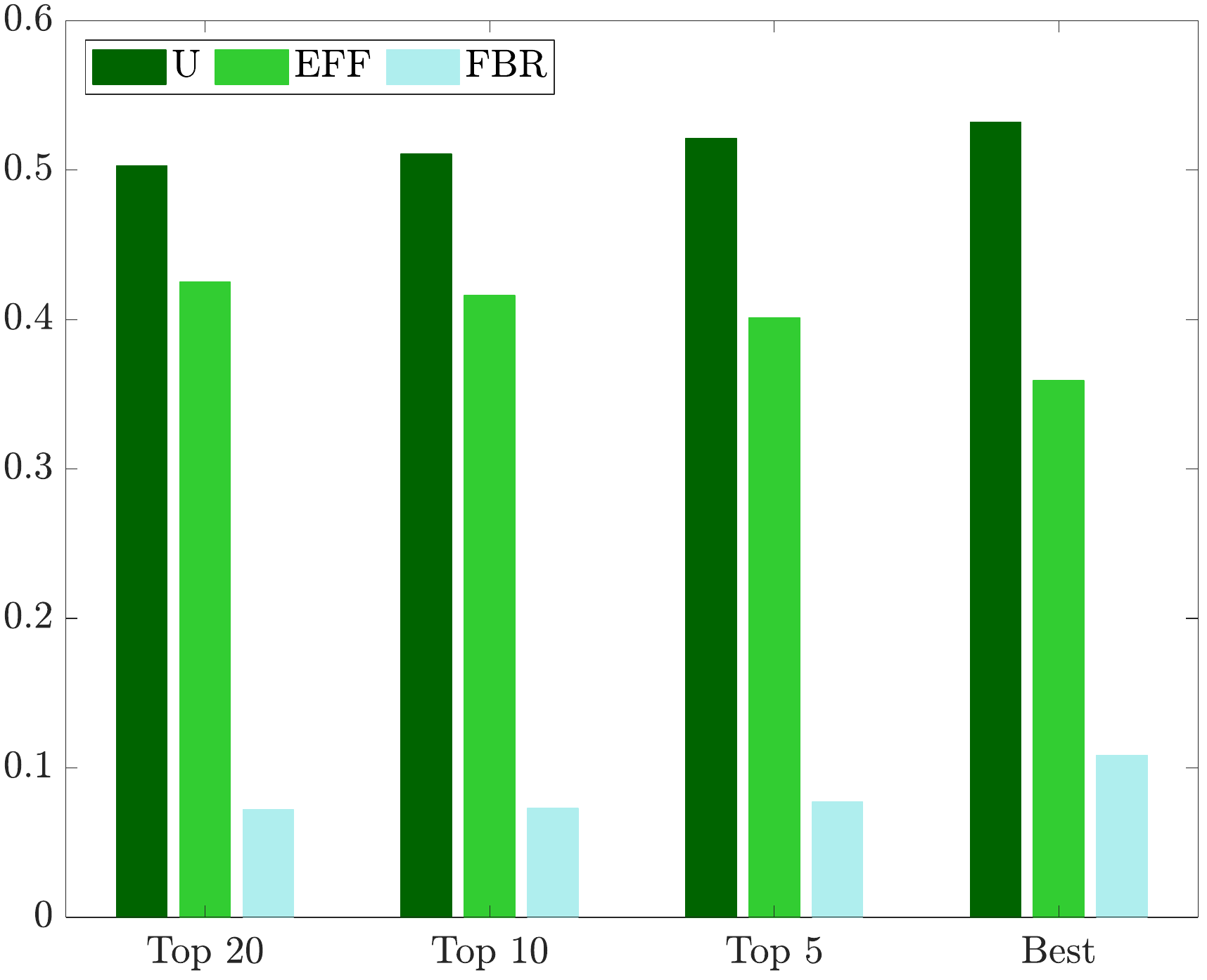}}%
	\hfill
	\subfloat[Stopping criterion]{\label{fig:RankingStacked_Relerr_d}\includegraphics[width=0.49\textwidth]{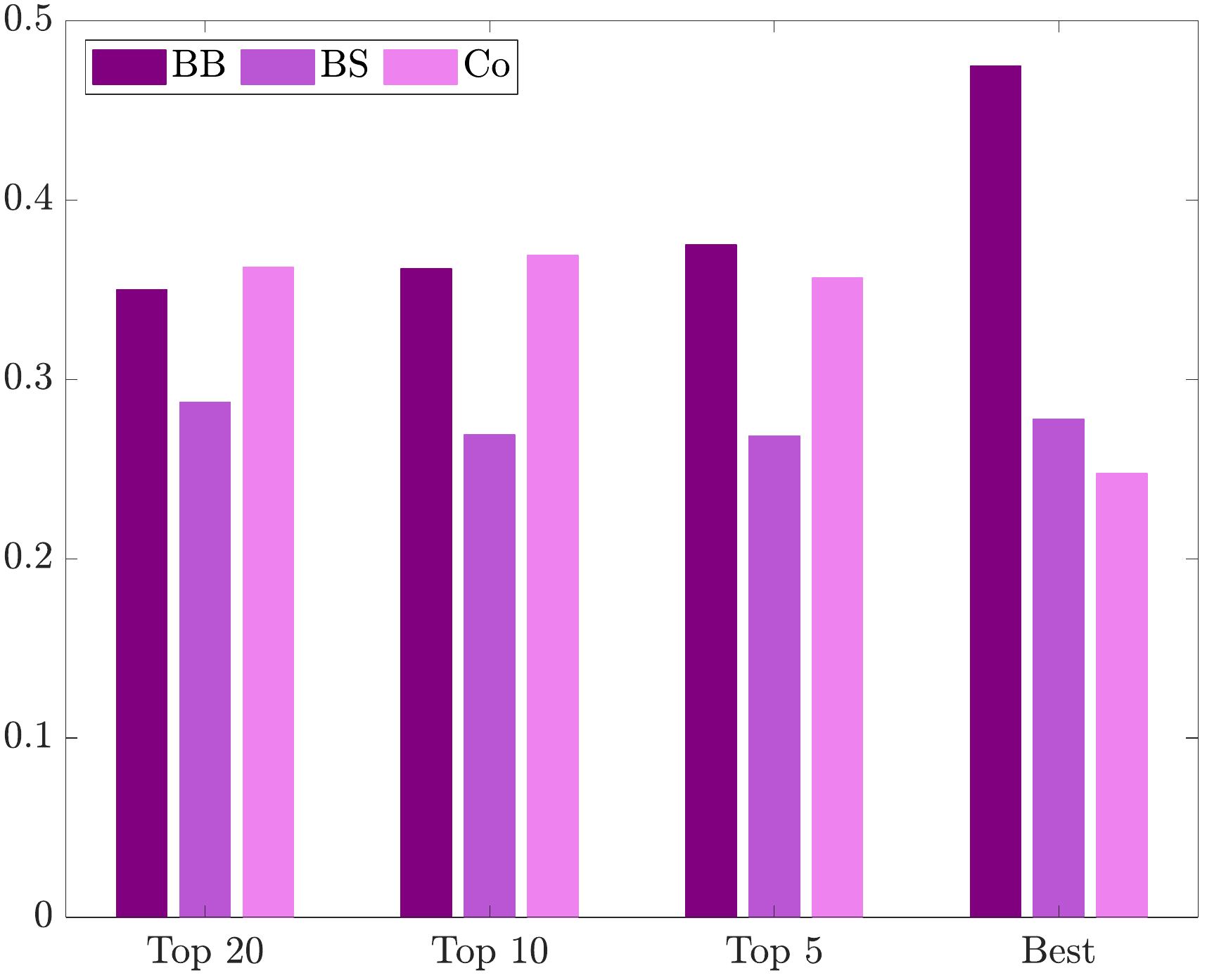}}%
	\caption{\rev{Relative number of times a given method is among the top 20, top 10, top 5 or is the best. Ranking is made w.r.t. the relative error $\varepsilon_{\beta}$}}
	\label{fig:RankingStacked_Relerr}
\end{figure}

In the previous section, we analyzed the strategies as a block, now we split them into their four components and perform the same statistical analysis. The strategies are once again  ranked for each problem and replication and we count the number of occurrences of each method in a given ranking. The results are summarized in Figures~\ref{fig:RankingStacked_Delta}~and~\ref{fig:RankingStacked_Relerr}, where the percentage of times a given method is the best is shown in \rev{the last group of bars} of each panel. To assess the variability in the ranking we also count the number of times a given method is within the first $5$, $10$ or $20$ positions. Despite some minor variations in the  share of each method in the top positions, the ranking remains unchanged if considering either the $\Delta$-criterion or the relative error. 
In terms of surrogate models, PC-Kriging is the best performing choice, as it accounts for roughly half the occurrence in the best rankings. 
The reliability module is the most balanced, even though subset simulation shows a slight margin over the two others. 
In terms of learning function, the deviation number $U$ outperforms both the expected feasibility function ($EFF$) and its PCE counterpart ($FBR$).

The convergence criterion is the only module, the results of which differ depending on the ranking criterion considered. 
The best method seems to be $\beta$-stability when it comes to the $\Delta$-criterion. 
However this turns to $\beta$-bounds when considering the relative error. 
The explanation is simply that $\beta$-stability converges faster than $\beta$-bounds. 
Hence, when accuracy is the prior concern, the second criterion is more suitable. 
However, when the computational budget is limited, $\beta$-stability is a more appropriate convergence criterion.

\subsection{Aggregation of the results w.r.t. the problem features}
Using the ranking from the previous two sections, it is clear now which methods lead on average to the best performance in terms of accuracy and efficiency. 
In this section, we go a step further and try to determine if these methods behave in the same way as a function of two selected features of the problem, namely input dimensionality and magnitude of the probability of failure, or if their performances are intrinsically linked to the type of problem at hand. 
We split the benchmark into low- ($M < 20$) and high-dimensional ($M \geq 20$) problems, as well as small ($\beta_{\textrm{ref}} < 3.5$) and high ($\beta_{\textrm{ref}} \geq 3.5$) reliability indices.

\subsubsection{Performance with respect to dimensionality}
\begin{figure}[!ht]
	\centering
	\subfloat[Surrogate model]{\label{fig:PerDimension_Delta_a}\includegraphics[width=0.49\textwidth]{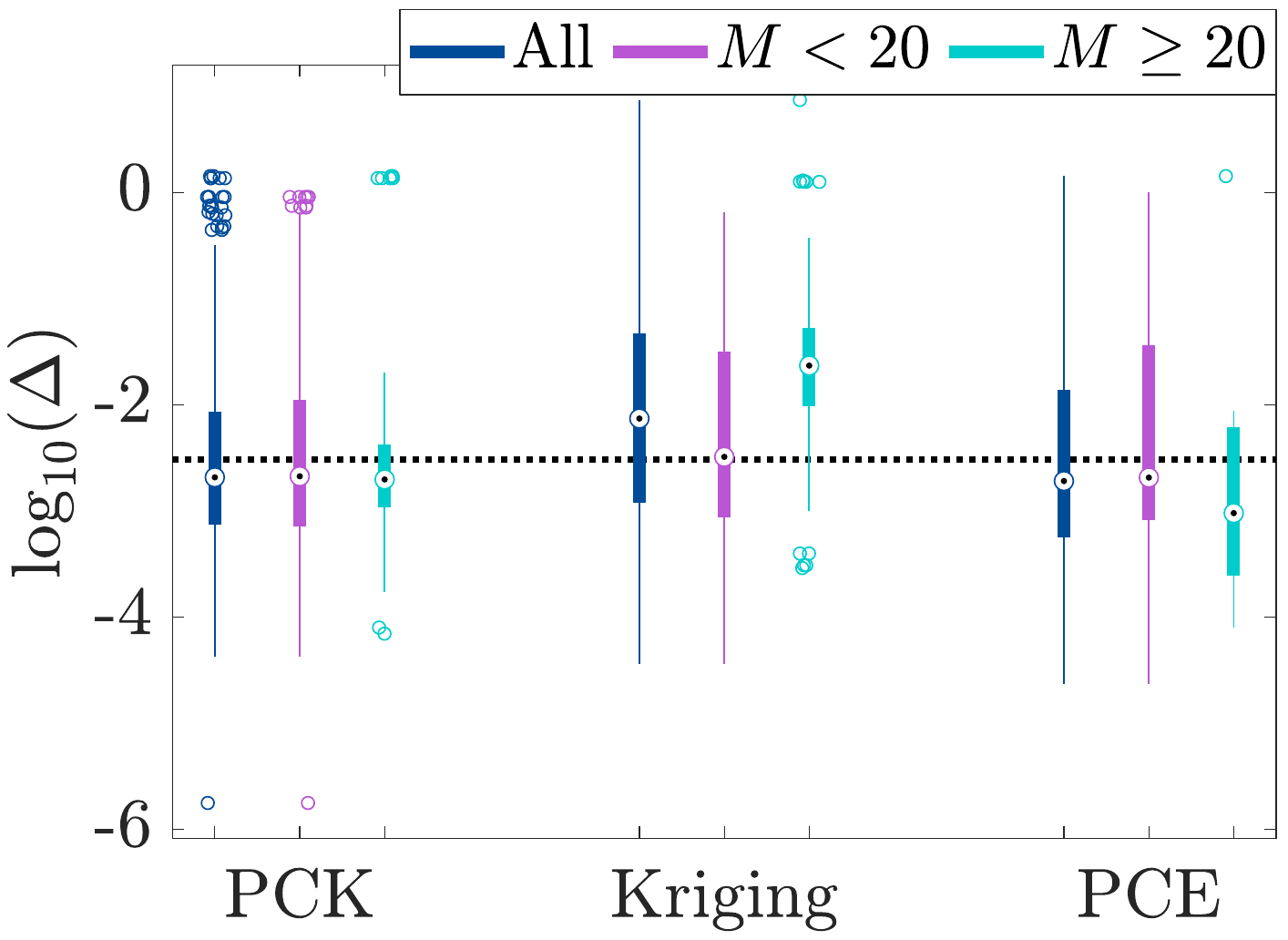}}%
	\hfill
	\subfloat[Reliability estimation algorithm]{\label{fig:PerDimension_Delta_b}\includegraphics[width=0.49\textwidth]{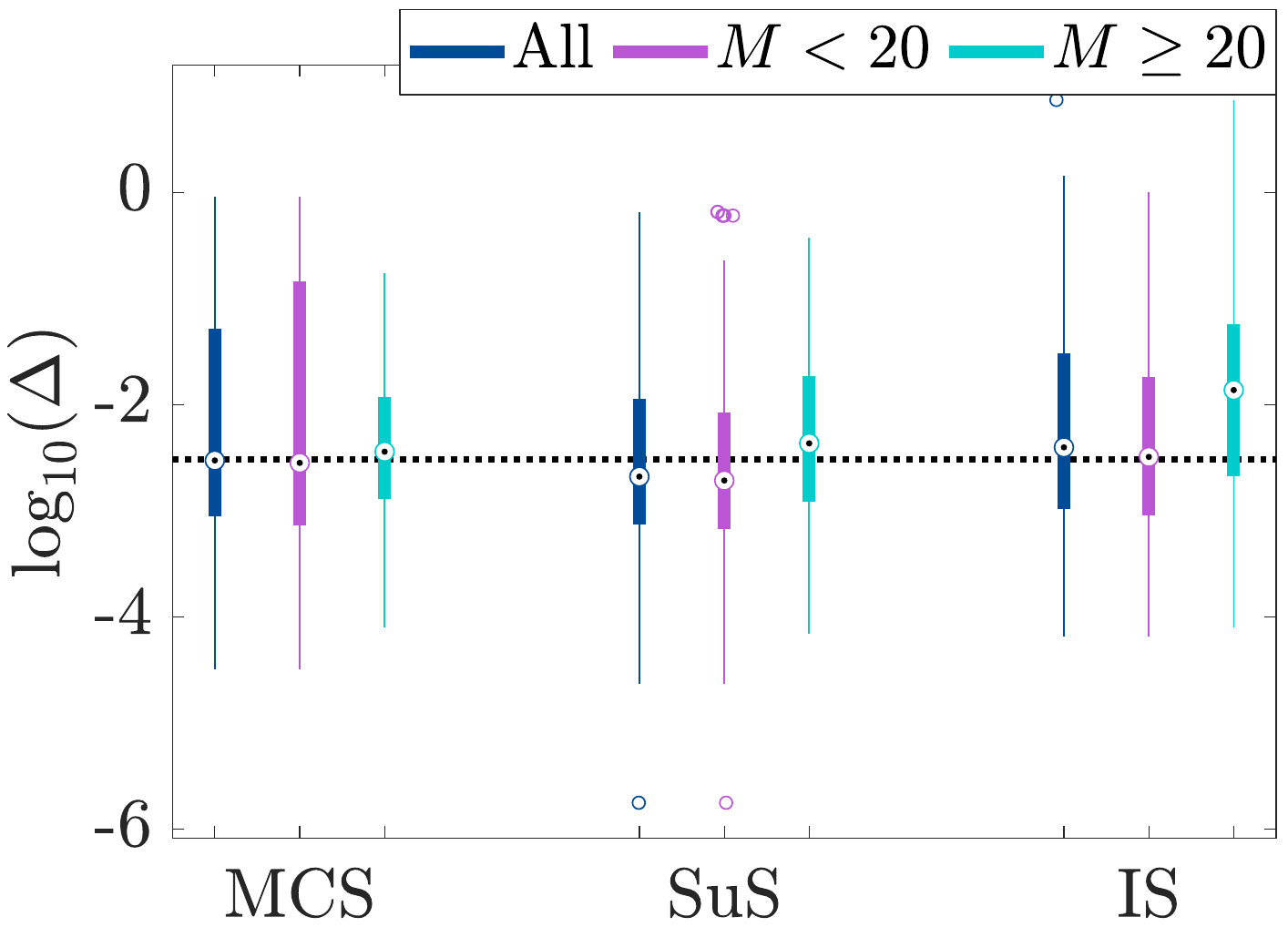}}%
	\\
	\subfloat[Learning function]{\label{fig:PerDimension_Delta_c}\includegraphics[width=0.49\textwidth]{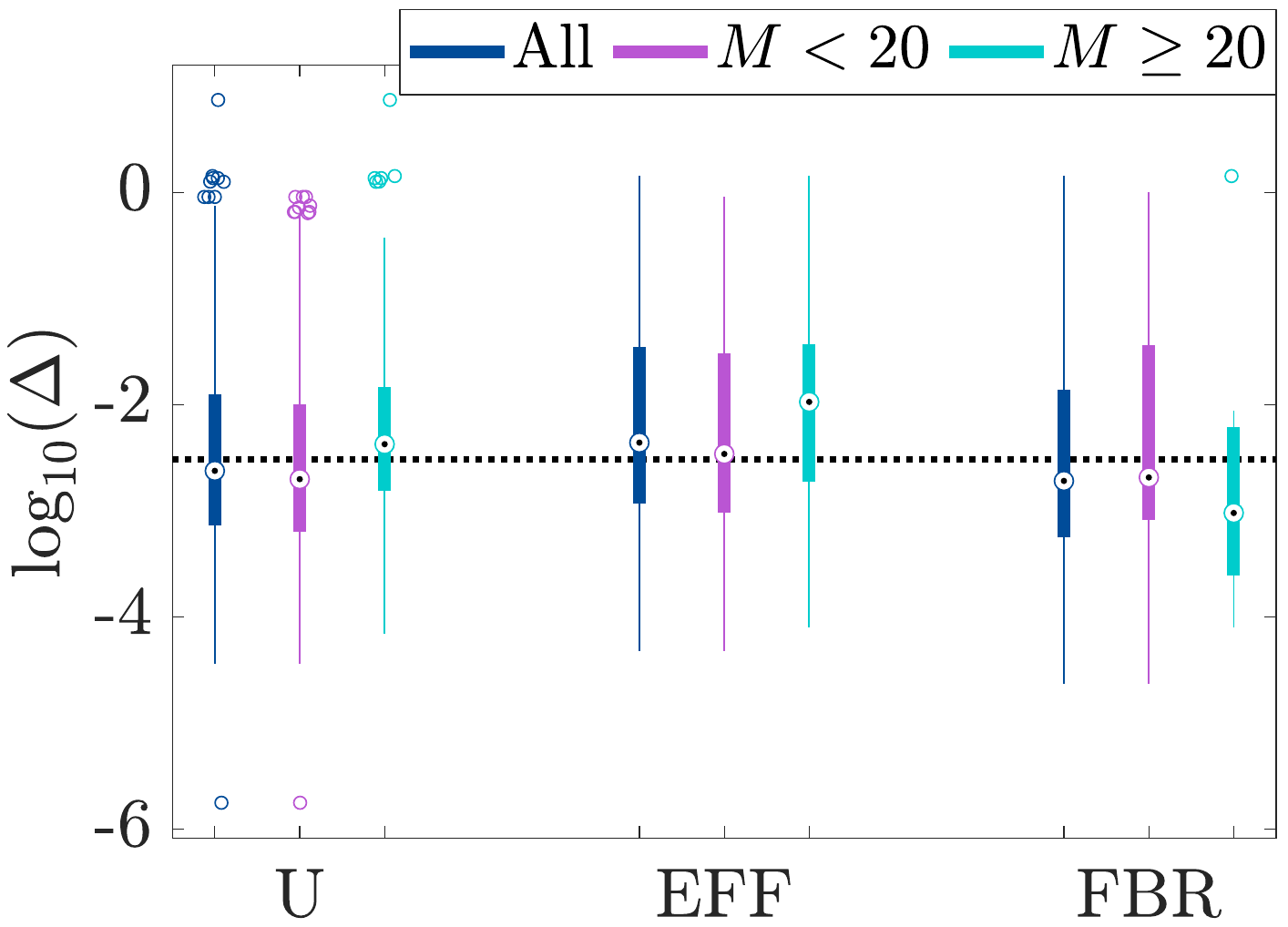}}%
	\hfill
	\subfloat[Stopping criterion]{\label{fig:PerDimension_Delta_d}\includegraphics[width=0.49\textwidth]{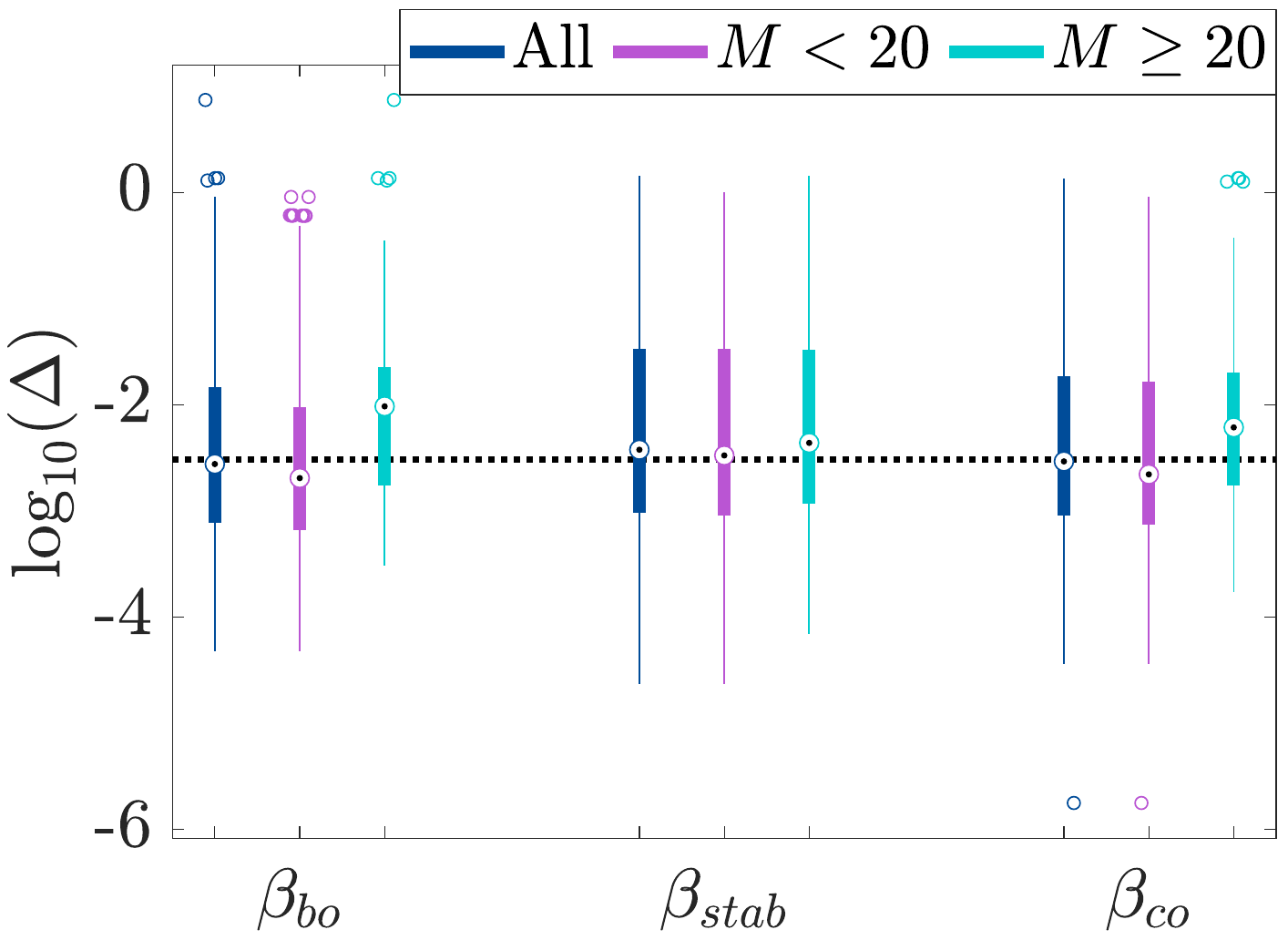}}%
	\caption{Different methods compared w.r.t. the combined criterion $\Delta$ with problems split in two: low- ($M<20$) and high-dimensional ($M \geq 20$).}
	\label{fig:PerDimension_Delta}
\end{figure}

\begin{figure}[!ht]
	\centering
	\subfloat[Surrogate model]{\label{fig:PerDimension_Relerr_a}\includegraphics[width=0.49\textwidth]{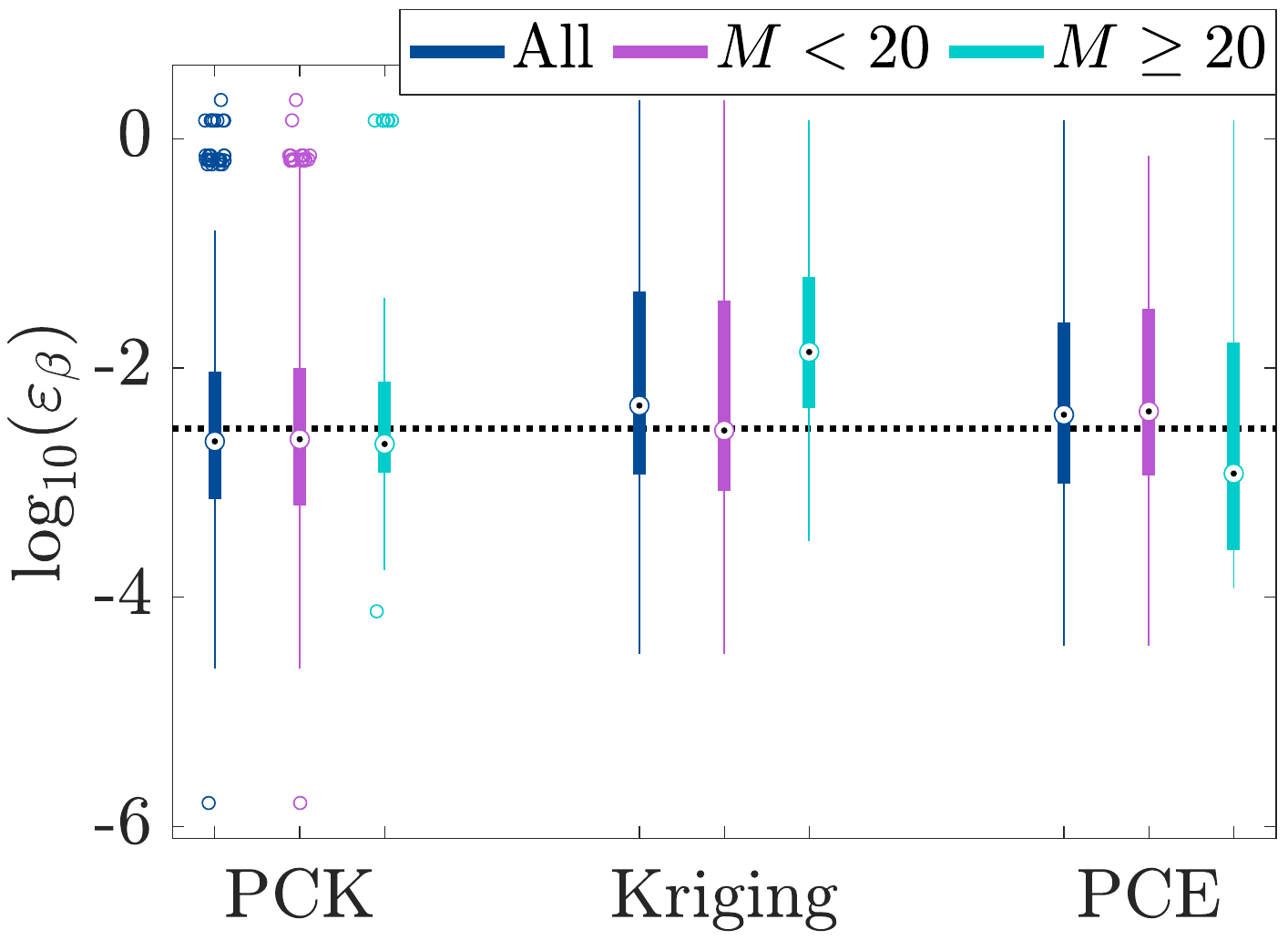}}%
	\hfill
	\subfloat[Reliability estimation algorithm]{\label{fig:PerDimension_Relerr_b}\includegraphics[width=0.49\textwidth]{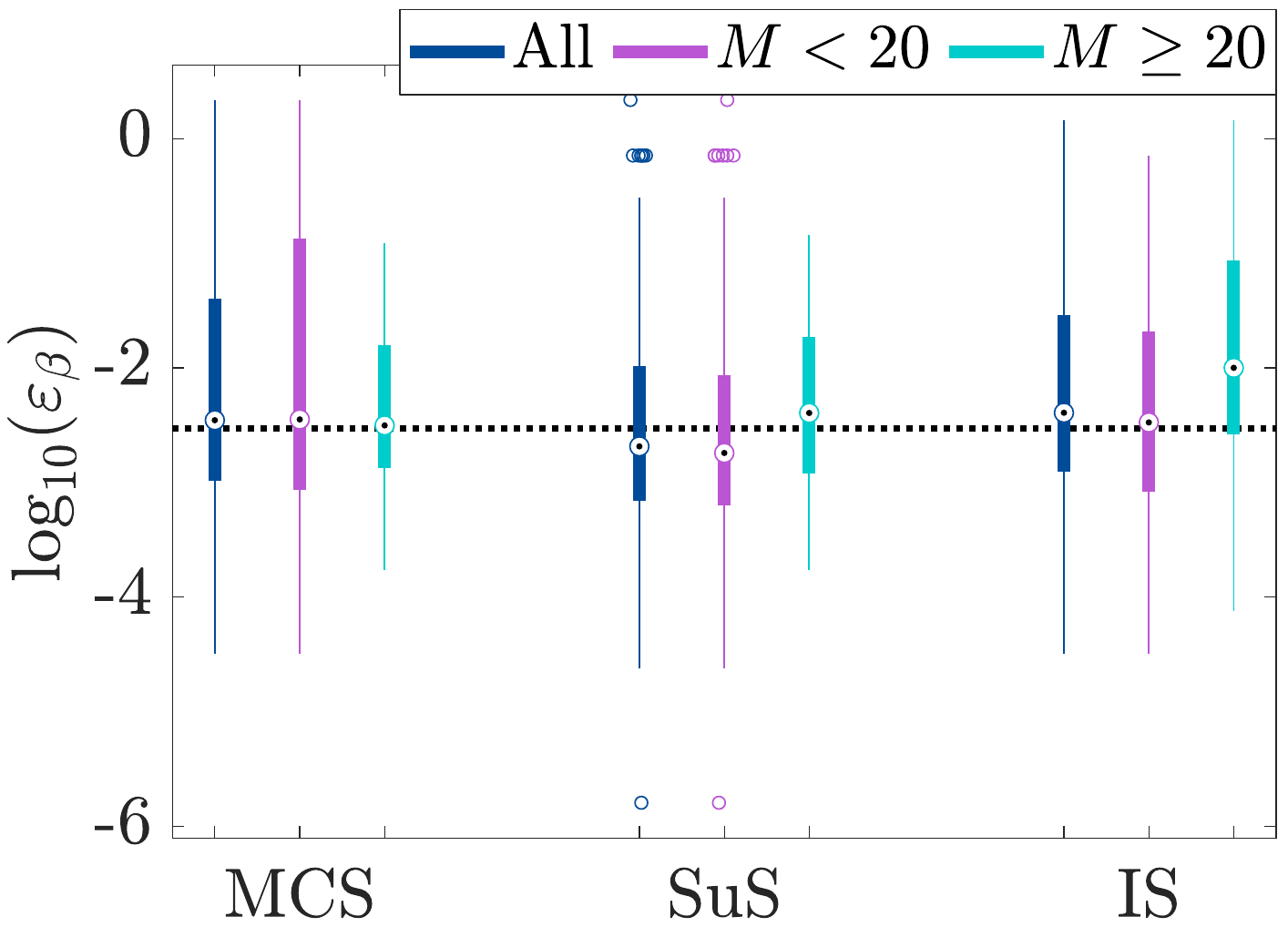}}%
	\\
	\subfloat[Learning function]{\label{fig:PerDimension_Relerr_c}\includegraphics[width=0.49\textwidth]{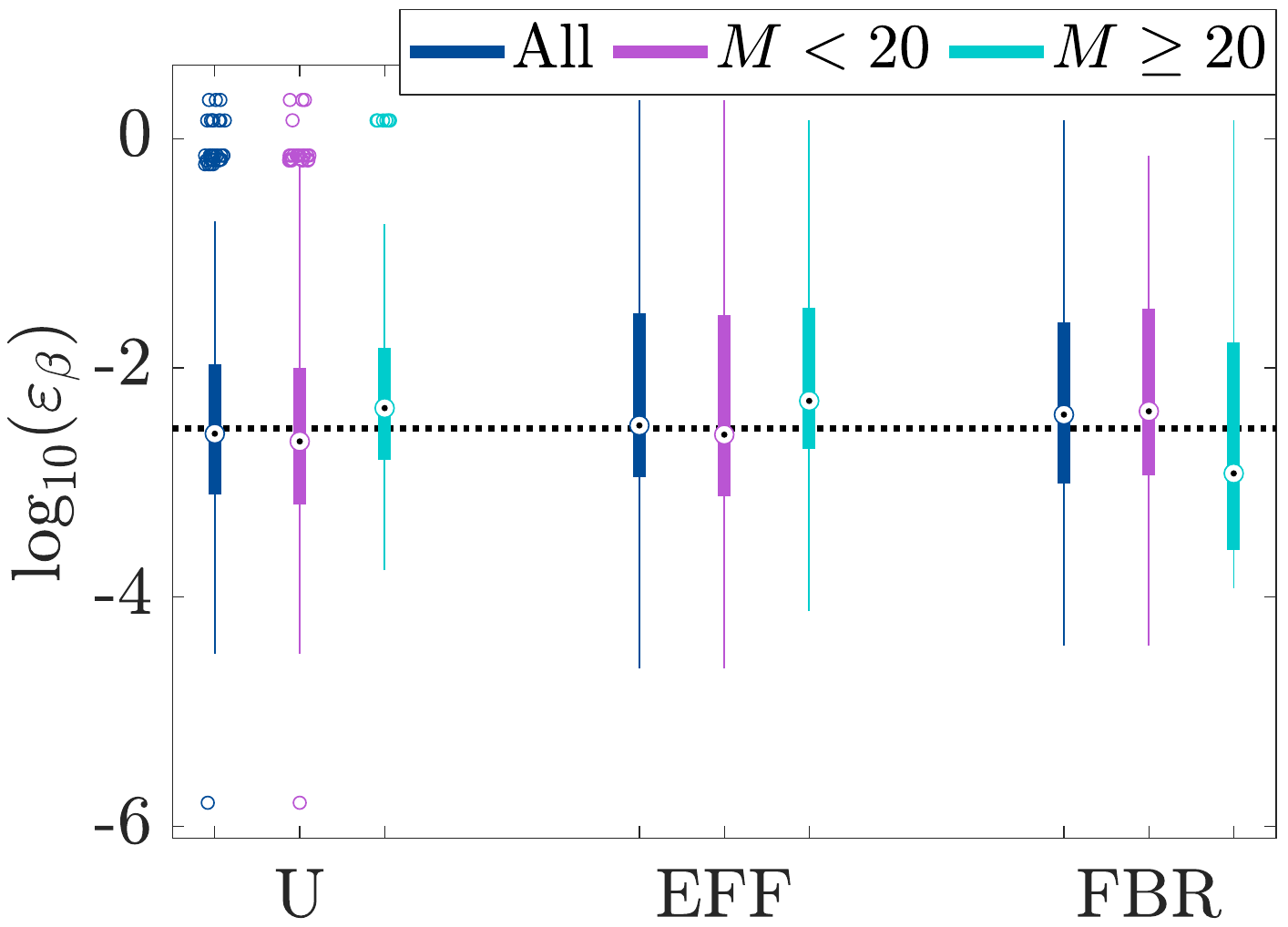}}%
	\hfill
	\subfloat[Stopping criterion]{\label{fig:PerDimension_Relerr_d}\includegraphics[width=0.49\textwidth]{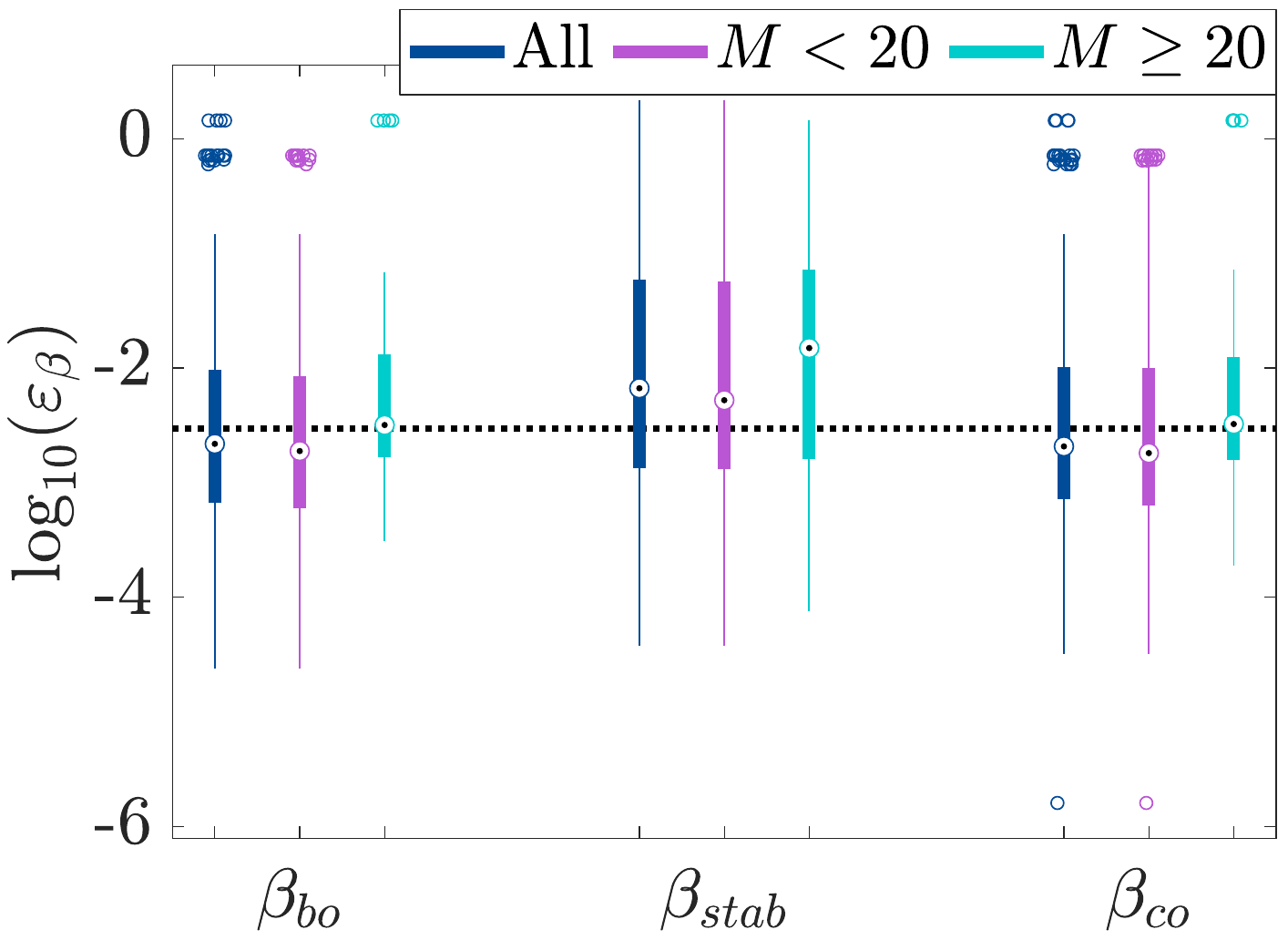}}%
	\caption{Different methods compared w.r.t. the relative error $\varepsilon_{\beta}$ with problems split in two: low- ($M<20$) and high-dimensional ($M \geq 20$).}
	\label{fig:PerDimension_Relerr}
\end{figure}

Figures~\ref{fig:PerDimension_Delta}~and~\ref{fig:PerDimension_Relerr} respectively show the $\Delta$-criterion values and relative errors aggregated over all problems and then split for each method. The horizontal dotted black line represents the median over all problems. In each panel, the boxplots represent the aggregated median results (over all $15$ replications) considering all (blue), low- to medium- (magenta) and high-dimensional problems (cyan). 

Starting with the surrogate models and looking at the $\Delta$-criterion (Figure~\ref{fig:PerDimension_Delta_a}), we observe that PC-Kriging does not seem to be strongly affected by the problem dimensionality, as in all cases the conditional median remains slightly below the overall median, while PCE seems to improve its relative performance in higher dimensional problems.
Kriging on the other hand performs noticeably poorly. 
The same trend is observed with the accuracy criterion $\varepsilon_{\beta}$ (Figure~\ref{fig:PerDimension_Relerr_a}).

Next are the reliability estimation algorithms (Figure~\ref{fig:PerDimension_Delta_b}) and as expected Monte Carlo simulation is essentially insensitive to the dimension. 
Subset simulation performs slightly worse in high-dimension but not as much as importance sampling. 
The latter becomes slightly worse when considering the purely accuracy-oriented criterion (Figure~\ref{fig:PerDimension_Relerr_b}).

Regarding the learning functions (Figure~\ref{fig:PerDimension_Delta_c}~and~\ref{fig:PerDimension_Relerr_c}), the deviation number ($U$) also does not seem to be noticeably affected by the dimension, contrary to the expected feasibility function ($EFF$) which gets considerably worse as the dimension increases. The fraction of bootstrap replicates ($FBR$) mirrors the behavior of PCE as there is a one-to-one mapping between the two. Their performance gets somehow better for high-dimensional problems when considering either of the criteria.

Finally, as already observed earlier, contradicting conclusions are obtained when considering either the $\Delta$-criterion value (Figure~\ref{fig:PerDimension_Delta_d}) or the relative error (Figure~\ref{fig:PerDimension_Relerr_d}) for the stopping criterion. In the former case, $\beta$-stability seems to be the best option when dimension increases whereas in the latter $\beta$-bounds, or even the combined criterion, seem to be the best option. This again can be explained by the fact that $\beta$-bounds is a stricter convergence criterion, especially in high-dimensional cases where the Kriging variance hardly shrinks as the experimental design is enriched.

\subsubsection{Performance with respect to the magnitude of the failure probability}
\begin{figure}[!ht]
	\centering
	\subfloat[Surrogate model]{\label{fig:PerBeta_Delta_a}\includegraphics[width=0.49\textwidth]{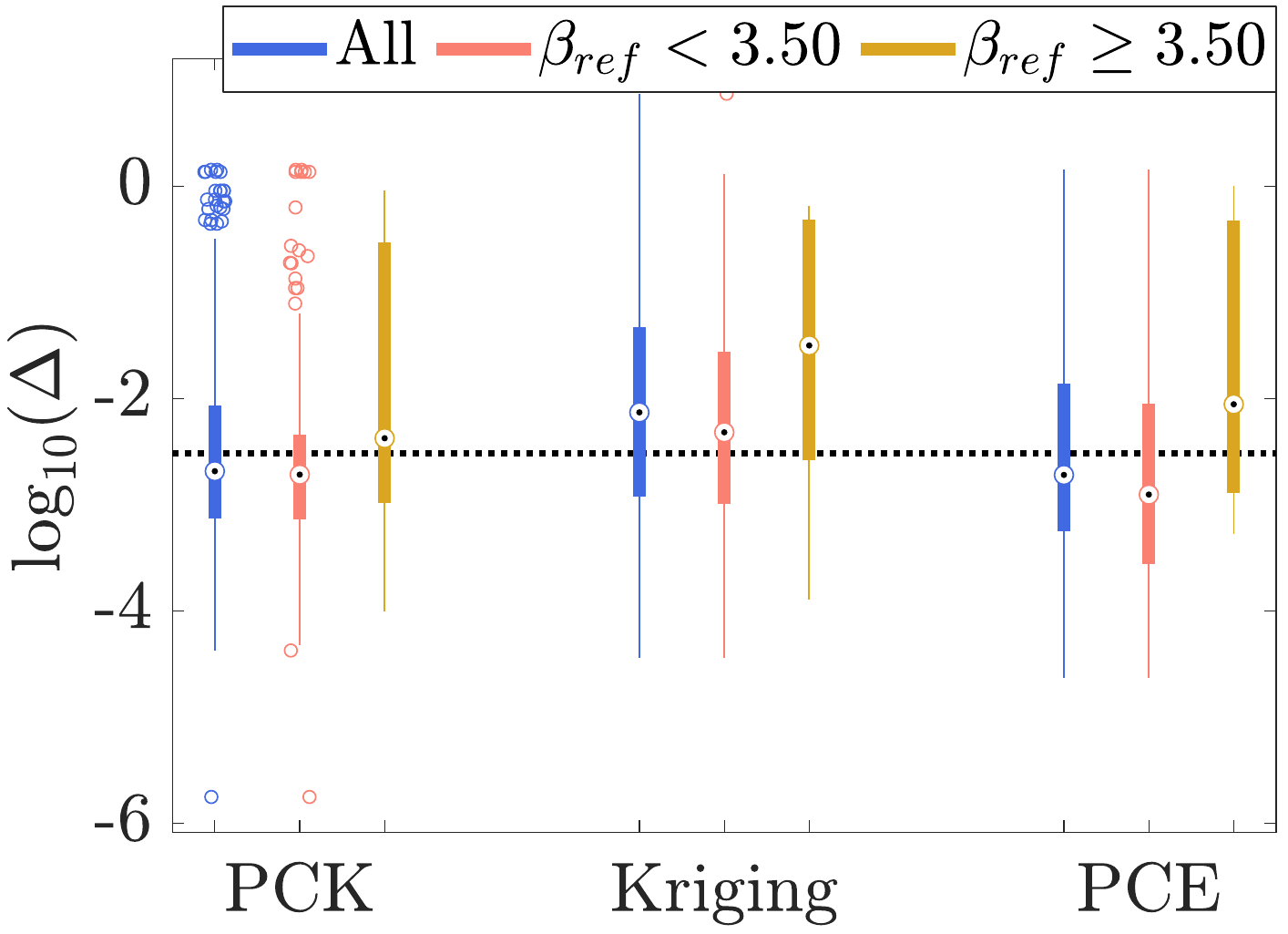}}%
	\hfill
	\subfloat[Reliability estimation algorithm]{\label{fig:PerBeta_Delta_b}\includegraphics[width=0.49\textwidth]{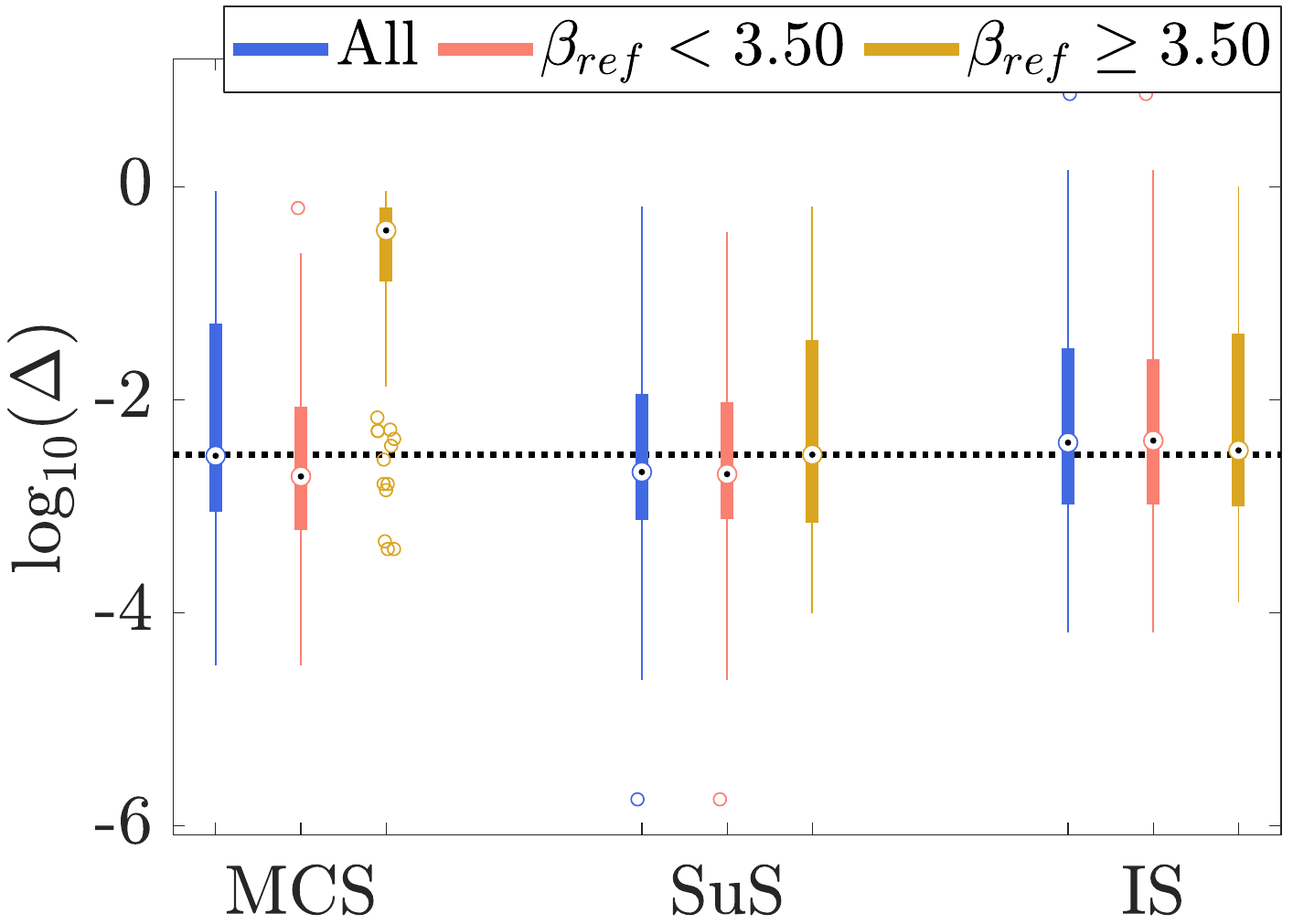}}%
	\\
	\subfloat[Learning function]{\label{fig:PerBeta_Delta_c}\includegraphics[width=0.49\textwidth]{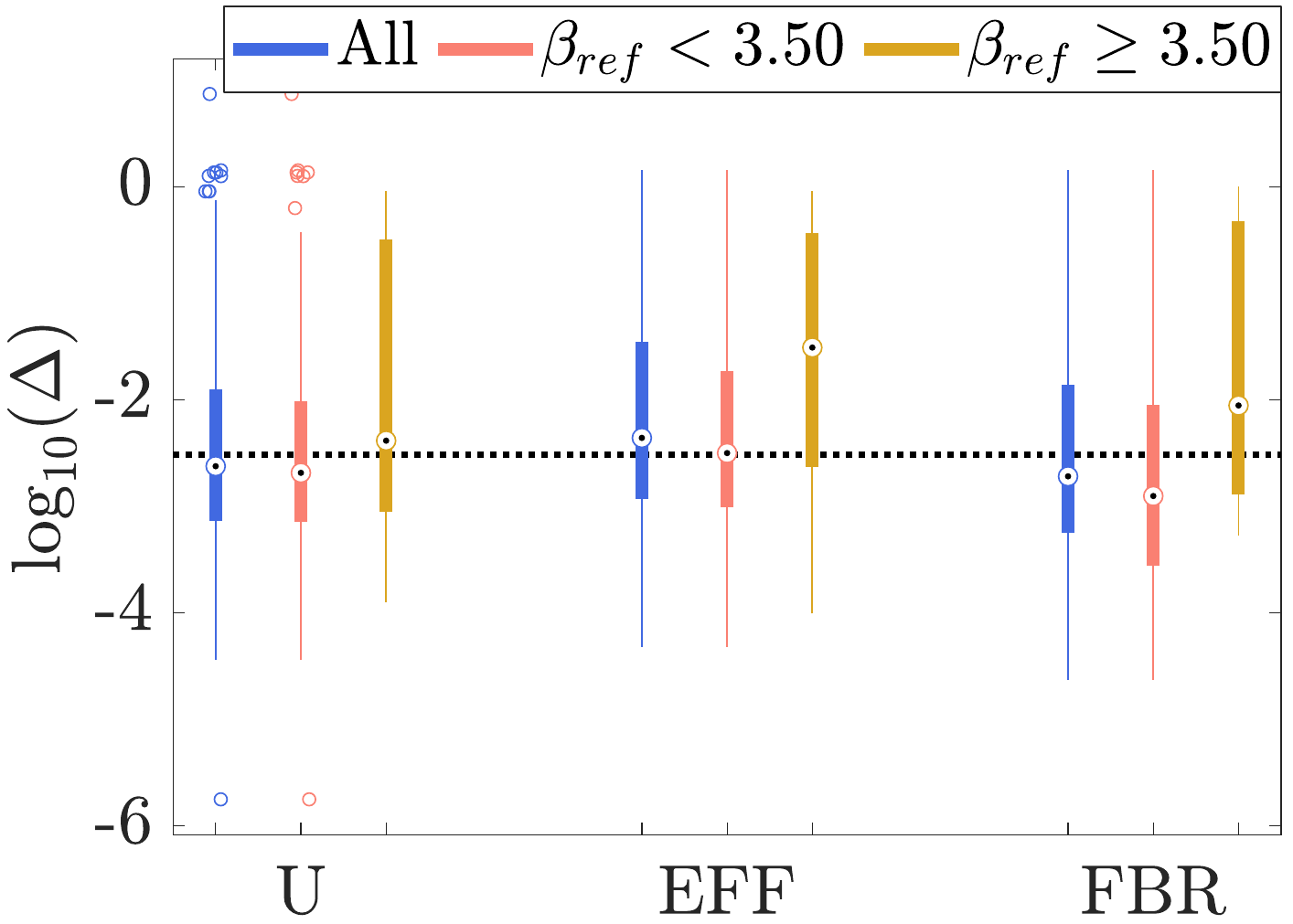}}%
	\hfill
	\subfloat[Stopping criterion]{\label{fig:PerBeta_Delta_d}\includegraphics[width=0.49\textwidth]{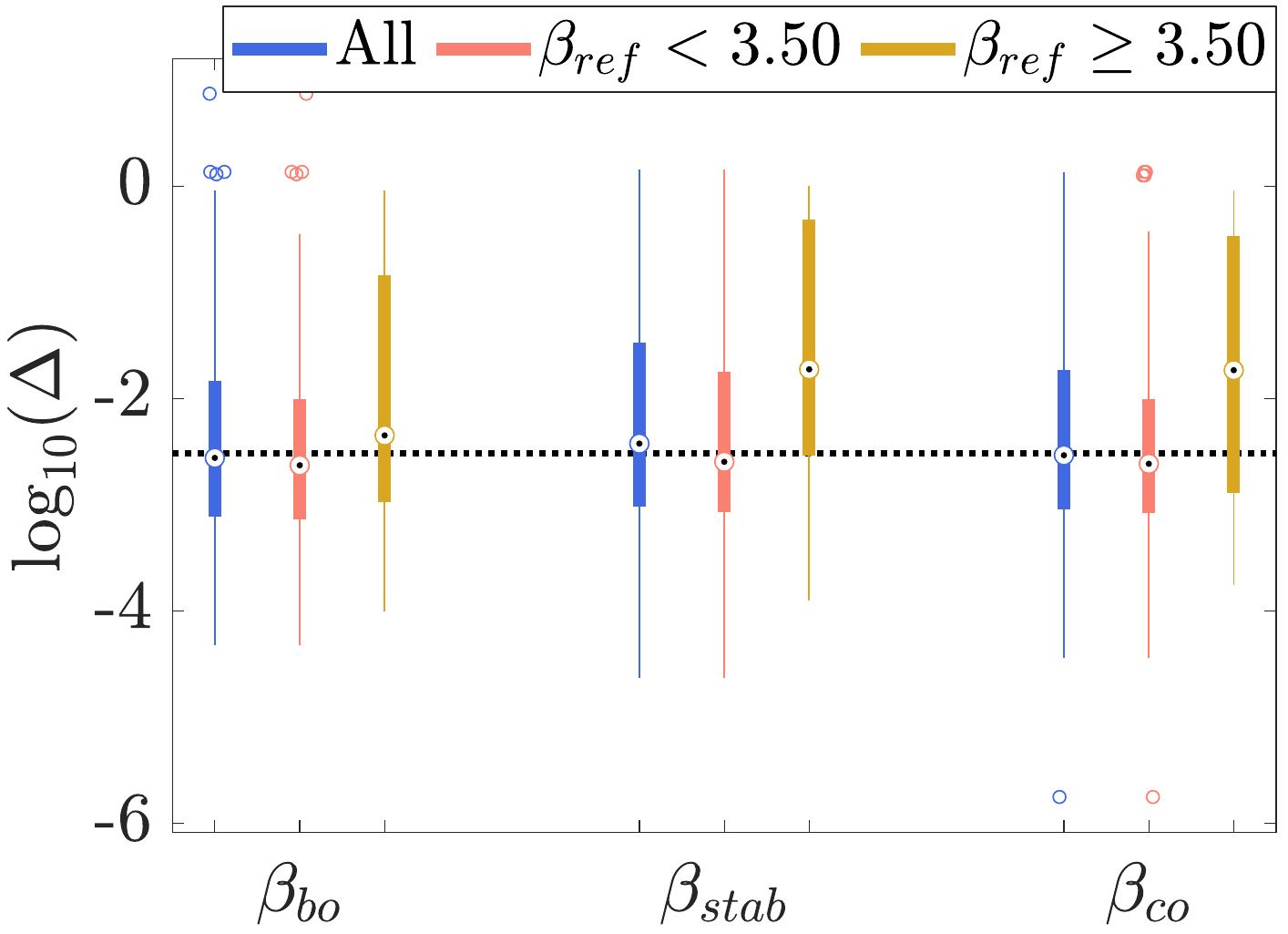}}%
	\caption{Different methods compared w.r.t. the combined criterion $\Delta$ with problems split in two: small ($\beta_{\textrm{ref}} < 3.5$) and large ($\beta_{\textrm{ref}} \geq 3.5$) reliability indices.}
	\label{fig:PerBeta_Delta}
\end{figure}

\begin{figure}[!ht]
	\centering
	\subfloat[Surrogate model]{\label{fig:PerBeta_Relerr_a}\includegraphics[width=0.49\textwidth]{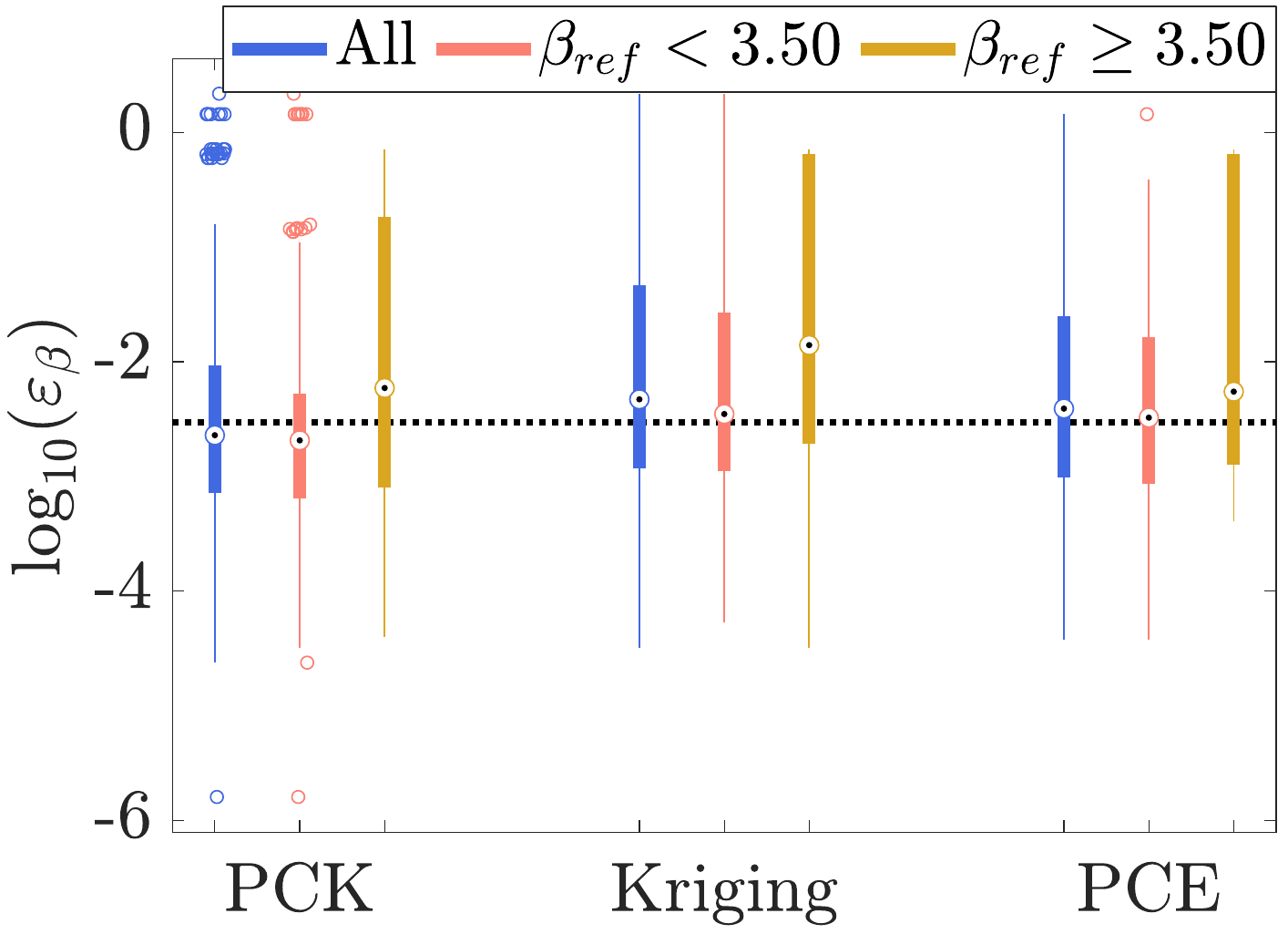}}%
	\hfill
	\subfloat[Reliability estimation algorithm]{\label{fig:PerBeta_Relerr_b}\includegraphics[width=0.49\textwidth]{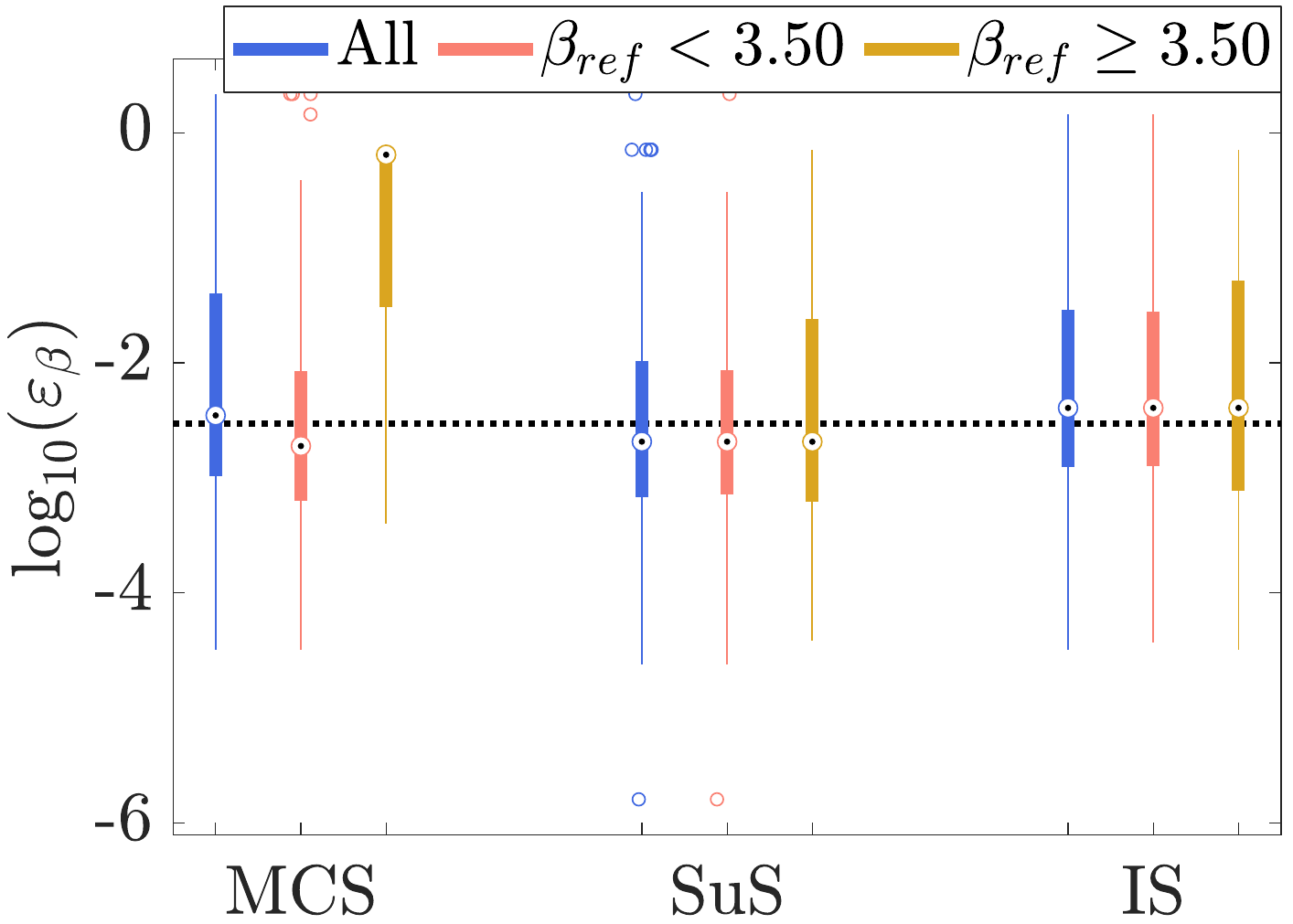}}%
	\\
	\subfloat[Learning function]{\label{fig:PerBeta_Relerr_c}\includegraphics[width=0.49\textwidth]{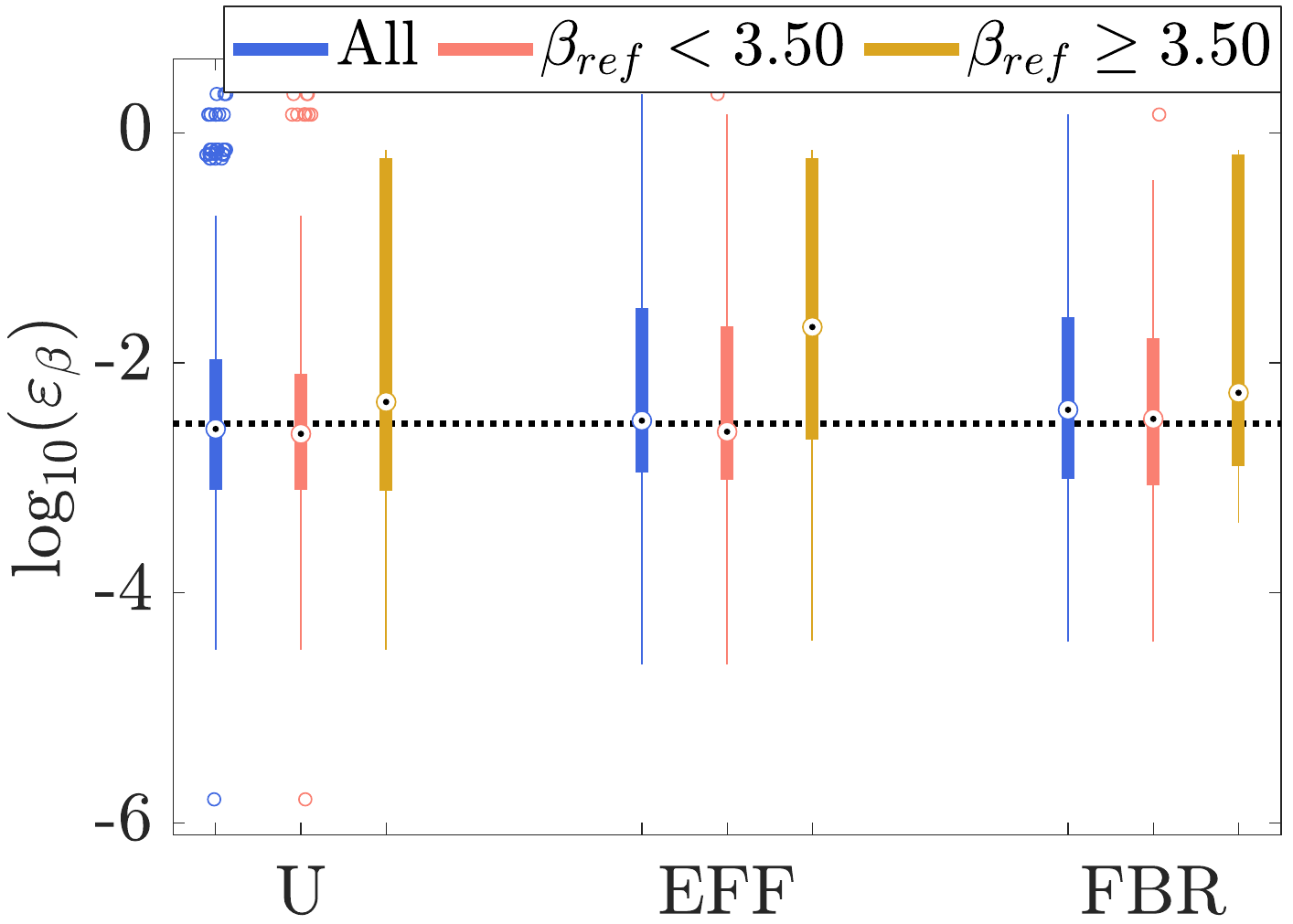}}%
	\hfill
	\subfloat[Stopping criterion]{\label{fig:PerBeta_Relerr_d}\includegraphics[width=0.49\textwidth]{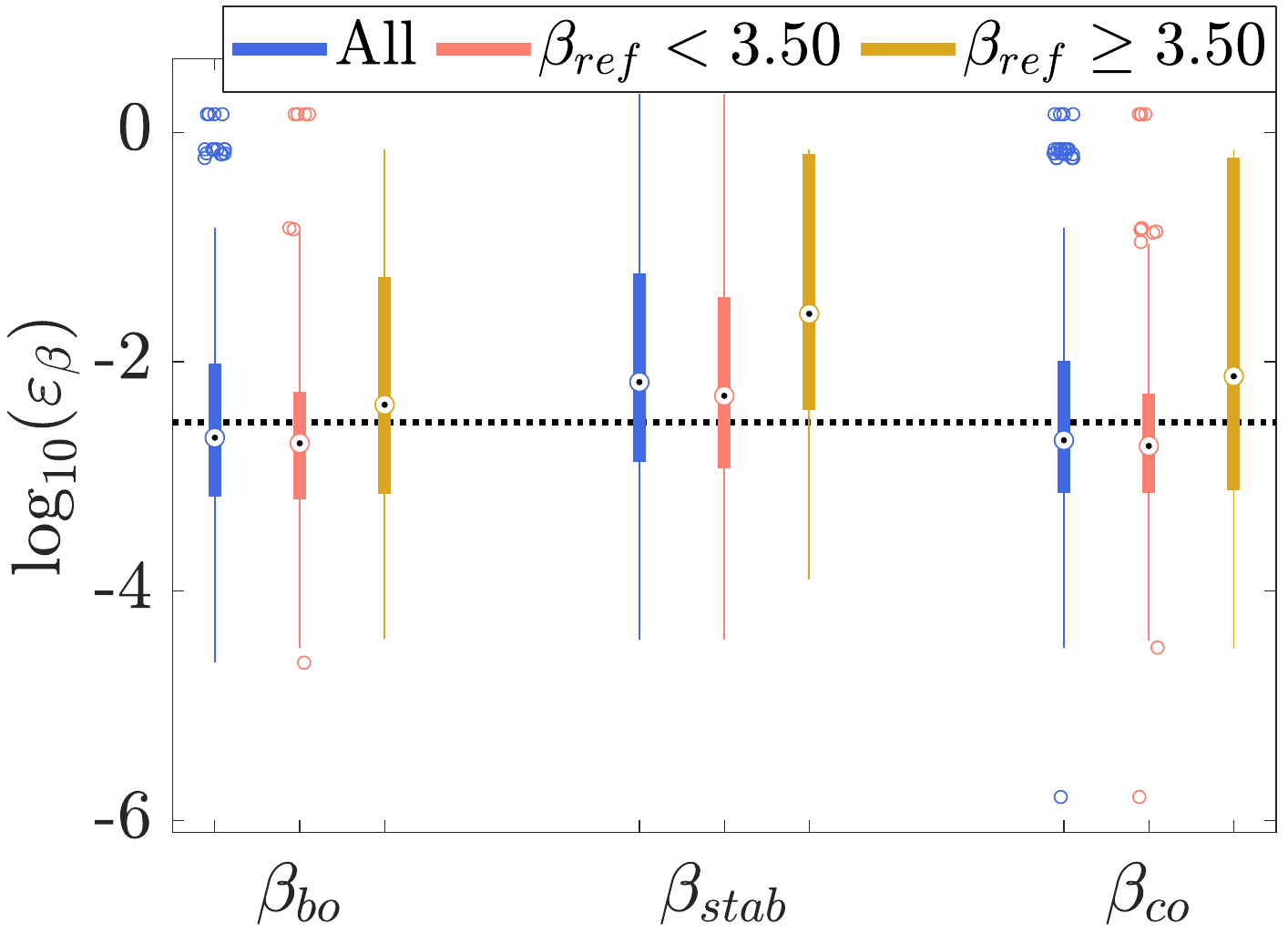}}%
	\caption{Different methods compared w.r.t. the relative error $\varepsilon_{\beta}$ with problems split in two: small ($\beta_{\textrm{ref}} < 3.5$) and large ($\beta_{\textrm{ref}} \geq 3.5$) reliability indices.}
	\label{fig:PerBeta_Relerr}
\end{figure}

In contrast to the dimensionality cases where some methods were not particularly affected by the increase in dimensions, the level of the reliability index seems to affect pretty much all the methods. 
More specifically, the extremely low failure probability cases result on average in systematically poorer performances both in terms of relative error and $\Delta$ (See Figures~\ref{fig:PerBeta_Delta}~and~\ref{fig:PerBeta_Relerr}). 
As expected, Monte Carlo simulation performs considerably worse than its alternatives. 
The reason is that even in its ``overkill" setting, the maximum number of model evaluations was set to $10^7$, hence limiting the statistical estimator variance for extremely low failure probability problems.

\subsection{Results with strategies aggregated over different problems}
\begin{figure}[!ht]
	\centering
	\subfloat[$\Delta$-criterion]{\label{fig:Problems_Beta_Relerr_b}\includegraphics[width=0.49\textwidth]{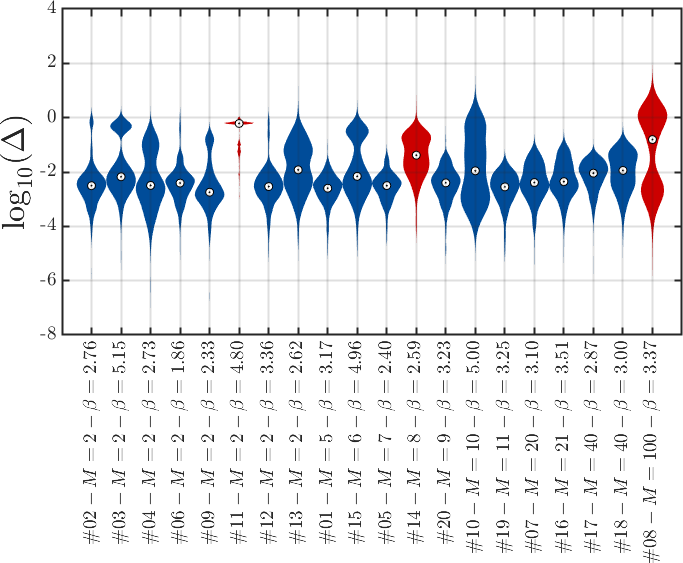}}%
	\hfill
	\subfloat[Relative error]{\label{fig:Problems_Beta_Relerrr_a}\includegraphics[width=0.49\textwidth]{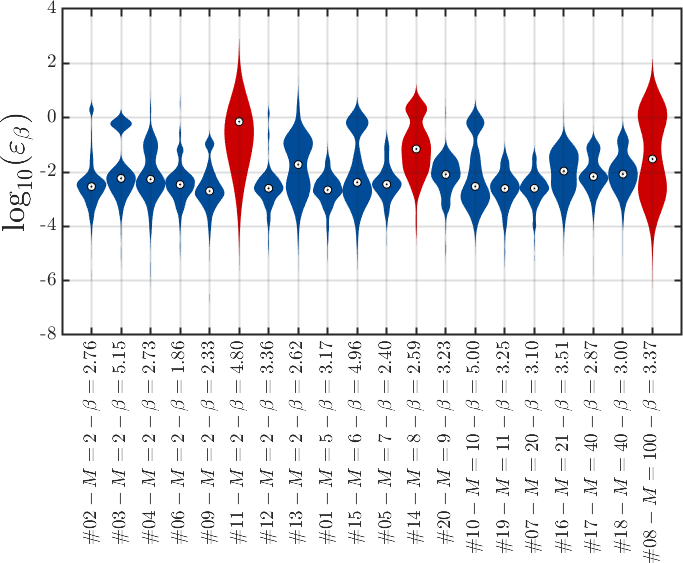}}%
	\caption{Results aggregated for all solutions strategies and shown for each problem considering the combined criterion and the relative error. The three problems with the worst median results are highlighted in red.}
	\label{fig:Problems_Beta_Relerr}
\end{figure}

Now that we have a clear picture of the performances of each strategy as a whole and more particularly of different methods w.r.t. different problems and their features, let us have a look at the overall performance of all methods for each problem. The aggregated relative error and $\Delta$ values for all $20$ problems (ordered in increasing dimensions) are shown in Figure~\ref{fig:Problems_Beta_Relerr} as violin plots, \emph{i.e.}, boxplots with an additional indication of the probability density of the data.
The three problems with the overall worst median performances are highlighted in red. The first observation from \rev{Figure~\ref{fig:Problems_Beta_Relerr}} is that, regardless of the problem, there are still at least a few strategies that lead to good performances. As a matter of fact, the median  values of both criteria are for most problems below the level arbitrarily set at $10^{-2}$.

\begin{figure}[!ht]
	\centering
	\subfloat[ $\Delta$ with PCK+SuS+U+Co]{\label{fig:Problems_Beta_Comparison_a}\includegraphics[width=0.70\textwidth]{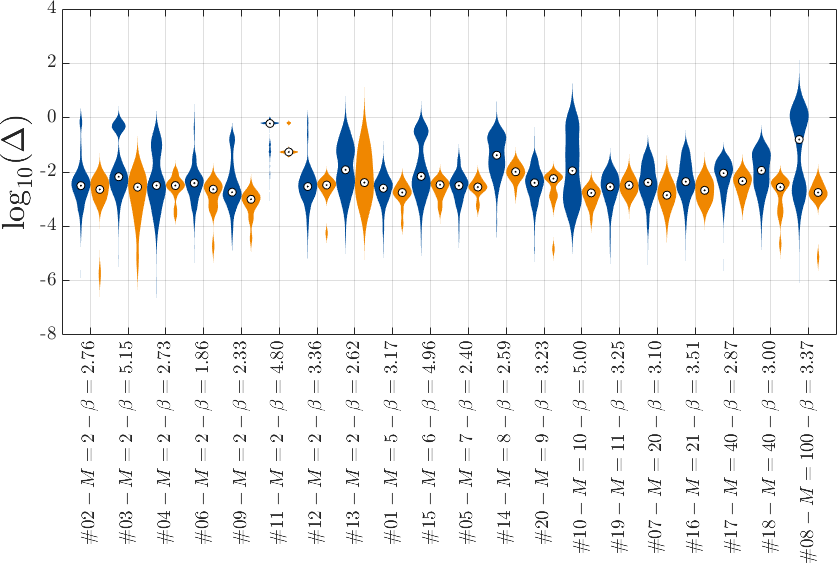}}%
	\\
	\subfloat[$\varepsilon_{\beta}$ with PCK+SuS+U+Co]{\label{fig:Problems_Beta_Comparison_b}\includegraphics[width=0.70\textwidth]{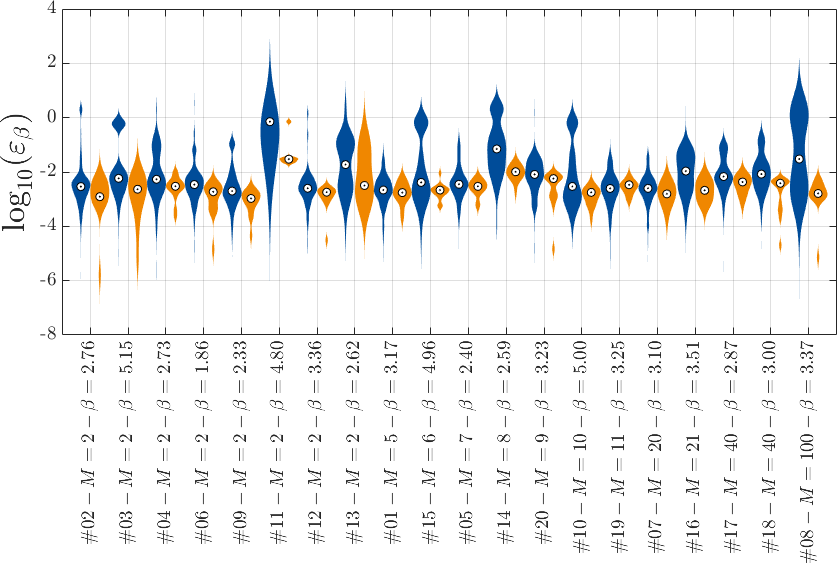}}%
	\caption{Results aggregated for all solutions strategies (blue) compared to those corresponding to the overall best strategy (orange) w.r.t. to $\Delta$ and $\varepsilon_{\beta}$}
	\label{fig:Problems_Beta_Comparison}
\end{figure}

To end this benchmark, the methods which were ranked most often best are compared with the overall pool of strategies. This is shown in Figure~\ref{fig:Problems_Beta_Comparison}. More specifically, Figures~\ref{fig:Problems_Beta_Comparison_a}~and~\ref{fig:Problems_Beta_Comparison_b} show comparison on the $\Delta$-criterion and relative error respectively, with respect to the combination of PC-Kriging with subset simulation, deviation number $U$ and Combined criterion. This is the approach that was consistently best both in the strategies and methods ranking. In both cases, we can observe that this choice of methods improves the performance on most problems and also reduce the scatter in the results. This graph also confirms the no-free-lunch principle exhibited by this benchmark as the overall best strategy is not necessarily the best for each problem.

\subsection{Investigation of the most difficult problems}
\rev{In this section, we closely look at the three problems that were most difficult to solve using the proposed ALR methods. By doing so, we note that some of these problems could not have been solved correctly even when considering the direct reliability estimation algorithms (without surrogates). Problem $\#11$ contains four disjoint failure regions, which makes it impossible to solve using the standard importance sampling configuration considered in this benchmark. 
	This also applies to the problem $\#8$, which is spherical along the variables $\acc{X_2, X_3 \enum X_{100}}$. 
	Furthermore, problem $\#11$ cannot be solved with direct Monte Carlo simulation due to the low reference failure probability $p_f = 7.83 \cdot 10^{-7}$, which would require $N_\text{MCS}~\approx~10^{9-10}$  samples to achieve sufficient accuracy.}

\begin{figure}[!ht]
	\centering
	\includegraphics[width=0.75\textwidth]{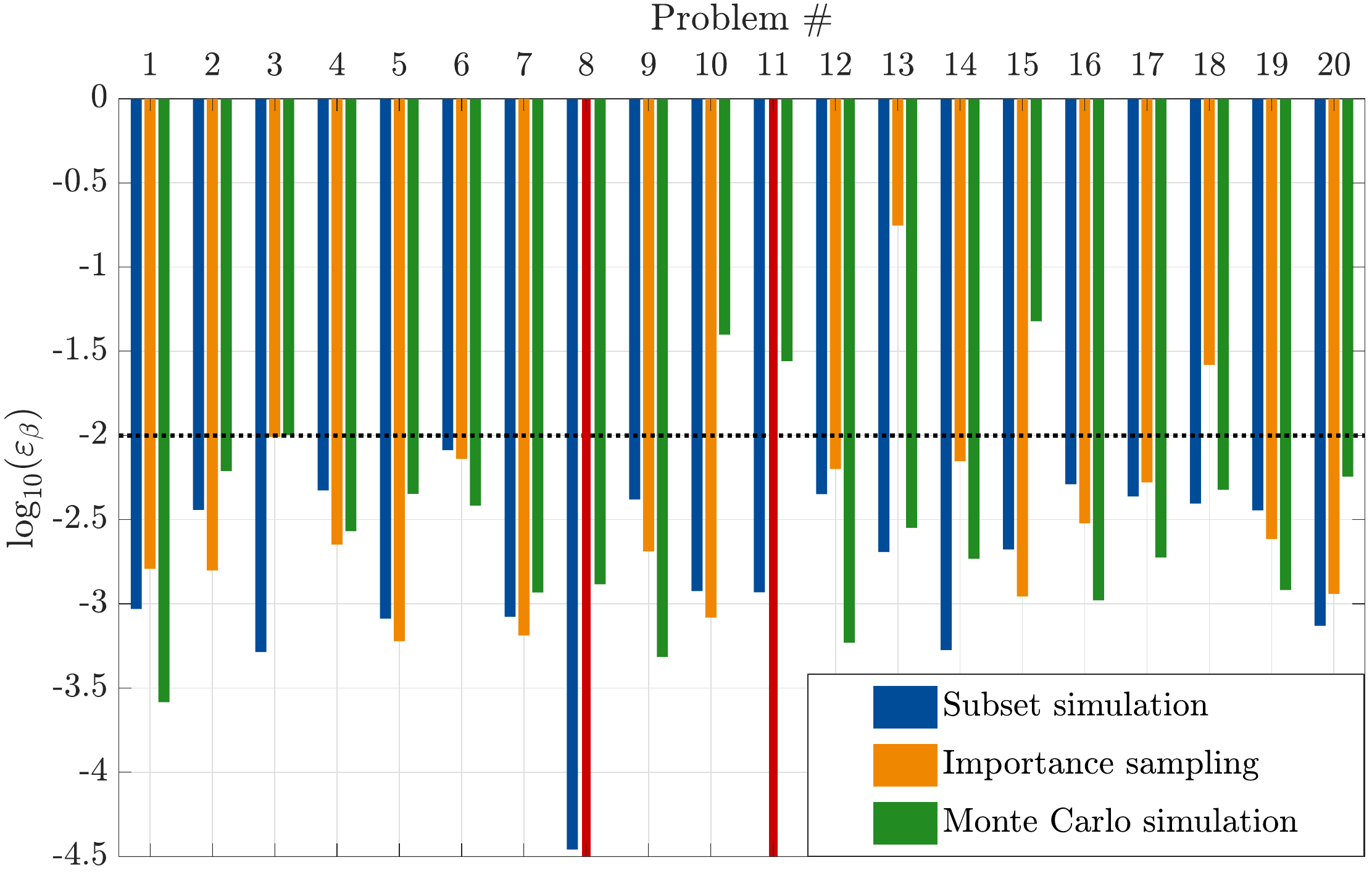}%
	\caption{Solution of the $20$ problems using the three ``overkill" reliability settings without surrogate models.}
	\label{fig:Overkill}
\end{figure}

These observations can be confirmed by Figure~\ref{fig:Overkill} which shows the results corresponding to the solution of the benchmark problems using the three reliability estimation  algorithms, with their overkill settings and without the aid of surrogate models. 
The red lines correspond to problems whose relative error is larger than $1$. 
Using the threshold for acceptable accuracy at $10^{-2}$ as in the previous sections, we can see that Monte Carlo simulation fails to solve problems $\#10, \, \#11,$ and $\#15$ while importance sampling fail with problems $\#8$, $\#11$, $\#13$, $\#18$.  This illustrates an important result of the benchmark, namely that the surrogate models were never the main cause of failure of the ALR strategies.

\section{Research questions and recommendations}\label{sec:Recommendations}
Our extensive benchmarking exercise allows us to showcase the framework introduced in this paper. 
We compared strategies built using this generalized active learning reliability framework with respect to various metrics. 
In this section, we summarize the findings from this benchmark and set up some guidelines as to the choice of the methods within the modules of the framework.

The first question that we set out to answer in this benchmark was whether there was one strategy that would consistently outperform the others with respect to all metrics and throughout all problems. 
The answer is clearly that there is no such strategy (no-free-lunch principle), yet using a surrogate model is always beneficial.
For each analysis, the best strategy would vary according to the metric of interest and the type of problems. 
A natural follow-up question is whether any pattern could be uncovered by the benchmark. 
Without prior knowledge about the problem and using a non-intrusive combination of different methods, the conclusion is sharp: the best results are most likely obtained by combining PC-Kriging, subset simulation, deviation number $U$ and $\beta$-bounds or combined stopping criterion. 

This conclusion is obviously only limited to the methods selected for the benchmark. 
However, we may in some cases extrapolate to general guidelines by considering the characteristics of the different methods in relation to various features of the problems solved in the benchmark. 
We made the following observations regarding each module of the framework:
\begin{itemize}
	\item \textbf{Surrogate model:} From this benchmark, it is clear that Kriging, which is the most used method in the literature, is not necessarily the best choice for active learning reliability. 
	The main reason is that it performs poorly in fairly high-dimensional problems. 
	PC-Kriging, which  combines the global and local approximations capabilities of PCE and Kriging respectively, has shown consistently high performance throughout this benchmark. 
	PCE has shown to \rev{perform} better than Kriging for high-dimensional problems and this also benefits PC-Kriging to some extent. 
	Finally, PC-Kriging possesses the same built-in error measure as Kriging, which makes it compatible with the various learning functions that take advantage of the Kriging variance. 
	\item \textbf{Reliability estimation algorithm:} From this benchmark, it is clear that the active learning scheme inherits the pros and cons of the reliability estimation algorithm it uses, although this is somewhat mitigated by the ability to use ``overkill" configurations.
	There is a direct correlation between the ability of a reliability estimation method to solve a given type of problems and the performance of the associated active learning scheme. 
	Once again, Monte Carlo simulation, which is widely used in the literature, has proven not to be the best choice of algorithm according to the benchmark results. 
	The introduction of surrogates does not eliminate the weakness of Monte Carlo simulation indeed when it comes to problems with extremely small failure probabilities.
	This is true also for the other algorithms, as importance sampling still showed limitations with problems where multiple failure regions exist. Note that other importance sampling densities may be used in practice when such a multiple-failure feature is known in advance (this was not considered in the benchmark).
	Regardless of the three algorithms used in the benchmark, it seems that the best course of action is to choose whatever algorithm the analyst thinks is best given his \emph{a priori} knowledge on the problem. 
	Finally, an important result highlighted by the benchmark is that over-calibrating the reliability estimation algorithm is beneficial and can lead to even better results than the non-surrogate equivalent using conventional settings. 
	Therefore, surrogate models should be used to fully harness the benefits of the most sophisticated reliability estimation methods. 
	\item \textbf{Learning function:} Considering this benchmark, the deviation number $U$ seems to outperform $EFF$. 
	The fraction of bootstrap replicates ($FBR$) is better than $U$ when it comes to high-dimensional problems but loses its advantage when considering problems with small failure probabilities. 
	All the methods selected for the benchmark are however very similar. It would have been interesting in this specific case to explore other type of learning functions, \emph{e.g.}, those that also include the PDF of the input random variables.   
	\item \textbf{Stopping criterion:} The stopping criteria selected for the benchmark are based on the accuracy of the $P_f$ (or $\beta$) estimate rather than on the learning function. 
	This finding is consistent with those identified in previous contributions \citep{SchoebiASCE2016}. 
	This benchmark has however highlighted the difference between  stopping criteria based on the local accuracy of the surrogate with those based on the stability of the limit-state surface. 
	The former are shown to provide good results overall but to perform poorly in high-dimensional problems. 
	This can be explained by the difficulty to sufficiently reduce the Kriging variance for high-dimensional problems. 
	In contrast, stability criteria are somewhat more efficient for high-dimensional problems. 
	However, they are prone to premature convergence. 
	Combining these two criteria did not lead to a noticeable increase in performance.
\end{itemize}

To further summarize the observations made by analyzing the results of the benchmark, Table~\ref{tab:Recommendations} gives a few recommendations on the basis of the benchmark.
\begin{table}[!ht]
	\centering
	\caption{General recommendations on the basis of the benchmark carried out in this paper. $\beta_{\textrm{bo}}$ stands for $\beta$-bounds convergence, $\beta_{\textrm{stab}}$ stands for $\beta$-stability convergence and $\beta_{\textrm{co}}$ is the combined criterion.}
	\label{tab:Recommendations}
	\begin{threeparttable}[t]
		\begin{tabular}{lcccc}
			\hline
			Module & \multicolumn{2}{c}{Dimensionality} & \multicolumn{2}{c}{Failure probability
				magnitude} \\
			& $M < 20$ &  $20 \leq M \leq 100$ & $\beta \leq 3.5$ & $\beta \geq 3.5$ \\ \hline
			Surrogate model & PCK & PCE & PCE/PCK & PCK \\
			Reliability estimation algorithm & SuS & SuS & SuS & SuS \\
			Learning function & U & FBR & FBR/U & U\\
			Stopping criterion & $\beta_{\textrm{bo}}$,$\beta_{\textrm{co}}$ & $\beta_{\textrm{bo}}$,$\beta_{\textrm{co}}$\tnote{1} / $\beta_{\textrm{stab}}$\tnote{2} & $\beta_{\textrm{bo}}$,$\beta_{\textrm{co}}$ & $\beta_{\textrm{bo}}$ \\ \hline
		\end{tabular}
		\begin{tablenotes}
			\item[1] When considering accuracy only $\varepsilon_{\beta}$.
			\item[2] When factoring in efficiency $\Delta$.
		\end{tablenotes}
	\end{threeparttable}
\end{table}

\section{Conclusions}\label{sec:Conclusion}
This paper investigated the use of active learning strategies for the solution of structural reliability problems. 
We first conducted a literature survey and identified an underlying and recurring scheme. 
This scheme was used to propose a global framework for active learning reliability which is made of four components non-intrusively linked to each other. 
The four modules of this framework are surrogate model, reliability analysis, learning function and stopping criterion. 
We then showed that it is possible to combine various methods from each of the modules to build a wide array of viable solution strategies. 
On this basis, $39$ solution strategies were built to solve a set of $20$ selected problems. 
The results of this benchmark allowed us to identify patterns regarding the generalization capability of each method. 

The first observation is that there is no strategy that consistently outperforms the others. We could however identify clear patterns to give recommendations on which types of methods should be preferred with regard to a given feature of the problem at hand. The flexibility of the presented framework is in this regard of great value as it allows the analyst to build tailored active learning reliability schemes.

The second observation is that there is essentially no drawbacks in using surrogate models. The latter allows one indeed to better exploit the reliability estimation algorithms through ``overkill" settings.

Even though we ran an extensive benchmark a few aspects still need to be investigated. 
For instance, we did not explore the effect of the thresholds in the stopping criterion, nor did we explore more advanced learning functions. 
This includes techniques that weigh in the input PDF or allow for multiple points enrichment. 
By design, the scope of the analysis was limited to problems with single limit-state functions. 
More aspects need to be taken into account when considering multiple limit-state functions.
Finally, extremely high-dimensional problems ($M$ in the other of hundreds or even thousands) were not considered as they also would require special treatment. Finally the proposed optimal framework has been applied blindly to a structural reliability context organized by TNO \citep{Rozsas2019}, whose \rev{results are available at \href{https://rprepo.readthedocs.io/en/latest/results.html}{https://rprepo.readthedocs.io/en/latest/results.html} (last accessed on October 4th, 2021)}. Out of $16$ component- and $11$ system-reliability problems, our approach was the most efficient for $24$ problems. A short summary of the results is provided in \ref{sec:TNO} for the sake of completeness.

It is worth mentioning that the codes and the set of examples used are made publicly available through the \textsc{UQLab} release R1.4, such that it will be easy to extend the results presented here to new problems. 

\bibliographystyle{chicago}
\bibliography{ALR_Paper_Biblio}

\begin{thebibliography}{}

\bibitem[\protect\citeauthoryear{Au and Beck}{Au and Beck}{2001}]{Au2001}
Au, S.~K. and J.~L. Beck (2001).
\newblock Estimation of small failure probabilities in high dimensions by
  subset simulation.
\newblock {\em Prob. Eng. Mech.\/}~{\em 16\/}(4), 263--277.

\bibitem[\protect\citeauthoryear{Au and Beck}{Au and Beck}{2003}]{Au2003}
Au, S.~K. and J.~L. Beck (2003).
\newblock Subset simulation and its application to seismic risk based on
  dynamic analysis.
\newblock {\em J. Eng. Mech.\/}~{\em 129\/}(8), 901--917.

\bibitem[\protect\citeauthoryear{Balesdent, Morio, and Marzat}{Balesdent
  et~al.}{2013}]{Balesdent2013}
Balesdent, M., J.~Morio, and J.~Marzat (2013).
\newblock {Kriging-based adaptive Importance Sampling algorithms for rare event
  estimation}.
\newblock {\em Structural Safety\/}~{\em 44}, 1--10.

\bibitem[\protect\citeauthoryear{Basudhar and Missoum}{Basudhar and
  Missoum}{2008}]{Basudhar2010}
Basudhar, A. and S.~Missoum (2008).
\newblock An improved adaptive sampling scheme for the construction of explicit
  boundaries.
\newblock {\em Struct. Multidisc. Optim.\/}~{\em 42\/}(4), 517--529.

\bibitem[\protect\citeauthoryear{Bect, Li, and Vazquez}{Bect
  et~al.}{2017}]{Bect2017}
Bect, J., L.~Li, and E.~Vazquez (2017).
\newblock Bayesian subset simulation.
\newblock {\em SIAM/ASA J. Unc. Quant.\/}~{\em 5}, 762--786.

\bibitem[\protect\citeauthoryear{Bichon, Eldred, Swiler, Mahadevan, and
  McFarland}{Bichon et~al.}{2008}]{Bichon2008}
Bichon, B.~J., M.~S. Eldred, L.~Swiler, S.~Mahadevan, and J.~McFarland (2008).
\newblock Efficient global reliability analysis for nonlinear implicit
  performance functions.
\newblock {\em AIAA Journal\/}~{\em 46\/}(10), 2459--2468.

\bibitem[\protect\citeauthoryear{Blatman and Sudret}{Blatman and
  Sudret}{2010}]{BlatmanPEM2010}
Blatman, G. and B.~Sudret (2010).
\newblock An adaptive algorithm to build up sparse polynomial chaos expansions
  for stochastic finite element analysis.
\newblock {\em Prob. Eng. Mech.\/}~{\em 25}, 183--197.

\bibitem[\protect\citeauthoryear{Blatman and Sudret}{Blatman and
  Sudret}{2011}]{BlatmanJCP2011}
Blatman, G. and B.~Sudret (2011).
\newblock {Adaptive sparse polynomial chaos expansion based on Least Angle
  Regression}.
\newblock {\em J. Comput. Phys\/}~{\em 230}, 2345--2367.

\bibitem[\protect\citeauthoryear{Bo and HuiFeng}{Bo and HuiFeng}{2018}]{Bo2017}
Bo, X. and T.~HuiFeng (2018).
\newblock A robust and efficient structural reliability method combining
  radial-based importance sampling and {K}riging.
\newblock {\em Sci. China Technol. Sci.\/}~{\em 61}, 724--734.

\bibitem[\protect\citeauthoryear{Bourinet}{Bourinet}{2017}]{Bourinet2017}
Bourinet, J.-M. (2017).
\newblock Anisotropic-kernel-based support vector regression for reliability
  assessment.
\newblock In {\em Proc. 12th~International Conference on Structural Safety and
  Reliability (ICOSSAR), August 6-10, 2017, Vienna, Austria}.

\bibitem[\protect\citeauthoryear{Bourinet}{Bourinet}{2018}]{BourinetHDR}
Bourinet, J.-M. (2018).
\newblock {\em Reliability analysis and optimal design under uncertainty -
  Focus on adaptive surrogate-based approaches}.
\newblock Universit\'e Blaise Pascal, Clermont-Ferrand, France.
\newblock Habilitation \`a diriger des recherches, 243 pages.

\bibitem[\protect\citeauthoryear{Bourinet, Deheeger, and Lemaire}{Bourinet
  et~al.}{2011}]{Bourinet2011}
Bourinet, J.-M., F.~Deheeger, and M.~Lemaire (2011).
\newblock Assessing small failure probabilities by combined subset simulation
  and support vector machines.
\newblock {\em Structural Safety\/}~{\em 33\/}(6), 343--353.

\bibitem[\protect\citeauthoryear{Bucher}{Bucher}{2009}]{Bucher2009}
Bucher, C. (2009).
\newblock Asymptotic sampling for high-dimensional reliability analysis.
\newblock {\em Prob. Eng. Mech.\/}~{\em 24}, 504--510.

\bibitem[\protect\citeauthoryear{Bucher and Bourgund}{Bucher and
  Bourgund}{1990}]{Bucher1990}
Bucher, C. and U.~Bourgund (1990).
\newblock A fast and efficient response surface approach for structural
  reliability problems.
\newblock {\em Structural Safety\/}~{\em 7}, 57--66.

\bibitem[\protect\citeauthoryear{Cadini, Santos, and Zio}{Cadini
  et~al.}{2014}]{Cadini2014}
Cadini, F., F.~Santos, and E.~Zio (2014).
\newblock {An improved adaptive Kriging-based importance technique for sampling
  multiple failure regions of low probability}.
\newblock {\em Reliab. Eng. Syst. Saf.\/}~{\em 139}, 109--117.

\bibitem[\protect\citeauthoryear{Cheng and Lu}{Cheng and Lu}{2020}]{Cheng2020}
Cheng, K. and Z.~Lu (2020).
\newblock Active learning polynomial chaos expansion for reliability analysis
  by maximizing expected indicator function predictor error.
\newblock {\em Structural Safety\/}~{\em 82}, 1--13.

\bibitem[\protect\citeauthoryear{Chojazyk, Teixeira, Neves, Cardoso, and
  Guedes~Soares}{Chojazyk et~al.}{2015}]{Chojazyk2015}
Chojazyk, A.~A., A.~P. Teixeira, L.~C. Neves, J.~B. Cardoso, and
  C.~Guedes~Soares (2015).
\newblock Review and application of artificial neural networks models in
  reliability analysis of steel structures.
\newblock {\em Structural Safety\/}~{\em 52}, 78--89.

\bibitem[\protect\citeauthoryear{Constantine, Dow, and Wang}{Constantine
  et~al.}{2014}]{Constantine2014}
Constantine, P.~G., E.~Dow, and Q.~Wang (2014).
\newblock Active subspace methods in theory and practice: Applications to {K}
  riging surfaces.
\newblock {\em SIAM J. Sci. Comput.\/}~{\em 36\/}(4), A1500--A1524.

\bibitem[\protect\citeauthoryear{Deheeger and Lemaire}{Deheeger and
  Lemaire}{2007}]{Deheeger2007}
Deheeger, F. and M.~Lemaire (2007).
\newblock Support vector machines for efficient subset simulations:
  {${}^2$SMART}~method.
\newblock In {\em Proc. 10th~Int. Conf. on Applications of Stat. and Prob. in
  Civil Engineering (ICASP10), Tokyo, Japan}.

\bibitem[\protect\citeauthoryear{{Der Kiureghian} and {de Stefano}}{{Der
  Kiureghian} and {de Stefano}}{1990}]{DerKiureghian1990}
{Der Kiureghian}, A. and M.~{de Stefano} (1990).
\newblock An efficient algorithm for second-order reliability analysis.
\newblock Technical Report UCB/SEMM-90/20, University of California, Berkeley.
\newblock Dept of Civil and Environmental Engineering, University of
  California, Berkeley.

\bibitem[\protect\citeauthoryear{Ditlevsen and Madsen}{Ditlevsen and
  Madsen}{1996}]{Ditlevsen1996}
Ditlevsen, O. and H.~Madsen (1996).
\newblock {\em Structural reliability methods}.
\newblock {J.~Wiley and Sons, Chichester}.

\bibitem[\protect\citeauthoryear{Ditlevsen, Melchers, and Gluver}{Ditlevsen
  et~al.}{1990}]{Ditlevsen1990}
Ditlevsen, O., R.~E. Melchers, and H.~Gluver (1990).
\newblock General multi-dimensional probability integration by directional
  simulation.
\newblock {\em Comput. Struct.\/}~{\em 36\/}(2), 355--368.

\bibitem[\protect\citeauthoryear{Dubourg, Sudret, and Bourinet}{Dubourg
  et~al.}{2012}]{Dubourg2012}
Dubourg, V., B.~Sudret, and J.-M. Bourinet (2012).
\newblock Meta-model-based importance sampling for reliability sensitivity
  analysis.
\newblock In {\em Proc. 11th~ASCE Specialty Conference on Probabilistic
  Mechanics and Structural Reliability, Notre Dame, USA}.

\bibitem[\protect\citeauthoryear{Dubourg, Sudret, and Deheeger}{Dubourg
  et~al.}{2013}]{Dubourg2013}
Dubourg, V., B.~Sudret, and F.~Deheeger (2013).
\newblock Metamodel-based importance sampling for structural reliability
  analysis.
\newblock {\em Prob. Eng. Mech.\/}~{\em 33}, 47--57.

\bibitem[\protect\citeauthoryear{Echard, Gayton, and Lemaire}{Echard
  et~al.}{2011}]{Echard2011}
Echard, B., N.~Gayton, and M.~Lemaire (2011).
\newblock {AK-MCS}: an active learning reliability method combining {K}riging
  and {M}onte {C}arlo simulation.
\newblock {\em Structural Safety\/}~{\em 33\/}(2), 145--154.

\bibitem[\protect\citeauthoryear{Echard, Gayton, Lemaire, and Relun}{Echard
  et~al.}{2013}]{Echard2013}
Echard, B., N.~Gayton, M.~Lemaire, and N.~Relun (2013).
\newblock {A combined importance sampling and Kriging reliability method for
  small failure probabilities with time-demanding numerical models}.
\newblock {\em Reliab. Eng. Syst. Safety\/}~{\em 111}, 232--240.

\bibitem[\protect\citeauthoryear{Efron, Hastie, Johnstone, and
  Tibshirani}{Efron et~al.}{2004}]{Efron2004}
Efron, B., T.~Hastie, I.~Johnstone, and R.~Tibshirani (2004).
\newblock Least angle regression.
\newblock {\em Ann. Stat.\/}~{\em 32}, 407--499.

\bibitem[\protect\citeauthoryear{Faravelli}{Faravelli}{1989}]{Faravelli1989}
Faravelli, L. (1989).
\newblock Response surface approach for reliability analysis.
\newblock {\em J. Eng. Mech.\/}~{\em 115\/}(12), 2763--2781.

\bibitem[\protect\citeauthoryear{Fauriat and Gayton}{Fauriat and
  Gayton}{2017}]{Fauriat2014}
Fauriat, W. and N.~Gayton (2017).
\newblock {AK-SYS: An application of the AK-MCS method for system reliability}.
\newblock {\em Reliab. Eng. Struct. Safety\/}~{\em 123}, 137--144.

\bibitem[\protect\citeauthoryear{Gaspar, Teixeira, and Guedes~Soares}{Gaspar
  et~al.}{2017}]{Gaspar2017}
Gaspar, B., A.~P. Teixeira, and C.~Guedes~Soares (2017).
\newblock Adaptive surrogate model with active refinement combining {K}riging
  and a trust region method.
\newblock {\em Reliab. Eng. Syst. Saf.\/}~{\em 165}, 277--291.

\bibitem[\protect\citeauthoryear{Geyer, Papaioannou, and Straub}{Geyer
  et~al.}{2019}]{GeyerSS2019}
Geyer, S., I.~Papaioannou, and D.~Straub (2019).
\newblock Cross entropy-based importance sampling using {Gaussian} densities
  revisited.
\newblock {\em Structural Safety\/}~{\em 76}, 15--27.

\bibitem[\protect\citeauthoryear{Gomes}{Gomes}{2019}]{Gomes2019}
Gomes, W. J.~S. (2019).
\newblock Structural reliability analysis using artificial neural networks.
\newblock {\em {ASCE-ASME J. of Risk Uncertain. Eng. Syst., Part B:
  Mech.Eng.}\/}~{\em 5}, 1--8.

\bibitem[\protect\citeauthoryear{Guo, Liu, Chen, and Zhao}{Guo
  et~al.}{2020}]{Guo2020}
Guo, Q., Y.~Liu, B.~Chen, and Y.~Zhao (2020).
\newblock An active learning {K}riging model combined with directional
  importance sampling method for efficient reliability analysis.
\newblock {\em Prob. Eng. Mech.\/}~{\em 60}, 1--9.

\bibitem[\protect\citeauthoryear{Hasofer and Lind}{Hasofer and
  Lind}{1974}]{Hasofer1974}
Hasofer, A.-M. and N.-C. Lind (1974).
\newblock Exact and invariant second moment code format.
\newblock {\em J. Eng. Mech.\/}~{\em 100\/}(1), 111--121.

\bibitem[\protect\citeauthoryear{Hu and Mahadevan}{Hu and
  Mahadevan}{2016}]{Hu2016}
Hu, Z. and S.~Mahadevan (2016).
\newblock {Global sensitivity analysis-enhanced surrogate (GSAS) modeling for
  reliability analysis}.
\newblock {\em Struct. Multidisc. Optim.\/}~{\em 53}, 501--521.

\bibitem[\protect\citeauthoryear{Huang, Chen, and Zhu}{Huang
  et~al.}{2016}]{Huang2016}
Huang, X., J.~Chen, and H.~Zhu (2016).
\newblock Assessing small failure probabilities by {AK–SS}: An active
  learning method combining {K}riging and subset simulation.
\newblock {\em Structural Safety\/}~{\em 59}, 86--95.

\bibitem[\protect\citeauthoryear{Hurtado}{Hurtado}{2004}]{Hurtado2004a}
Hurtado, J.~E. (2004).
\newblock An examination of methods for approximating implicit limit state
  functions from the viewpoint of statistical learning theory.
\newblock {\em Structural Safety\/}~{\em 26}, 271--293.

\bibitem[\protect\citeauthoryear{Jian, Zhili, Qiang, and Rui}{Jian
  et~al.}{2017}]{Jian2017}
Jian, W., S.~Zhili, Y.~Qiang, and L.~Rui (2017).
\newblock Two accuracy measures of the {K}riging model for structural
  reliability analysis.
\newblock {\em Reliab. Eng. Sys. Safety\/}~{\em 167}, 494--505.

\bibitem[\protect\citeauthoryear{Jiang, Qiu, Yang, Chen, Gao, and Li}{Jiang
  et~al.}{2019}]{Jiang2019}
Jiang, C., H.~Qiu, Z.~Yang, L.~Chen, L.~Gao, and P.~Li (2019).
\newblock A general failure-pursuing sampling framework for surrogate-based
  reliability analysis.
\newblock {\em Reliab. Eng. Struct. Safety\/}~{\em 183}, 47--59.

\bibitem[\protect\citeauthoryear{Jones, Schonlau, and Welch}{Jones
  et~al.}{1998}]{Jones1998}
Jones, D.~R., M.~Schonlau, and W.~J. Welch (1998).
\newblock Efficient global optimization of expensive black-box functions.
\newblock {\em J. Global Optim.\/}~{\em 13\/}(4), 455--492.

\bibitem[\protect\citeauthoryear{Koutsourelakis, Pradlwarter, and
  Schu\"eller}{Koutsourelakis et~al.}{2004}]{Koutsourelakis2004}
Koutsourelakis, P.~S., H.~J. Pradlwarter, and G.~I. Schu\"eller (2004).
\newblock Reliability of structures in high dimensions, part {I}: algorithms
  and applications.
\newblock {\em Prob. Eng. Mech.\/}~{\em 19}, 409--417.

\bibitem[\protect\citeauthoryear{Kroetz, Moustapha, Beck, and Sudret}{Kroetz
  et~al.}{2020}]{Kroetz2020}
Kroetz, H.~M., M.~Moustapha, A.~T. Beck, and B.~Sudret (2020).
\newblock A two-level {K}riging-based approach with active learning for solving
  time-variant risk optimization problems.
\newblock {\em Reliab. Eng. Syst. Saf.\/}~{\em 203}, 107033.

\bibitem[\protect\citeauthoryear{Kroetz, Tessari, and Beck}{Kroetz
  et~al.}{2017}]{Kroetz2017}
Kroetz, H.~M., R.~K. Tessari, and A.~T. Beck (2017).
\newblock Performance of global metamodeling techniques in solution of
  structural reliability problems.
\newblock {\em Adv. Eng. Softw.\/}~{\em 114}, 394--404.

\bibitem[\protect\citeauthoryear{Lacaze and Missoum}{Lacaze and
  Missoum}{2014}]{Lacaze2014}
Lacaze, S. and S.~Missoum (2014).
\newblock A generalized “max-min” sample for surrogate update.
\newblock {\em Struct. Multidisc. Optim.\/}~{\em 49}, 683--687.

\bibitem[\protect\citeauthoryear{Lataniotis, Marelli, and Sudret}{Lataniotis
  et~al.}{2017}]{UQdoc_10_105}
Lataniotis, C., S.~Marelli, and B.~Sudret (2017).
\newblock {UQLab} user manual -- {K}riging.
\newblock Technical report, Chair of Risk, Safety \& Uncertainty
  Quantification, ETH Zurich.
\newblock Report \# UQLab-V1.0-105.

\bibitem[\protect\citeauthoryear{Lataniotis, Marelli, and Sudret}{Lataniotis
  et~al.}{2020}]{Lataniotis2020}
Lataniotis, C., S.~Marelli, and B.~Sudret (2020).
\newblock Extending classical surrogate modeling to high dimensions through
  supervised dimensionality reduction: a data-driven approach.
\newblock {\em Int. J. Uncertainty Quantification\/}~{\em 10\/}(1), 55--82.

\bibitem[\protect\citeauthoryear{Leli\`evre, Beaurepaire, Mattrand, and
  Gayton}{Leli\`evre et~al.}{2018}]{Lelievre2018}
Leli\`evre, N., P.~Beaurepaire, C.~Mattrand, and N.~Gayton (2018).
\newblock {AK-MCSi}: {A K}riging-based method to deal with small failure
  probabilities and time-consuming models.
\newblock {\em Structural Safety\/}~{\em 73}, 1--11.

\bibitem[\protect\citeauthoryear{Lemaire}{Lemaire}{1998}]{LeMaire1998}
Lemaire, M. (1998).
\newblock Finite element and reliability : combined methods by response
  surfaces.
\newblock In G.~Frantziskonis (Ed.), {\em Probamat-21st Century, Probabilities
  and Materials~: Tests, Models and Applications for the 21st century}, pp.\
  317--331. Kluwer Academic Publishers.

\bibitem[\protect\citeauthoryear{Lemaire}{Lemaire}{2009}]{LeMaire2009}
Lemaire, M. (2009).
\newblock {\em Structural reliability}.
\newblock Wiley.

\bibitem[\protect\citeauthoryear{Leonel, Beck, and Venturini}{Leonel
  et~al.}{2011}]{Leonel2011}
Leonel, E.~D., A.~T. Beck, and W.~S. Venturini (2011).
\newblock On the performance of response surface and direct coupling
  approachesin solution of random crack propagation problems.
\newblock {\em Struct. Saf.\/}~{\em 33}, 261--274.

\bibitem[\protect\citeauthoryear{Li, Bect, and Vazquez}{Li
  et~al.}{2012}]{Li2012}
Li, L., J.~Bect, and E.~Vazquez (2012).
\newblock Bayesian subset simulation: a {K}riging-based subset simulation
  algorithm for the estimation of small failure probabilities.
\newblock In {\em 11th International Probabilistic Assessment and Management
  Conference (PSAM11) and The Annual European Safety and Reliability Conference
  (ESREL 2012), Helsinki : Finland (2012)}. Curran.

\bibitem[\protect\citeauthoryear{Li and Wang}{Li and Wang}{2020}]{Li2020}
Li, M. and Z.~Wang (2020).
\newblock Deep learning for high-dimensional reliability analysis.
\newblock {\em Mechanical Systems and Signal Processing\/}~{\em 139}, 1--18.

\bibitem[\protect\citeauthoryear{Li, Gong, Gu, Gao, Jing, and Su}{Li
  et~al.}{2018}]{Li2018}
Li, X., C.~Gong, L.~Gu, W.~Gao, Z.~Jing, and H.~Su (2018).
\newblock A sequential surrogate method for reliability analysis based on
  radial basis function.
\newblock {\em Structural Safety\/}~{\em 73}, 42--53.

\bibitem[\protect\citeauthoryear{Ling, Lu, Feng, and Zhang}{Ling
  et~al.}{2019}]{Ling2019}
Ling, C., Z.~Lu, K.~Feng, and X.~Zhang (2019).
\newblock A coupled subset simulation and active learning {K}riging reliability
  analysis method for rare failure events.
\newblock {\em Struct. Multidisc. Optim.\/}~{\em 60}, 2325--2341.

\bibitem[\protect\citeauthoryear{Liu, Wei, Zhou, and Yue}{Liu
  et~al.}{2019}]{Liu2019}
Liu, F., P.~Wei, C.~Zhou, and Z.~Yue (2019).
\newblock Reliability and reliability sensitivity analysis of structure by
  combining adaptive linked importance sampling and {K}riging reliability
  method.
\newblock {\em Chinese Journal of aeronautics\/}~{\em 33}, 1218--1227.

\bibitem[\protect\citeauthoryear{Lv, Lu, and Wang}{Lv et~al.}{2015}]{Lv2015}
Lv, Z., Z.~Lu, and P.~Wang (2015).
\newblock {A new learning function for Kriging and its applications to solve
  reliability problems in engineering}.
\newblock {\em Comput. Math. Appl.\/}~{\em 70}, 1182--1197.

\bibitem[\protect\citeauthoryear{Marelli and Sudret}{Marelli and
  Sudret}{2014}]{MarelliUQLab2014}
Marelli, S. and B.~Sudret (2014).
\newblock {UQLab}: A framework for uncertainty quantification in {Matlab}.
\newblock In {\em Vulnerability, Uncertainty, and Risk (Proc. 2nd Int. Conf. on
  Vulnerability, Risk Analysis and Management {(ICVRAM2014)}, Liverpool, United
  Kingdom)}, pp.\  2554--2563.

\bibitem[\protect\citeauthoryear{Marelli and Sudret}{Marelli and
  Sudret}{2016}]{MarelliAPSRRA2016}
Marelli, S. and B.~Sudret (2016).
\newblock Bootstrap-polynomial chaos expansions and adaptive designs for
  reliability analysis.
\newblock In H.~Huang, J.~Li, J.~Zhang, and J.~Chen (Eds.), {\em Proc. {6th}
  {Asian-Pacific} {S}ymp. Struct. Reliab. {(APSSRA'2016)}, Tongji University,
  Shanghai (China), May 27-31.}

\bibitem[\protect\citeauthoryear{Marelli and Sudret}{Marelli and
  Sudret}{2017}]{UQdoc_10_104}
Marelli, S. and B.~Sudret (2017).
\newblock {UQLab} user manual -- {P}olynomial chaos expansions.
\newblock Technical report, Chair of Risk, Safety \& Uncertainty
  Quantification, ETH Zurich.
\newblock Report \# UQLab-V1.0-104.

\bibitem[\protect\citeauthoryear{Marelli and Sudret}{Marelli and
  Sudret}{2018}]{MarelliSS2018}
Marelli, S. and B.~Sudret (2018).
\newblock An active-learning algorithm that combines sparse polynomial chaos
  expansions and bootstrap for structural reliability analysis.
\newblock {\em Structural Safety\/}~{\em 75}, 67--74.

\bibitem[\protect\citeauthoryear{McKay, Beckman, and Conover}{McKay
  et~al.}{1979}]{McKay1979}
McKay, M.~D., R.~J. Beckman, and W.~J. Conover (1979).
\newblock A comparison of three methods for selecting values of input variables
  in the analysis of output from a computer code.
\newblock {\em Technometrics\/}~{\em 2}, 239--245.

\bibitem[\protect\citeauthoryear{Melchers}{Melchers}{2018}]{Melchers2018}
Melchers, R. E. amd~Beck, A.~T. (2018).
\newblock {\em Structural reliability analysis and prediction}.
\newblock John Wiley \& Sons.

\bibitem[\protect\citeauthoryear{Melchers}{Melchers}{1989}]{Melchers1989}
Melchers, R.~E. (1989).
\newblock Importance sampling in structural systems.
\newblock {\em Structural Safety\/}~{\em 6}, 3--10.

\bibitem[\protect\citeauthoryear{Moustapha, Sudret, Bourinet, and
  Guillaume}{Moustapha et~al.}{2016}]{MoustaphaSMO2016}
Moustapha, M., B.~Sudret, J.-M. Bourinet, and B.~Guillaume (2016).
\newblock Quantile-based optimization under uncertainties using adaptive
  {K}riging surrogate models.
\newblock {\em Struct. Multidisc. Optim.\/}~{\em 54\/}(6), 1403--1421.

\bibitem[\protect\citeauthoryear{Pan and Dias}{Pan and Dias}{2017}]{Pan2017}
Pan, Q. and D.~Dias (2017).
\newblock {An efficient reliability method combining adaptive support vector
  machines and Monte Carlo Simulation}.
\newblock {\em Structural Safety\/}~{\em 67}, 85--95.

\bibitem[\protect\citeauthoryear{Pan, Qu, Liu, and Dias}{Pan
  et~al.}{2020}]{Pan2020}
Pan, Q., X.~Qu, L.~Liu, and D.~Dias (2020).
\newblock A sequential sparse polynomial chaos expansion using {B}ayesian
  regression for geotechnical reliability estimations.
\newblock {\em Int. J. Numer. Anal. Methods Geomech.\/}~{\em 44}, 874--889.

\bibitem[\protect\citeauthoryear{Papaioannou, Papadimitriou, and
  Straub}{Papaioannou et~al.}{2016}]{Papaioannou2016}
Papaioannou, I., C.~Papadimitriou, and D.~Straub (2016).
\newblock Sequential importance sampling for structural reliability analysis.
\newblock {\em Structural Safety\/}~{\em 62}, 66--75.

\bibitem[\protect\citeauthoryear{Peijuan, Ming, Zhouhong, and Liqi}{Peijuan
  et~al.}{2017}]{Peijuan2017}
Peijuan, Z., W.~Ming, Z.~Zhouhong, and W.~Liqi (2017).
\newblock {A new active learning method based on the learning function U of the
  AK-MCS reliability analysis method}.
\newblock {\em Eng. Struct.\/}~{\em 148}, 185--194.

\bibitem[\protect\citeauthoryear{Rackwitz and Fiessler}{Rackwitz and
  Fiessler}{1978}]{Rackwitz78}
Rackwitz, R. and B.~Fiessler (1978).
\newblock Structural reliability under combined load sequences.
\newblock {\em Comput. Struct.\/}~{\em 9}, 489--494.

\bibitem[\protect\citeauthoryear{Rajashekhar and Ellingwood}{Rajashekhar and
  Ellingwood}{1993}]{Rajashekhar1993}
Rajashekhar, M.-R. and B.-R. Ellingwood (1993).
\newblock A new look at the response surface approach for reliability analysis.
\newblock {\em Structural Safety\/}~{\em 12}, 205--220.

\bibitem[\protect\citeauthoryear{Ranjan, Bungham, and Michailidis}{Ranjan
  et~al.}{2008}]{Ranjan2008}
Ranjan, P., D.~Bungham, and G.~Michailidis (2008).
\newblock Sequential experiment design for contour estimation from complex
  computer codes.
\newblock {\em Technometrics\/}~{\em 50}, 527--541.

\bibitem[\protect\citeauthoryear{Rasmussen and Williams}{Rasmussen and
  Williams}{2006}]{Rasmussen2006}
Rasmussen, C.~E. and C.~K.~I. Williams (2006).
\newblock {\em Gaussian processes for machine learning\/} ({Internet} ed.).
\newblock Adaptive computation and machine learning. Cambridge, Massachusetts:
  MIT Press.

\bibitem[\protect\citeauthoryear{Razaaly and Congedo}{Razaaly and
  Congedo}{2018}]{Razaaly2018}
Razaaly, N. and P.~M. Congedo (2018).
\newblock Novel algorithm using active metamodel learning and importance
  sampling: {A}pplication to multiple failure regions of low probability.
\newblock {\em J. Comput. Phys.\/}~{\em 368}, 92--114.

\bibitem[\protect\citeauthoryear{Roussouly, Petitjean, and Salaun}{Roussouly
  et~al.}{2013}]{Roussouly2013}
Roussouly, N., F.~Petitjean, and M.~Salaun (2013).
\newblock A new adaptive response surface method for reliability analysis.
\newblock {\em Prob. Eng. Mech.\/}~{\em 32}, 103--115.

\bibitem[\protect\citeauthoryear{Rozsas and Slobbe}{Rozsas and
  Slobbe}{2019}]{Rozsas2019}
Rozsas, A. and A.~Slobbe (2019).
\newblock Repository and black-box reliability challenge 2019.
\newblock \textsf{https://gitlab.com/rozsasarpi/rprepo/}.
\newblock Accessed: 2021-05-04.

\bibitem[\protect\citeauthoryear{Sadoughi, Li, Hi, and Mackenzie}{Sadoughi
  et~al.}{2017}]{Sadoughi2017}
Sadoughi, M.~K., M.~Li, C.~Hi, and C.~A. Mackenzie (2017).
\newblock High-dimensional reliability analysis of engineered systems involving
  computationally expensive black-box simulations.
\newblock In {\em Proc. ASME 2017 International Design Engineering Technical
  Conferences and Computers and Information in Engineering Conference, August
  6-9, 2017, Cleveland, Ohia, USA}.

\bibitem[\protect\citeauthoryear{Santner, Williams, and Notz}{Santner
  et~al.}{2003}]{Santner2003}
Santner, T.~J., B.~J. Williams, and W.~I. Notz (2003).
\newblock {\em The {D}esign and {A}nalysis of {C}omputer {E}xperiments}.
\newblock Springer, New York.

\bibitem[\protect\citeauthoryear{Sch\"obi, Marelli, and Sudret}{Sch\"obi
  et~al.}{2017}]{UQdoc_10_109}
Sch\"obi, R., S.~Marelli, and B.~Sudret (2017).
\newblock {UQLab} user manual -- {PC-Kriging}.
\newblock Technical report, Chair of Risk, Safety \& Uncertainty
  Quantification, ETH Zurich.
\newblock Report \# UQLab-V1.0-109.

\bibitem[\protect\citeauthoryear{Sch\"obi, Sudret, and Marelli}{Sch\"obi
  et~al.}{2016}]{SchoebiASCE2016}
Sch\"obi, R., B.~Sudret, and S.~Marelli (2016).
\newblock Rare event estimation using {Polynomial-Chaos-Kriging}.
\newblock {\em ASCE-ASME J. Risk Uncertainty Eng. Syst., Part A: Civ.
  Eng.\/}~{\em 3\/}(2).
\newblock D4016002.

\bibitem[\protect\citeauthoryear{Shi, B., and Ibrahim}{Shi
  et~al.}{2019}]{Shi2019}
Shi, L., S.~B., and D.~S. Ibrahim (2019).
\newblock An active learning reliability method with multiple kernel functions
  based on radial basis function.
\newblock {\em Struct. Multidisc. Optim.\/}~{\em 60}, 211--229.

\bibitem[\protect\citeauthoryear{{Sobol'}}{{Sobol'}}{1967}]{Sobol1967}
{Sobol'}, I.~M. (1967).
\newblock Distribution of points in a cube and approximate evaluation of
  integrals.
\newblock {\em U.S.S.R Comput. Maths. Math. Phys.\/}~{\em 7}, 86--112.

\bibitem[\protect\citeauthoryear{Sun, Wang, Li, and Tong}{Sun
  et~al.}{2017}]{Sun2017}
Sun, Z., J.~Wang, R.~Li, and C.~Tong (2017).
\newblock {LIF: A new Kriging based learning function and its application to
  structural reliability analysis}.
\newblock {\em Reliab. Eng. Syst. Saf.\/}~{\em 157}, 152--165.

\bibitem[\protect\citeauthoryear{Sundar and Shields}{Sundar and
  Shields}{2016}]{Sundar2016}
Sundar, V. and M.~D. Shields (2016).
\newblock Surrogate-enhanced stochastic search algorithms to identify
  implicitly defined functions for reliability analysis.
\newblock {\em Structural Safety\/}~{\em 62}, 1--11.

\bibitem[\protect\citeauthoryear{Teixeira, Nogal, and O'Connor}{Teixeira
  et~al.}{2021}]{Teixeira2021}
Teixeira, R., M.~Nogal, and A.~O'Connor (2021).
\newblock Adaptive approaches in metamodel-based reliability analysis: {A}
  review.
\newblock {\em Structural Safety\/}~{\em 89}, 102019.

\bibitem[\protect\citeauthoryear{Tong, Sun, Zhao, Wang, and Wang}{Tong
  et~al.}{2015}]{Tong2015}
Tong, C., Z.~Sun, Q.~Zhao, Q.~Wang, and S.~Wang (2015).
\newblock A hybrid algorithm for reliability analysis combining {K}riging and
  subset simulation importance sampling.
\newblock {\em J. Mech sci. Tech.\/}~{\em 29}, 3183--3193.

\bibitem[\protect\citeauthoryear{Tong, Wang, and J.}{Tong
  et~al.}{2019}]{Tong2019}
Tong, C., J.~Wang, and L.~J. (2019).
\newblock A {K}riging-based active learning algorithm for mechanical
  reliability analysis with time-consuming and nonlinear response.
\newblock {\em Math. Probl. Eng.\/}~{\em 2019}, 1--14.

\bibitem[\protect\citeauthoryear{Wagner, Marelli, Papaioannou, Straub, and
  Sudret}{Wagner et~al.}{2021}]{Wagner2021}
Wagner, P.-R., S.~Marelli, I.~Papaioannou, D.~Straub, and B.~Sudret (2021).
\newblock Rare event estimation using stochastic spectral embedding.
\newblock {\em Structural Safety\/}.

\bibitem[\protect\citeauthoryear{Wang, Broccardo, and Song}{Wang
  et~al.}{2019}]{Wang2019}
Wang, Z., M.~Broccardo, and J.~Song (2019).
\newblock {Hamiltonian Monte Carlo methods for subset simulation in reliability
  analysis}.
\newblock {\em Structural Safety\/}~{\em 76}, 51--67.

\bibitem[\protect\citeauthoryear{Wen, Pei, Liu, and Yue}{Wen
  et~al.}{2016}]{Wen2016}
Wen, Z., H.~Pei, H.~Liu, and Z.~Yue (2016).
\newblock A sequential {K}riging reliability analysis method with
  characteristics of adaptive sampling regions and parallelizability.
\newblock {\em Reliab. Eng. Sys. Saf.\/}~{\em 153}, 170--179.

\bibitem[\protect\citeauthoryear{Xiao, Zuo, and Guo}{Xiao
  et~al.}{2018}]{Xiao2018a}
Xiao, N.-C., M.~J. Zuo, and W.~Guo (2018).
\newblock {Efficient reliability analysis based on adaptive sequential sampling
  design and cross-validation}.
\newblock {\em Appl. Math. Model.\/}~{\em 58}, 404--420.

\bibitem[\protect\citeauthoryear{Xiu and Karniadakis}{Xiu and
  Karniadakis}{2002}]{Xiu2002}
Xiu, D. and G.~E. Karniadakis (2002).
\newblock {The Wiener-Askey polynomial chaos for stochastic differential
  equations}.
\newblock {\em SIAM J. Sci. Comput.\/}~{\em 24\/}(2), 619--644.

\bibitem[\protect\citeauthoryear{Yang, Liu, Mi, and Wang}{Yang
  et~al.}{2018}]{Yangetal2018}
Yang, X., Y.~Liu, C.~Mi, and X.~Wang (2018).
\newblock Active learning {K}riging model combining with kernel-density
  estimation-based importance sampling method for the estimation of low failure
  probability.
\newblock {\em J. Mech. Des.\/}~{\em 140}, 1--9.

\bibitem[\protect\citeauthoryear{Zhang and Taflanidis}{Zhang and
  Taflanidis}{2018}]{Zhang2018}
Zhang, J. and A.~A. Taflanidis (2018).
\newblock Adaptive {K}riging stochastic sampling and density approximation and
  its application to rare-event estimation.
\newblock {\em {ASCE-ASME J. Risk Uncertainty Eng. Syst., Part A: Civ.
  Eng.}\/}~{\em 4}, 1--17.

\bibitem[\protect\citeauthoryear{Zhang, Xiao, and Gao}{Zhang
  et~al.}{2019}]{Zhang2019b}
Zhang, J., M.~Xiao, and L.~Gao (2019).
\newblock An active learning reliability method combining {K}riging constructed
  with exploration and exploitation of failure region and subset simulation.
\newblock {\em Reliab. Eng. Syst. Saf.\/}~{\em 188}, 90--102.

\bibitem[\protect\citeauthoryear{Zhang, Wang, and S{\o}rensen}{Zhang
  et~al.}{2019}]{Zhang2019a}
Zhang, X., L.~Wang, and J.~D. S{\o}rensen (2019).
\newblock {REIF: A novel active-learning function toward adaptive Kriging
  surrogate models for structural reliability analysis}.
\newblock {\em Reliab. Eng. Syst. Saf.\/}~{\em 185}, 440--454.

\bibitem[\protect\citeauthoryear{Zhang, Wang, and Sorensen}{Zhang
  et~al.}{2020}]{Zhang2020}
Zhang, X., L.~Wang, and J.~D. Sorensen (2020).
\newblock {AKOIS}: An adaptive {K}riging oriented importance sampling method
  for structural system reliability analysis.
\newblock {\em Structural Safety\/}~{\em 82}, 1--13.

\bibitem[\protect\citeauthoryear{Zhao, yue, Liu, Gao, and Zhang}{Zhao
  et~al.}{2015}]{Zhao2015}
Zhao, H., Z.~yue, H.~Liu, Z.~Gao, and Y.~Zhang (2015).
\newblock {An efficient reliability method combining adaptive importance
  sampling and Kriging metamodel}.
\newblock {\em Appl. Math. Model.\/}~{\em 39}, 1853--1866.

\bibitem[\protect\citeauthoryear{Zheng and Xue}{Zheng and Xue}{}]{Zheng2009}
Zheng, N. and J.~Xue.
\newblock {\em Manifold Learning}, Chapter~4, pp.\  87--119.
\newblock Springer, London.

\bibitem[\protect\citeauthoryear{Zhou and Peng}{Zhou and Peng}{2020}]{Zhou2020}
Zhou, T. and Y.~Peng (2020).
\newblock Structural reliability analysis via dimension reduction, adaptive
  sampling, and {Monte Carlo} simulation.
\newblock {\em Struct. Multidisc. Optim.\/}~{\em 62}, 2629--2651.

\end{thebibliography}

\appendix

\section{TNO Benchmark}\label{sec:TNO}
The TNO benchmark is a truly black-box benchmark of structural reliability analysis methods organized by TNO (Netherlands) in 2019 \citep{Rozsas2019}. It is a two-part challenge which \rev{consists} in a set of $16$ component- and $11$ system-reliability problems. It aims at assessing the efficiency and accuracy of various structural reliability methods. The limit-state functions were not known to the participants and were only accessible via an anonymous server API, \emph{i.e.}, the participants could only submit a set of sample points and the server would return the corresponding model evaluations.

The methods highlighted by the benchmark in this paper, \emph{i.e.}, a combination of PC-Kriging, subset simulation with overkill settings, deviation number $U$ and combined stopping criterion, were used to participate in the challenge. The results were disclosed in terms of accuracy and efficiency.
\begin{figure}[!ht]
	\centering
	\subfloat[Part 1: Component-reliability problems]{\label{fig:TNO_a_b}\includegraphics[width=0.75\textwidth]{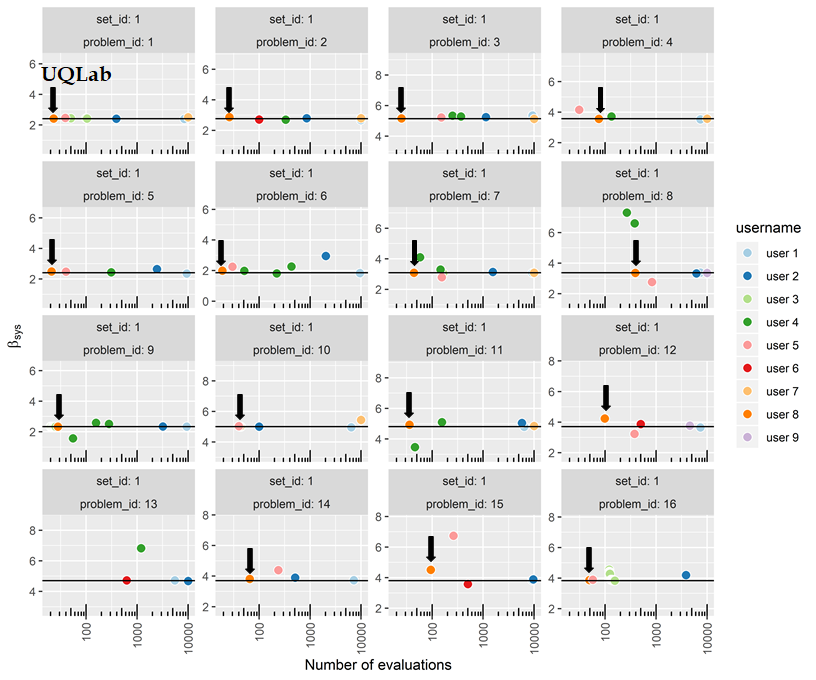}}%
	\\
	\subfloat[Part 2: System-reliability problems]{\label{fig:TNO_b}\includegraphics[width=0.75\textwidth]{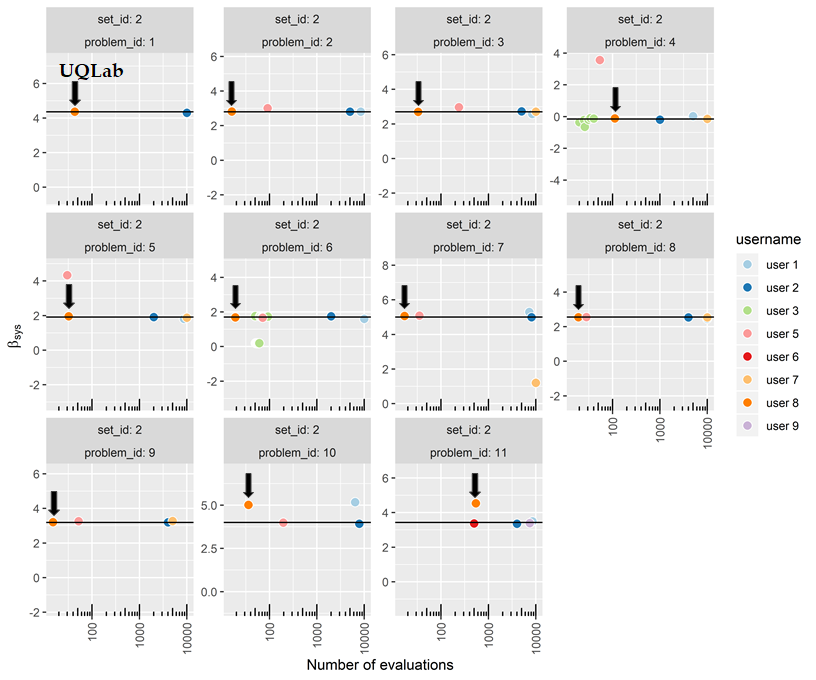}}%
	\caption{Results of the black-box reliability challenge as disclosed by \citet{Rozsas2019}. The results submitted using the approach highlighted in this paper are marked by the black arrow.}
	\label{fig:TNO}
\end{figure}

Figure~\ref{fig:TNO} shows the results submitted anonymously by the nine research groups that took part in the challenge. The black arrows point to the results submitted using our approach. Our approach turned out to be both the most accurate and efficient in $24$ out of $27$ problems. This confirms the potential of such a flexible framework for the solution of a wide variety of structural reliability problems.

\section{Supplementary materials}\label{sec:Suppl}

\subsection{Detailed list of problems}\label{app:Ref_Problems}
This section presents the different problems used in the benchmark. Only the transmission tower problems which were specifically developed for this benchmark are presented in details. The reader is referred to the mentioned references for further details on the other problems.
{\small
	\begin{table}[!ht]
		\centering
		\caption{Summary of the benchmark problems ($\#01$ to $\#11$ are from \citep{Rozsas2019}. $\#19$ \& $\#20$ are based on finite element models.)}
		\label{tab:ProbSum}
		\begin{tabular}{lccc}
			\hline 
			Problem & Dimension & $P_{f,\textrm{ref}}$ & Reference \\
			\hline
			01	(TNO RP 14) &	$5$	& $7.69 \, 10^{-4}$ & \citet{Rozsas2019}\\
			02	(TNO RP 24) & 	$2$ & 	$2.90 \, 10^{-3}$ & \citet{Rozsas2019} \\
			03	(TNO RP 28) & 	$2$ & 	$1.31 \, 10^{-7}$ & \citet{Rozsas2019} \\
			04	(TNO RP 31) & 	$2$	& $3.20 \, 10^{-3}$ & \citet{Rozsas2019} \\
			05	(TNO RP 38) &	 $7$	& $8.20 \cdot 10^{-3}$ & \citet{Rozsas2019}\\
			06	(TNO RP 53) & 	$2$	& $3.14 \cdot 10^{-2}$ & \citet{Rozsas2019}\\
			07	(TNO RP 54) & 	$20$ & 	$9.79 \cdot 10^{-4}$ & \citet{Rozsas2019} \\
			08	(TNO RP 63) & 	$100$ &	$3.77 \cdot 10^{-4}$ & \citet{Rozsas2019} \\
			09	(TNO RP 75) & 	$2$	& $9.80 \cdot 10^{-3}$ & \citet{Rozsas2019} \\
			10	(TNO RP 107) & 	$10$ & 	$ 2.85 \cdot 10^{-7}$ & \citet{Rozsas2019}\\
			11	(TNO RP 111) & 	$2$	& $7.83 \cdot 10^{-7}$ & \citet{Rozsas2019} \\
			12	(4-branch series) & 	$2$ & 	$3.85 \cdot 10^{-4}$ & \citet{Echard2011} \\
			13	(Hat function) & 	$2$	& $4.40 \cdot 10^{-3}$ & \citet{SchoebiASCE2016} \\
			14	(Damped oscillator)	& $8$	& $4.80 \cdot 10^{-3}$ & \citet{DerKiureghian1990} \\
			15	(Non-linear oscillator)	& $6$ & $3.47 \cdot 10^{-7}$ & \citet{Echard2011,Echard2013} \\
			16	(Frame)	& $21$	& $2.25 \cdot 10^{-4}$ & \citet{Echard2013}\\
			17	(HD function)	& $40$	& $2.00 \cdot 10^{-3}$ & \\
			18	(VNL function)	& $40$	& $1.40 \cdot 10^{-3}$ & \citet{Bichon2008,Sadoughi2017}\\
			19	(Transmission tower 1) & 	$11$	& $5.76 \cdot 10^{-4}$ &  See Section~\ref{app:TransTow}\\
			20	(Transmission tower 2) & 	$9$	& $6.27 \cdot 10^{-4}$ & See Section~\ref{app:TransTow}\\
			\hline
		\end{tabular}
	\end{table}
}

\subsubsection{Transmission tower}\label{app:TransTow}
The transmission tower example is originally developped within this paper. It is a three-dimensional finite element model consisting of $51$ nodes and $172$ bars. The bars are split in four groups, each characterized by their cross-sectional areas ($A_1$ to $A_4$) and constitutive materials Young's moduli ($E_1$ to $E_4$). The truss is subjected at its tip to a horizontal wind load $F$ whose deviation $\alpha$ from the lateral axis of the hands is random. At the extremity of the hands two vertical loads due to cables weight are added. All these parameters are random and described in Table~\ref{tab:TransTow}.

The finite element analysis is carried out in \textsc{Matlab} and failure is assumed:
\begin{itemize}
	\item for problem $\#19$, when the displacement at the tip of the tower is larger than $0.07$ m and;
	\item for problem $\#20$, when the maximum stress in any of the bars is larger than the yield stress $f_y$ defined in Table~\ref{tab:TransTow}.
\end{itemize}
\begin{figure}[!ht]
	\centering
	\includegraphics[trim={0.3cm 0.8cm 0.3cm 0.7cm},clip,width=0.35\textwidth]{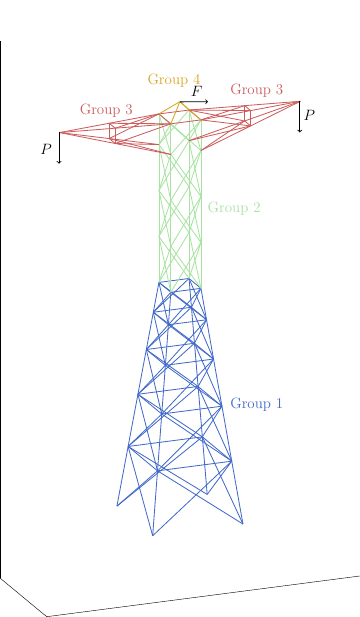}%
	\caption{Illustration of the transmission tower used for problems $\#19$ and $\#20$.}
	\label{fig:TransTow}
\end{figure}
\begin{table}[!ht]
	\centering
	\caption{Probabilistic model for the transmission tower problems.}
	\label{tab:TransTow}
	\begin{threeparttable} 
		\begin{tabular}{lccc}
			\hline 
			Parameter &	Distribution	& Mean & C.o.V\\
			\hline
			$A_1$ (m$^2$) & 	Gaussian & 	$10^{-3}$ & $0.05$ \\
			$A_2$ (m$^2$) & 	Gaussian & 	$10^{-3}$ & $0.05$ \\
			$A_3$ (m$^2$) & 	Gaussian & 	$5 \cdot 10^{-3}$ & $0.05$ \\
			$A_4$ (m$^2$) & 	Gaussian & 	$5 \cdot 10^{-3}$ & $0.05$ \\
			$E_1$ (GPa)\tnote{a} & 	Gaussian & 	$210$ & $0.05$ \\
			$E_2$(GPa)\tnote{a} & 	Gaussian & 	$210$ & $0.05$ \\
			$E_3$(GPa)\tnote{a} & 	Gaussian & 	$210$ & $0.05$ \\
			$E_4$(GPa)\tnote{a} & 	Gaussian & 	$210$ & $0.05$ \\
			$F$ (N) & 	Gumbel & 	$3.5 \cdot 10^4$ & $0.30$ \\
			$P$ (GPa) & 	Gumbel & 	$10^4$ & $0.30$ \\
			$\alpha$ (degrees) & 	Uniform\tnote{b} & 	$-30$ & $30$ \\
			$f_y$ (MPa)\tnote{c} & 	Lognormal & 	$355$ & $0.2$ \\
			\hline
		\end{tabular}
		{\footnotesize	
			\begin{tablenotes}
				\item[a] For problem $\#20$, $E= E_1 = E_2 = E_3 = E_4$;
				\item[b]  The values in the next columns are parameters of the Uniform distribution (minimum and maximum);
				\item[c] Only used for Problem $\#20$.
			\end{tablenotes}
		}
	\end{threeparttable}
\end{table}

\subsection{Algorithmic settings}\label{app:AlSet}
\subsubsection{Surrogate models}
\paragraph{Kriging}
Kriging also known as Gaussian process modelling considers the model to approximate as a realization of a stochastic Gaussian process made up of two parts \citep{Santner2003,Rasmussen2006}:
\begin{equation}\label{eq:KRG}
	\mathcal{M}\prt{\ve{x}}=  \ve{f}^T\prt{\ve{x}} \ve{\beta} + \sigma^2 \ve{Z}\prt{\ve{x}},
\end{equation}
where $\ve{f}\prt{\ve{x}}$ is a vector of regressors with their corresponding coefficients $\ve{\beta}$, $\sigma^2$ is the process constant variance and $\ve{Z}\prt{\ve{x}}$ a zero-mean, unit-variance stationnary Gaussian process.

This parametric form is calibrated by learning over an experimental design and the prediction for any unknown sample is given by the following analytical formula:
\begin{equation}\label{eq:KRG_MEAN}
	\mu_{\widehat{g}}\prt{\ve{x}} = \ve{f}^T \prt{\ve{x}} \widehat{\ve{\beta}} + \ve{r}^T\prt{\ve{x}} \mat{R}^{-1} \prt{\ve{y} - \mat{F}^T \widehat{\ve{\beta}}},
\end{equation}
where $\ve{f}^T \prt{\ve{x}} \widehat{\ve{\beta}}$ is a polynomial trend calibrated through least-square regression, $R$ is a parametric auto-correlation matrix, $\ve{r}\prt{\ve{x}}$ is a vector of cross-correlations between the point $\ve{x}$ and the experimental design points and $\mat{F}$ is the so-called observation matrix.
Interestingly, Kriging not only allows for prediction but can also provide a measure of its own accuracy through the following variance:
\begin{equation}\label{eq:KRG_STD}
	\sigma^2_{\widehat{g}} \prt{\ve{x}} = \sigma^2   \prt{1 - \ve{r}^T\prt{\ve{x}} \mat{R}^{-1} \ve{r}\prt{\ve{x}}  + \ve{u}^T\prt{\ve{x}} \prt{\mat{F}^T \mat{R}^{-1}\mat{F}}^{-1} \ve{u}\prt{\ve{x}} }
\end{equation}
where $\ve{u}\prt{\ve{x}} = \mat{F}^T \mat{R}^{-1} \ve{r}\prt{\ve{x}} - \ve{f}\prt{\ve{x}}$.

In this benchmark, we consider ordinary Kriging, meaning that the trend is an unknown constant as it is a common practice in GP modelling. Furthermore, we consider an anisotropic Gaussian auto-correlation function, the parameters of which are calibrated using maximum likehood estimation. The optimization is carried out using a genetic algorithm whose results are refined by a gradient-based solver.

The implementations in the Kriging module \citep{UQdoc_10_105} of \textsc{UQLab} \citep{MarelliUQLab2014} were used for the applications in this paper. 

\paragraph{Polynomial chaos expansions}
Polynomial chaos expansions result from a spectral expansion of a random variable $\ve{Y}$ onto a set of orthonormal polynomials \citep{Xiu2002}:
\begin{equation}\label{eq:PCE}
	Y = \mathcal{M}\prt{\ve{X}} = \sum_{\alpha \in \mathbb{N}^{M}} y_{\alpha} \Psi_{\alpha} \prt{\ve{X}}
\end{equation}
where $\Psi_{\alpha}$ are a set of multivariate polynomials orthornormal with respect to $f_{\ve{X}}$ and $y_{\alpha}$ are the coefficients representing the coordinates of the individual components of $\Psi_{\alpha}$ indexed by $\alpha \in \mathcal{N}^M$.

Building a PCE approximation requires two main steps. The first is to truncate the infinite expansions into a finite sum, which is achieved in this work using hyperbolic truncation ($q$-norm with $q=0.75)$ (\rev{See \citet{BlatmanPEM2010} for details}). Furthermore, the maximum degree is limited to $20$ and the interactions to the second order. The second step is to estimate the coefficients, which is achieved here using a regularized least-square minimization problem whose formulation allows for sparsity in the PC expansion. Practically, this problem is solved by the hybrid least-angle regression algorithm (LARS, \citet{Efron2004}) as proposed in \citet{BlatmanJCP2011}.

The implementations in the PCE module \citep{UQdoc_10_104} of \textsc{UQLab} \citep{MarelliUQLab2014} were used for the applications in this paper.

\paragraph{PC-Kriging}
PC-Kriging is a metamodellling technique obtained by combining polynomial chaos expansions and Kriging. More specifically, a PC-Kriging model is simply a universal Kriging model whose trend is a set of orthonormal polynomials \citep{SchoebiASCE2016}:
\begin{equation}\label{eq:PCK}
	\mathcal{M}\prt{\ve{x}}= \sum_{\alpha \in \mathcal{A}} y_{\alpha} \Psi_{\alpha} \prt{\ve{x}} + \sigma^2 \ve{Z}\prt{\ve{x}},
\end{equation}
where $\mathcal{A}$ is a finite set of multi-indices.

The calibration of the PC-Kriging model is carried out sequentially: first the terms of the (sparse) polynomial trend are detected with LARS then the Kriging metamodel is fitted (both the trend coefficients and the hyperparameters of the covariance kernel). The same algorithms and settings as in the two previous paragraphs are used. The only difference is the maximum polynomial degree, which is set here equal to $3$. 

The implementations in the PC-Kriging module \citep{UQdoc_10_109} of \textsc{UQLab} \citep{MarelliUQLab2014} were used for the applications in this paper.

\subsubsection{Reliability analysis}
\paragraph{Monte Carlo simulation}
Monte Carlo simulation is a direct integration of Eq.~\ref{eq:Pf} by sampling the probability density function $f_{\ve{X}}$. Given a population of size $N$, the MC estimate of $P_f$ reads:
\begin{equation}
	\widehat{P}_f = \sum_{k=1}^{N} \ve{1}_{\acc{\ve{x}:g\prt{\ve{x}} \leq 0}}\prt{\ve{x}^{(k)}} = \frac{N_{\textrm{fail}}}{N},
\end{equation}
where $N_{\textrm{fail}}$ is the number of failed samples.
The only parameter to calibrate here is the sample size. In this work, the simulation is carried out in batches of size $10^5$ 
until the coefficient of variation of the estimate:
\begin{equation}
	CoV = \sqrt{\frac{1-\widehat{P}_f}{N \widehat{P}_f }} 
\end{equation}
is smaller than a threshold arbitrarily set to $0.025$ or until the maximum sample size of $10^7$ is reached.

\paragraph{Importance sampling}
Importance sampling is a variance reduction simulation technique where the samples are generated following a proposal distribution rather than the original random variables PDF. The associated failure probability estimate therefore reads:
\begin{equation}
	\widehat{P}_f = \frac{1}{N}\sum_{k=1}^{N} \ve{1}_{\acc{\ve{x}:g\prt{\ve{x}} \leq 0}}\prt{\ve{x}^{(k)}} \frac{f_{\ve{X}}\prt{\ve{x}^{(k)}}}{h\prt{\ve{x}^{(k)}}},
\end{equation}
where the set $\acc{\ve{x}^{(1)} \enum \ve{x}^{(N)}}$ is sampled following the distribution $h\prt{\ve{x}}$.

In this work, the proposal distribution is simply a standard Gaussian centered around the most probable failure point as estimated using FORM. The maximum sample size is set to $10^4$.

\paragraph{Subset simulation} Subset simulation is another popular variance reduction technique obtained by solving a series of reliability problems with a relatively large target failure probability. Considering a set of nested events $\mathcal{D}_1 \supset \mathcal{D}_2 \enum \supset \mathcal{D}_m = \mathcal{D}_f$ defined such that $\mathcal{D}_k = \acc{\ve{x}: g\prt{\ve{x}} \leq t_k}$ with $t_1 > t_2 > \enum t_m = 0$, the failure probability can be recast as:
\begin{equation}
	P_f = \Prob{\mathcal{D}_f} = \Prob{\cap_{k=1}^{m} \mathcal{D}_k} = \Prob{\mathcal{D}_1} \prod_{i=1}^{m-1}\Prob{\mathcal{D}_{i+1}| \mathcal{D}_i} 
\end{equation}
While the initial failure probability $\Prob{\mathcal{D}_1}$ is estimated using crude Monte Carlo simulation, the remaining conditional failure probabilities are computed using Markov Chain Monte Carlo. The intermediate thresholds are set on-the-fly such that the $\Prob{\mathcal{D}_{i+1}| \mathcal{D}_i} = p_0$ is large enough. Traditionnally $p_0$ is set to $0.1$ but in this work, we consider $p_0 = 0.25$ and a batch size of $10^5$ for each subset level. This overkill setting allows us to better explore the random input space and results in a relatively small coefficient of variation of $\widehat{P}_f$.
\subsubsection{Learning functions}
In this work, we consider three different learning functions. The first two are associated to Kriging and PC-Kriging while the third one is only used with PCE. They respectively read:
\paragraph{Deviation number}
\begin{equation} 
	U\prt{\ve{x}} = \frac{\abs{\mu_{\widehat{g}}\prt{\ve{x}}}}{\sigma_{\widehat{g}}\prt{\ve{x}}}
\end{equation}
\paragraph{Expected feasibility function}
\begin{multline} \label{eqn:SRA:Theory:AK:LF:bichon}
	EFF(\ve{x}) = \mu_{\widehat{g}}(\ve{x}) \bra{ 2\Phi\prt{\frac{-\mu_{\widehat{g}}(\ve{x})}{\sigma_{\widehat{g}}(\ve{x})}} - \Phi\prt{ \frac{-\epsilon - \mu_{\widehat{g}}(\ve{x})}{\sigma_{\widehat{g}}(\ve{x})} } -\Phi\prt{ \frac{\epsilon - \mu_{\widehat{g}}(\ve{x})}{\sigma_{\widehat{g}}(\ve{x})} }  } \\
	- \sigma_{\widehat{g}}(\ve{x}) \bra{ 2\varphi\prt{\frac{-\mu_{\widehat{g}}(\ve{x})}{\sigma_{\widehat{g}}(\ve{x})}} - \varphi\prt{ \frac{-\epsilon - \mu_{\widehat{g}}(\ve{x})}{\sigma_{\widehat{g}}(\ve{x})} } - \varphi\prt{ \frac{\epsilon-\mu_{\widehat{g}}(\ve{x})}{\sigma_{\widehat{g}}(\ve{x})} } } \\
	+ \epsilon \bra{ \Phi\prt{ \frac{\epsilon - \mu_{\widehat{g}}(\ve{x})}{\sigma_{\widehat{g}}(\ve{x})} } - \Phi\prt{ \frac{-\epsilon-\mu_{\widehat{g}}(\ve{x})}{\sigma_{\widehat{g}}(\ve{x})} } },
\end{multline}
where $\epsilon = 2 \sigma_{\widehat{g}}(\ve{x})$ and $\phi$ and $\Phi$ are respectively the probability density function (PDF) and cumulative distribution function (CDF) of a Gaussian random variable.
\paragraph{Fraction of bootstrap replicates} This learning function, developed by \citet{MarelliAPSRRA2016}, is based on bootstrap replicates of a PCE model. First, $B$ PCE models are built using $B$ new experimental designs $\acc{\mathcal{E}^{(b)}, b = 1 \enum B}$ where each new ED is constructed by drawing samples with replacement from $\mathcal{E}$. Following this procedure, the learning function reads:
\begin{equation}
	U_{\textrm{FBR}} (\ve{x}) = \frac{\abs{B_s\prt{\ve{x}}-B_f\prt{\ve{x}}}}{B},
\end{equation}
where $B_s\prt{\ve{x}}$ and $B_f\prt{\ve{x}} \in \bra{0, \enum, B}$ are respectively the number of safe and failed PC-bootstrap replicates at the point $\ve{x}$.

No specific algorithmic settings are considered for these learning functions. Considering a candidate pool made up of samples drawn within the previous iterations of the reliability algorithm, the next point to add to the ED is simply chosen by selecting the one that minimizes $U$ (or $U_{\textrm{FBR}}$) and maximizes $EFF$.

\end{document}